\documentclass[aps,prd,10pt,notitlepage,nofootinbib]{revtex4-1}

\usepackage[utf8]{inputenc}
\usepackage{amsmath,amssymb,amsfonts}
\usepackage{newtxtext,newtxmath}

\usepackage{bbold}
\usepackage{bm}
\usepackage{graphicx}
\usepackage[usenames,dvipsnames]{xcolor}
\usepackage{color}
\usepackage[colorlinks=true,linkcolor=blue,urlcolor=blue,citecolor=blue]{hyperref}

\usepackage{slashed}
\usepackage[english]{babel}
\usepackage{dcolumn}
\usepackage{pifont}
\usepackage{dsfont,mathrsfs}
\usepackage{cancel}
\usepackage{bigints}
\usepackage{accents}
\usepackage{soul}
\usepackage{multirow}

\usepackage{amsmath, amssymb}
\usepackage{bbold}
\usepackage{makecell}
\usepackage{simpler-wick}

\usepackage{orcidlink} 

\newcommand{\nn}{\nonumber}

\newcommand{\MB}[1]{\left|#1\right|}
\newcommand{\mb}[1]{|#1|}
\newcommand{\FB}[1]{\left(#1\right)}
\newcommand{\fb}[1]{(#1)}
\newcommand{\SB}[1]{\left\{#1\right\}}
\newcommand{\TB}[1]{\left[#1\right]}
\newcommand{\AB}[1]{\left<#1\right>}

\newcommand{\scrL}{\mathscr{L}}

\newcommand{\munu}{{\mu\nu}}

\newcommand{\Psibar}{\overline{\Psi}}

\newcommand{\Phibar}{\overline{\Phi}}

\newcommand{\del}{\partial}

\newcommand{\fsl}{\slashed}

\newcommand{\intzinf}{\int_{-\infty}^{\infty}}

\newcommand{\half}{\dfrac{1}{2}}

\newcommand{\kzint}[1]{\int_{-\infty}^\infty \dfrac{d{#1}_z}{2\pi}}

\newcommand{\diag}{\text{diag}}

\newcommand{\potU}{\mathcal{U}\big(\Phi,\Phibar;T\big)}

\newcommand{\fnppbT}{\big(\Phi,\Phibar,T\big)}

\newcommand{\F}{\Phi}

\newcommand{\Fb}{\overline{\Phi}}

\newcommand{\paroneder}[2]{\frac{\partial {#1}}{\partial {#2}} }

\newcommand{\DerivativePartialOne}[2]{\frac{\partial {#1} }{\partial {#2}}}
\newcommand{\DerivativePartialTwo}[2]{\frac{\partial^2 {#1} }{\partial {#2}^2}}

\newcommand{\PL}{P_\parallel}
\newcommand{\PT}{P_\perp}
\newcommand{\Mag}{\mathcal{M}}

\begin{document}
	\title{Anisotropic pressure of magnetized quark matter with anomalous magnetic moment}

	\author{Nilanjan Chaudhuri\orcidlink{0000-0002-7776-3503}$^{a,d}$}
	\email{sovon.nilanjan@gmail.com}
	\email{n.chaudhri@vecc.gov.in}

	\author{Snigdha Ghosh\orcidlink{0000-0002-2496-2007}$^{b}$}
	\email{snigdha.physics@gmail.com}
	\thanks{Corresponding Author}
	
	\author{Pradip Roy$^{c,d}$}
	\email{pradipk.roy@saha.ac.in}
	
	\author{Sourav Sarkar\orcidlink{0000-0002-2952-3767}$^{a,d}$}
	\email{sourav@vecc.gov.in}
	
	\affiliation{$^a$Variable Energy Cyclotron Centre, 1/AF Bidhannagar, Kolkata - 700064, India}
	\affiliation{$^b$Government General Degree College Kharagpur-II, Paschim Medinipur - 721149, West Bengal, India}
	\affiliation{$^c$Saha Institute of Nuclear Physics, 1/AF Bidhannagar, Kolkata - 700064, India}
	\affiliation{$^d$Homi Bhabha National Institute, Training School Complex, Anushaktinagar, Mumbai - 400085, India}

\begin{abstract}
	 We investigate magnetic field $(eB)$ dependence of constituent quark mass, the longitudinal and transverse pressure as well as the magnetization and magnetic susceptibility of strongly interacting quark matter. We employ the two-flavour Polyakov Nambu--Jona-Lasinio model with the inclusion of the anomalous magnetic moment (AMM) of the quarks at finite temperature $(T)$ and finite quark chemical potential $(\mu_q)$ capturing different stages of chiral phase transition. We find that the transverse pressure, magnetization and magnetic susceptibility become highly oscillatory for large values of $eB$ in the chiral symmetry broken phase. However the oscillations cease to occur at higher values of $T$ and $\mu_q$ when chiral symmetry is (partially) restored and the anisotropic nature of the pressure becomes significant even at smaller values of $eB$. As the inclusion of AMM of the quarks leads to inverse magnetic catalysis of the transition temperature we observe that the variations of transverse pressure, magnetization and magnetic susceptibility are significantly modified in the vicinity of the chiral transition temperature. Furthermore, above the chiral transition temperature the magnetic susceptibility is found to remain positive for a wide range of $eB$ indicating a paramagnetic character of the strongly interacting quark matter. Finally, we have also examined the magnetism of strongly interacting matter in the quarkyonic phase. The obtained results could be useful for a magnetohydrodynamic evolution of hot and dense matter created in heavy-ion collisions.
\end{abstract}

\maketitle
\section{Introduction}
Study of hot and/or dense matter in the presence of strong magnetic field has attracted a wide spectrum of researchers from both theoretical as well as experimental domain in the last few decades~\cite{Kharzeev:2013jha,Miransky:2015ava,Andersen:2014xxa,Friman:2011zz,Bali:2011qj,STAR:2021mii,An:2021wof,Milton:2021wku,Kharzeev:2022hqz}. Numercal estimations suggest that, in non-central or asymmetric collisions of two heavy nuclei, very strong magnetic fields of the order $ \sim 10^{18} $ Gauss or larger might be generated due to the receding spectators~\cite{Kharzeev:2007jp,Skokov:2009qp}. These fields are, in principle, time dependent and their decay process gets sufficiently delayed due to the presence of a large electrical conductivity of the hot and dense magnetized medium~~\cite{Tuchin:2013apa,Tuchin:2015oka,Tuchin:2013ie,Gursoy:2014aka}. Apart from this, strong magnetic fields can also exist in several other physical environments. For example, in the interior of certain astrophysical objects called magnetars~\cite{Duncan:1992hi,Thompson:1993hn}, magnetic field $\sim 10^{15}$ Gauss can be present. Moreover, it is conjectured that, primodial magnetic fields as high as $ \sim 10^{23}$ Gauss might have been produced in the early universe during the electroweak phase transition driven by chiral anomaly~\cite{Vachaspati:1991nm,Campanelli:2013mea}. Since the strength of these magnetic fields is equivalent to the typical Quantum Chromodynamics (QCD) energy scale ($eB\sim \Lambda_\text{QCD}^2$), various microscopic and bulk properties of the strongly interacting matter could be significantly modified (see Refs.~\cite{Miransky:2015ava,Kharzeev:2013jha,Friman:2011zz} for recent reviews). Furthermore, the presence of a strong background magnetic field results in a large number of interesting physical phenomena~\cite{Kharzeev:2012ph,Kharzeev:2007tn,Chernodub:2010qx, Chernodub:2012tf} owing to the rich vacuum structure of the underlying QCD, e.g. the Chiral Magnetic Effect (CME)~\cite{Fukushima:2008xe,Kharzeev:2007jp,Kharzeev:2009pj,Bali:2011qj}, 	Magnetic Catalysis (MC)~\cite{Shovkovy:2012zn,Gusynin:1994re,Gusynin:1995nb,Gusynin:1999pq}, Inverse Magnetic Catalysis (IMC)~\cite{Preis:2010cq,Preis:2012fh}, Chiral Vortical Effect (CVE), vacuum superconductivity and superfluidity~\cite{Chernodub:2011gs,Chernodub:2011mc} {\it etc}.

The spontaneous breakdown of chiral symmetry and color confinement are the most fundamental characteristics of QCD vacuum in the low energy region. The majority of our current understanding of these non-perturbative aspects is obtained from Latiice QCD simulations in absence of baryon density~\cite{deForcrand:2006pv,Aoki:2006br,Aoki:2009sc,Bazavov:2009zn,Cheng:2007jq,Muroya:2003qs}. The situation is less explored at finite chemical potentials owing to the (in)famous sign problem in the Monte Carlo simulation~\cite{Muroya:2003qs,Splittorff:2006fu,Fukushima:2006uv}. Thus, as an alternative, one has to rely on the phenomenological models, which capture the basic aspects of QCD and are useful to evaluate the constituent quark mass, the pion mass, and so on. The Nambu$ - $Jona-Lasinio (NJL) model~\cite{Nambu:1961fr,Nambu:1961tp,Klevansky:1992qe,Vogl:1991qt,Buballa:2003qv,Volkov:2005kw} is one such example which respects the global symmetries of QCD, most importantly the chiral symmetry. The NJL model is known to be non-renormalizable because of point-like interaction among the quarks which appears as a consequence of integrating out the gluonic degrees of freedom in this effective description~\cite{Klevansky:1992qe,Bijnens:1995ww}. Thus, to tame the divergent integrals a proper regularization scheme has to be adopted which will also fix the model parameters by reproducing a set of phenomenological quantities, such as, the pion-decay constant ($ f_\pi $), quark condensate, pion mass ($ m_\pi $) and so on. However, the NJL model lacks confinement. Thus, in order to acquire a simultaneous description of the spontaneous breaking of chiral symmetry as well as confinement of the quarks within the effective model approach, the Polyakov loop extended Nambu$ - $Jona-Lasinio (PNJL) model model has been proposed by introducing interactions between quarks with a temporal, static and homogeneous gluon-like field~\cite{Ratti:2005jh,Ratti:2006wg}.

Significant literature can be found where the PNJL model has been employed to investigate the deconfinement transition and chiral symmetry restoration in the strongly interacting hot and dense magnetized matter~\cite{Kharzeev:2012ph,Andersen:2014xxa,Chaudhuri:2020lga,Gatto:2010qs,Fukushima:2008wg,Mattos:2021alf,Mattos:2021tmz,Wang:2022xxp}. It has been observed  that,  the presence of uniform background magnetic field is likely to catalyze (or strengthen) the chiral condensate implying magnetic catalysis (MC)~\cite{Gusynin:1994re,Gusynin:1999pq}. Consequently an increase in the transition temperature from chiral symmetry broken to the restored phase is also observed. The LQCD results agree at small values of temperature, however, at higher values of  the temperature an opposite trend is observed which leads to an overall decrease in the transition temperature.  This phenomena is known as inverse magnetic catalysis (IMC)~\cite{Bali:2011qj,Bali:2012av}. An extensive amount of effort has been dedicated to explain this discrepancy by incorporating appropriate modifications in the NJL-type models (see Refs.~\cite{Preis:2012fh,Bandyopadhyay:2020zte} for a review). For example, IMC is observed in~\cite{Ferreira:2013tba,Ferreira:2014kpa,Farias:2016gmy,Avancini:2018svs,Sheng:2021evj} with a magnetic-field-dependent coupling, which is determined using the results from LQCD simulations. In~\cite{Mao:2016fha}, it is demonstrated that, employing Pauli-Villiars regularization scheme and considering beyond mean field approximation, IMC can be achieved. In Refs.~\cite{Fayazbakhsh:2014mca,Chaudhuri:2019lbw,Chaudhuri:2020lga,Chaudhuri:2021skc,Chaudhuri:2021lui,Ghosh:2020xwp,Xu:2020yag,Mei:2020jzn,Aguirre:2021ljk,Ghosh:2021dlo,Farias:2021fci}, it is observed  that the consideration of nonzero values of the anomalous magnetic moment (AMM) of the quarks results in a decrease in chiral transition temperature implying IMC.

It is well known that, in the presence of a magnetic field ($ eB $), the energy-momentum tensor (EMT) shows anisotropies owing to the breaking of the spatial rotational symmetry. Now, in the local rest frame, if the spatial elements of the EMT are interpreted as the pressures due to the response of the thermodynamic potential of the system against compressions in the corresponding directions, then there is a difference induced by the orientation of the magnetic field~\cite{Chatterjee:2014qsa}. In Ref.~\cite{Bali:2013esa}, it has been demonstrated that, the derivatives of the partition function obtained at constant magnetic flux correspond to the spatial elements of the energy–momentum tensor where the directional difference becomes manifest.  Often these different elements are termed as longitudinal ($ P_\parallel $) and transverse ($ P_\perp $) pressures. These quantities in turn affect the equation of state (EoS) of the strongly interacting matter. Many papers in the literature have been dedicated to study the effects of magnetic field on the EoS of certain compact stars such as neutron star, quark star, hybrid star and so on, incorporating the anisotropic nature of the pressure leading to many  important consequences~\cite{Chatterjee:2014qsa,Canuto:1968apg,Martinez:2003dz,Noronha:2007wg,Huang:2009ue,Ferrer:2010wz,Strickland:2012vu,Dexheimer:2012mk,Sinha:2013dfa,PeresMenezes:2015ukv,Menezes:2015fla,Ferrer:2015wca,Avancini:2017gck,Ferrer:2019xlr}.  However, in the above studies, the system is considered without boundaries and the currents generating the magnetic field are not taken into account and whether the anisotropy can or cannot be compensated by the currents generated by the rotation of the star, a surface magnetization produced by boundary effects, etc., are the topics of ongoing discussion~\cite{Blandford_1982,Potekhin:2011eb,Ferrer:2011ze,Chatterjee:2014qsa}.

In the case of heavy ion collisions, the EoS is also a quantity of major importance for studying the thermodynamic properties as well as the time evolution of hot and dense matter created in the relativistic heavy ion collisions. During hydrodynamic evolution in absence of background magnetic field, one usually employs as EoS a parametrization from LQCD simulations~\cite{Huovinen:2009yb,Mamo:2012kqw,Guenther:2017hnx,Bazavov:2009zn,Bazavov:2017dsy,Bazavov:2017dus} for vanishing baryon density. However, at non-zero baryon density, one has to rely on an effective description of the strongly interacting matter to obtain the EoS. In this case, at finite $eB$ a magneto-hydrodynamical description is necessary to obtain the space-time integrated observables which has gained a lot of research interest in recent times~\cite{Roy:2015kma,Hernandez:2017mch,Denicol:2018rbw,Denicol:2019iyh}. Moreover, since the speed of sound in the medium is associated with oscillations in pressure, it is expected that the anisotropy in the pressure also affects the transmission of the sound along and perpendicular direction to the field. Presence of background field can also change the transport properties e.g. electrical conductivity~\cite{Hattori:2016cnt,Feng:2017tsh,Ghosh:2019ubc,Kalikotay:2020snc}, shear and bulk viscosities~\cite{Tuchin:2011jw,Hattori:2017qih,Kurian:2018dbn,Kurian:2018qwb,Kurian:2020kct} and so on. In~\cite{Karmakar:2019tdp}, anisotropic pressure of deconfined QCD matter in presence of strong magnetic field has been evaluated in hard thermal loop perturbation theory within one loop using lowest Landau level (LLL) approximations at high values of temperature.

The $ eB $-dependence of the EoS is also encoded in magnetic susceptibility ($ \chi_B $) which provides the knowledge about the strength of the induced magnetiztion of the QCD matter. Its sign distinguishes between diamagnet ($ \chi_B < 0 $), which expels the external field and paramagnet ($ \chi_B > 0 $), which favours the exposure to the background field energetically. Recently LQCD results show that the magnetized QCD matter exhibits diamagnetism at small values of temperature and paramagnetism  at high temperature~\cite{Bali:2012jv,Bali:2020bcn}. In Refs.~\cite{Xu:2020yag,Lin:2022ied}, efforts have been made to evaluate temperature dependence of $ \chi_B $ using NJL model and it is found that the properly normalized susceptibility follows the lattice results qualitatively at high temperature but it fails to reproduce the diamagnetic character  of the QCD matter in the chiral symmetry broken phase. 

In this work we shall numerically evaluate the transverse and longitudinal pressures, magnetization and magnetic susceptibility of a `strongly' interacting matter for three different stages of chiral phase transition in which we take into account  all the Landau levels. We show that the pressure (both the longitudinal and the transverse) is strongly dependent on both the magnetic field and the AMM of the quarks. It will also be shown 	that similar behaviour is obtained in case of magnetization and magnetic susceptibility. We will show that, the strong oscillatory behaviour seen in these previously mentioned quantities for high values of magnetic field at lower values of temperature, vanishes in the (partially) chiral symmetry restored phase. Finally we will also study the magnetic properties of newly proposed quarkyonic phase. The formalism part is described in Secs.~\ref{sec_Formalism} and \ref{sec_M_TD} followed by numerical results in Sec.~\ref{sec.results}. We summarize in Sec.~\ref{sec.summary} and in the Appendix, few expressions for important derivatives are provided.

\section{2-flavour Polyakov Nambu-Jona--Lasinio model in a hot and dense magnetized medium} 
\label{sec_PNJL}
\subsection{Formalism}\label{sec_Formalism}
The Lagrangian (density) for the two-flavour PNJL model considering the AMM of the quarks in presence of constant background magnetic field with quark chemical potential $ \mu_q $ is given by\cite{Fayazbakhsh:2014mca,Chaudhuri:2019lbw,Chaudhuri:2020lga}
\begin{equation}
\scrL = \Psibar(x)\Big(i\fsl{D}-m_0 + \gamma_0 \mu_q +\half\hat{a} e\sigma^\munu F_\munu  \Big)\Psi(x)  
+ G \Big\{ \big(\Psibar(x) \Psi (x)\big)^2 + \big( \Psibar(x) i\gamma_5 \vec{\tau} \Psi(x)\big)^2\Big\} - \mathcal{U} \big(\Phi,\Phibar;T\big) \label{PNJL_lagrangian}
\end{equation}
where the flavour $ f \in \{u,d \}$ and colour $ c\in \{r,g,b\} $ indices from the Dirac field $\Psi^{fc}(x)$ are dropped for conveience. In Eq.~\eqref{PNJL_lagrangian}, $m_0$ is current quark mass representing the explicit breaking of chiral  symmetry. We have considered $ m_u= m_d = m_0 $ to ensure isospin symmetry of the theory in absence of background magnetic field. The constituent quarks interact with the Abelian gauge field $ A^{\rm ext}_\mu  $ corresponding to the external magnetic field and the $ {\rm SU_C(3)} $
gauge field $\mathscr{A}_\mu  $ representing the non-trivial background due to the Polyakov loop by means of the covariant derivative
\begin{eqnarray}
D_\mu = \partial_\mu  -i\hat{Q}eA^{\rm ext}_\mu- i \mathscr{A}_\mu^a 
\end{eqnarray}
where $ e>0 $ denotes the absolute value of charge of an electron and $ \hat{Q} = \diag ( 2/3, -1/3) $ is the charge matrix in flavour space. 
We will choose the Landau gauge for which $A^{\rm ext}_\mu \equiv \FB{0,0,xB,0}$ that generates constant magnetic field $B\hat{z}$ along $\hat{z}$ direction. In Eq.~\eqref{PNJL_lagrangian}, the factor $\hat{a}= \hat{Q}\hat{\kappa }$,  where $ \hat{\kappa}={\diag}(\kappa_u,\kappa_d) $, is a matrix in the flavour space containing the AMM of the quarks; $ F_\munu= (\partial_\mu A^{\rm ext}_\nu- \partial_\nu A^{\rm ext}_\mu)  $ is the anti-symmetric electromagnetic field tensor and  $ \sigma^\munu=i[\gamma^\mu,\gamma^\nu]/2$. We will use the metric tensor $g^\munu = {\diag}\FB{1,-1,-1,-1}$ throughout this article.

The potential $ \mathcal{U}\big(\Phi,\Phibar;T\big) $ in the Lagrangian in Eq.~\eqref{PNJL_lagrangian} governs the dynamics of the traced Polyakov loop and its conjugate and is given by~\cite{Roessner:2006xn}
\begin{eqnarray}
\frac{\mathcal{U}\big(\Phi,\Phibar;T\big)}{T^4} = -\frac{1}{2}A(T)  \Phibar \Phi + B(T) \ln \Big\{ 1 - 6 \Phibar \Phi   + 4  \big(\Phibar^3 + \Phi^3\big)  - 3 \big(\Phibar \Phi\big)^2  \Big\}
\label{polyakov_potential}
\end{eqnarray} 
where, 
\begin{eqnarray}
A(T) = a_0 + a_1 \FB{\frac{T_0}{T}}+ a_2 \FB{\frac{T_0}{T}}^2 ~~~\text{and, } ~~~
B(T) = b_3 \FB{\frac{T_0}{T}}^3.
\label{eq.A.B}
\end{eqnarray}
Values of the different coefficients present in Eq.~\eqref{eq.A.B}, and the value of the scalar coupling $G$ appearing in Eq.~\eqref{PNJL_lagrangian} are provided in Table~\ref{Table_parameters} below~\cite{Roessner:2006xn,Fukushima:2010fe}:
\begin{table}[h]
	\caption{Parameter set for Polyakov potential}
	\begingroup
	\renewcommand*{\arraystretch}{1.5}
	\setlength{\tabcolsep}{8pt} 
	\begin{tabular} {cccccc}
		\hline \hline
		$ a_0 $ & $ a_1 $ &  $ a_2 $   & $ b_3 $ & $ T_0 $ (MeV)  \\
		\hline \hline
		$ 3.51 $ & $ -2.47$ & $ 15.2 $ & $ -1.75 $ & $ 270 $  \\
		\hline
	\end{tabular}
	\endgroup
	\label{Table_parameters}
\end{table}
\subsection{Constituent Quark Mass and Thermodynamics}\label{sec_M_TD}
Employing the mean field approximation on the Lagrangian in Eq.~\eqref{PNJL_lagrangian}, one can show that, the thermodynamic potential $(\Omega)$ for a  two-flavour PNJL model can be expressed as
\begin{eqnarray}\label{eq_Eff_Pot}
\Omega &=&   \frac{(M- m_0)^2}{4 G} + {\frac{B^2}{2}}+ \potU - 3 \sum_{nfs}  \frac{\mb{e_f B}}{2\pi } \kzint{p}\omega_{nfs}  \nn \\&& 
- T \sum_{nfs} \frac{\mb{e_f B}}{2\pi } \kzint{p}\TB{\ln g^{(+)}\big(\Phi,\Phibar,T\big) + \ln g^{(-)}\big(\Phi,\Phibar,T\big) }
\end{eqnarray}
where, $n \in \{0,\mathds{Z}^+\} $, $f\in\{u,d\}$ , $s \in \{\pm1\}$, $e_u = 2e/3$, $e_d = -e/3$ and
\begin{eqnarray}
g^{(+)}\big(\Phi,\Phibar,T,\mu_q\big) &=& 1 + 3 \big(  \Phi + \Fb e^ { -  ( \omega_{nfs} -\mu_q )/T} \big) e^ { -  ( \omega_{nfs} -\mu_q )/T} + e^ { - 3 ( \omega_{nfs} -\mu_q )/T}, \label{eq.gp}\\
g^{(-)}\big(\Phi,\Phibar,T,\mu_q\big) &=& g^{(+)}\big(\Phibar,\Phi,T,-\mu_q\big). \label{eq.gm}
\end{eqnarray}
Here $ \omega_{nfs} >0  $ represents the energy eigenvalues of the quarks considering the finite values of AMM and is given by~\cite{Fayazbakhsh:2014mca,Chaudhuri:2021lui}
\begin{eqnarray} \label{energy}
\omega_{nfs}^2 = p_z^2 + \Big\{\sqrt{   M^2  + (2n + 1 -s)  |e_f B|  }  - s \kappa_f e_f B\Big\}^2.
\label{eq.wnfs}
\end{eqnarray}

The constituent quark mass ($ M $) and the expectation values of the Polyakov loops $ \Phi  $ and $\Phibar$ can be evaluated self-consistently by using the following stationary conditions
\begin{equation}
\paroneder{\Omega}{M} = 0~~ ; ~~~~ \paroneder{\Omega}{\F} = 0 ~~;~~~~ \paroneder{\Omega}{\Fb} =0~.
\label{eq.stationary}
\end{equation} 
Now using Eq.~\eqref{eq_Eff_Pot} in Eq.~\eqref{eq.stationary}, we arrive at the so called gap equations:
\begin{eqnarray}
	M &=& m_0 + 6G \sum_{nfs} \frac{\mb{e_f B}}{2\pi } \intzinf \frac{dp_z }{2\pi}\frac{M}{\omega_{nfs} }~ \FB{ 1 - \frac{s \kappa_f e_f B}{M_{nfs}} } \SB{ \frac{}{}1-  f^{(+)} \fnppbT  - f^{(-)} \fnppbT} \label{Gap_M}, \\
	 \frac{\del\mathcal{U}}{\del\Phi} &=&  3T  \sum_{nfs}  \frac{\mb{e_f B}}{2\pi } \intzinf \frac{dp_z }{2\pi}\TB{\frac{e^{- ( \omega_{nfs} -\mu_q)/T }}{g^{(+)}}  + \frac{e^{- 2( \omega_{nfs} +\mu_q)/T }}{g^{(-)}}} \label{Gap_p}, \\
	\frac{\del\mathcal{U}}{\del\Phibar} &=&  3T \sum_{nfs}  \frac{\mb{e_f B}}{2\pi } \intzinf \frac{dp_z }{2\pi} \TB{\frac{e^{- 2( \omega_{nfs} -\mu_q)/T }}{g^{(+)}}  + \frac{e^{- ( \omega_{nfs} +\mu_q)/T }}{g^{(-)}} }\label{Gap_pb},
\end{eqnarray}
where, $M_{nfs} = \sqrt{   M^2  + (2n + 1 -s)|e_f B|  }$, 
\begin{eqnarray}
f^{(+)} \big(\Phi,\Phibar,T,\mu_q\big) &=& \frac{1}{g^{(+)}}\TB{ \big(\F + 2\Fb e^{-( \omega_{nfs}  -\mu_q ) /T}\big) e^{-( \omega_{nfs}  -\mu_q )/T } + e^{-3( \omega_{nfs}  -\mu_q )/T } }  ~,\label{eq.fplus} \\
f^{(-)} \big(\Phi,\Phibar,T,\mu_q\big) &=& f^{(+)} \big(\Phibar,\Phi,T,-\mu_q\big)~, \label{eq.fminus} 
\end{eqnarray}
and the arguments of the functions $g^{(\pm)} = g^{(\pm)}\big(\Phi,\Phibar,T,\mu_q\big)$ in Eqs.~\eqref{Gap_p}-\eqref{eq.fminus} are suppressed for brevity. The quantities $\frac{\del\mathcal{U}}{\del\Phi}$ and $\frac{\del\mathcal{U}}{\del\Phibar}$ appearing in the left hand side of Eqs.~\eqref{Gap_p} and \eqref{Gap_pb} are calculated from Eq.~\eqref{polyakov_potential} as
\begin{eqnarray}
\frac{\del\mathcal{U}}{\del\Phi} &=& -T^4 \SB{ \frac{1}{2}A(T)\Fb + 6B(T)\frac{\Fb  - 2 \F^2 + \big( \Fb \F \big) \Fb}{  1 - 6 \Fb \F + 4 \big( \F^3 + \Fb^3\big) - 3 \big( \Fb \F\big)^2 } }, \\
\frac{\del\mathcal{U}}{\del\Phibar} &=& -T^4 \SB{ \frac{1}{2}A(T)\F + 6B(T)\frac{\F  - 2 \Fb^2 + \big( \Fb \F \big) \F}{  1 - 6 \Fb \F + 4 \big( \F^3 + \Fb^3\big) - 3 \big( \Fb \F\big)^2 } }.
\end{eqnarray}

Notice that in Eqs.~\eqref{eq_Eff_Pot} and ~\eqref{Gap_M}, the medium independent integrals are ultraviolet divergent. It is well-known that, PNJL model is non-renormalizable owing to the point-like interaction between the quarks~\cite{Klevansky:1992qe}. Hence we have to choose a regularization method to get rid of these divergent integrals. In this work, we will choose the smooth cutoff regularization procedure following~\cite{Fukushima:2010fe} and introduce a multiplicative form factor 
\begin{equation}
f_\Lambda (p_z,B) = {\frac{\Lambda^{2N}}{\Lambda^{2N} + \SB{p_z^2 + (2n +1 -s)|e_f B|}^N }}
\label{FormFactor}
\end{equation}
in the diverging vacuum integrals leaving the convergent medium dependent part unchanged. In Eq.~\eqref{FormFactor}, one has to choose the parameters $ N, \Lambda $  correctly such that the phenomenological vacuum values of pion decay constant $f_\pi$, chiral condensate and pion mass are reproduced.

We now specify the expressions for several thermodynamic quantities in the following. The longitudinal pressure is given by 
\begin{equation}
 P_\parallel(T,\mu_q)= - \Omega( M, \F, \Fb; T, \mu_q ).
 \label{Pressure}
\end{equation}
Using the identities given in Appendix~\ref{Appendix1}, the magnetization $\mathcal{M} =  -\frac{\partial \Omega}{\partial B}$ of the system comes out to be 
\begin{eqnarray}
\mathcal{M} &=&  - B+ 3 \sum_{nfs}  \frac{|e_f|}{2\pi } \kzint{p}\omega_{nfs} 
+ T \sum_{nfs} \frac{\mb{e_f }}{2\pi } \kzint{p}\TB{\ln g^{(+)} + \ln g^{(-)} } \nn \\ && 
+ 3\sum_{nfs} \frac{\mb{e_f B}}{2\pi } \intzinf \frac{dp_z }{2\pi}\frac{1}{\omega_{nfs} }~ \FB{ 1 - \frac{s \kappa_f e_f B}{M_{nfs}} }  \SB{ \FB{\frac{2n + 1 -s}{2}} \MB{e_f} - s \kappa_f e_f M_{nfs}  }\SB{ \frac{}{}1-  f^{(+)}   - f^{(-)} }  \nn \\ &&
 - 3 \sum_{nfs} \frac{\mb{e_f}}{2\pi } \kzint{p}\omega_{nfs} \dfrac{2 n N \mb{e_f}\SB{p_z^2 + (2n + 1 -s )\mb{e_f B} }^{N-1} \Lambda^{2N} }{ \SB{\FB{p_z^2 + (2n + 1 -s )\mb{e_f B} }^{N} + \Lambda^{2N}}^2}.
 \label{Eq_Mag}
\end{eqnarray}
Using magnetization, now one can define the transverse pressure as
\begin{equation}\label{PT}
P_\perp  = P_\parallel - \mathcal{M} B.
\end{equation}

The knowledge about the strength of the induced magnetiztion is provided by the magnetic susceptibility which is defined as $ \chi_B =-  \DerivativePartialTwo{\Omega}{B}  = \DerivativePartialOne{\Mag}{B} $.  Now starting from  Eq.~\eqref{Eq_Mag} it can be shown that
\begin{eqnarray}
\chi_B = \chi_B^{(0)} + \mathcal{A}_{M,B} \FB{\DerivativePartialOne{M}{B}}+ \mathcal{A}_{\Phi,B} \FB{\DerivativePartialOne{\Phi}{B}}+ \mathcal{A}_{\Phibar,B} \FB{\DerivativePartialOne{\Phibar}{B}}
\label{eq.sus}
\end{eqnarray}
where the explicit expressions of the coefficients $\chi_B^{(0)}$, $\mathcal{A}_{M,B}$, $\mathcal{A}_{\Phi,B}$ and $\mathcal{A}_{\Phibar,B}$ are provided in Appendix~\ref{Appendix2}.

Derivatives of $ M, \F, \Fb $ with respect to $ B $ can be evaluated by taking derivatives of the gap equations given by Eqs.~\eqref{Gap_M}, \eqref{Gap_p} and \eqref{Gap_pb} respectively. The similar exercise of finding $ T $-derivatives of several quantities are discussed in~\cite{Chaudhuri:2020lga}.
\section{Numerical Results}
\label{sec.results}
In this section, we present numerical results for constituent quark mass and several other thermodynamical quantities i different physical situations. As already pointed out in Sec. II B, owing to the four-fermion contact interaction among the quarks, PNJL model is known to be non-renormalizable and we have used a smooth cutoff scheme following Ref.~\cite{Fukushima:2010fe} to regularize the divergent vacuum integrals. Our model parameters are the following : $ \Lambda =  $ 620 MeV, $ G\Lambda^2 = 2.2 $ and $ m_0 $ = 5.5 MeV, which are chosen such that we get vacuum values of pion decay constant $ f_\pi = 92.4 $~MeV, mass of pions $ m_\pi = 138 $ MeV and chiral condensate $ \AB{\bar{u}u}^{1/3} = -245.7  $ MeV. The values of AMM of the quarks are $ \kappa_u $ = 0.29016, $ \kappa_d $ = 0.35986 in units of GeV$^{-1} $ following~\cite{Fayazbakhsh:2014mca}.

\begin{figure}[h]
		\includegraphics[angle=-90,scale=0.35]{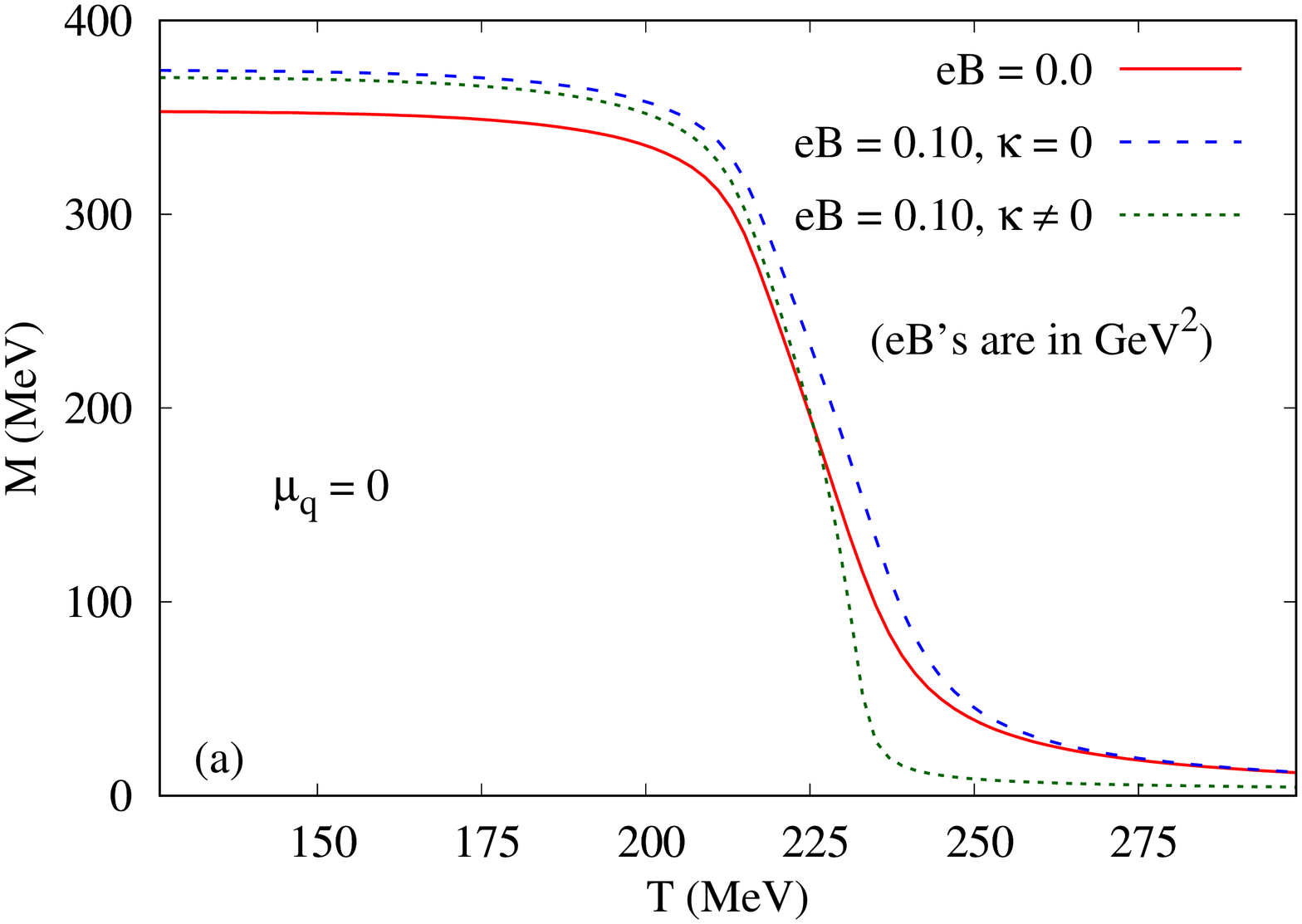} 
		\includegraphics[angle=-90,scale=0.35]{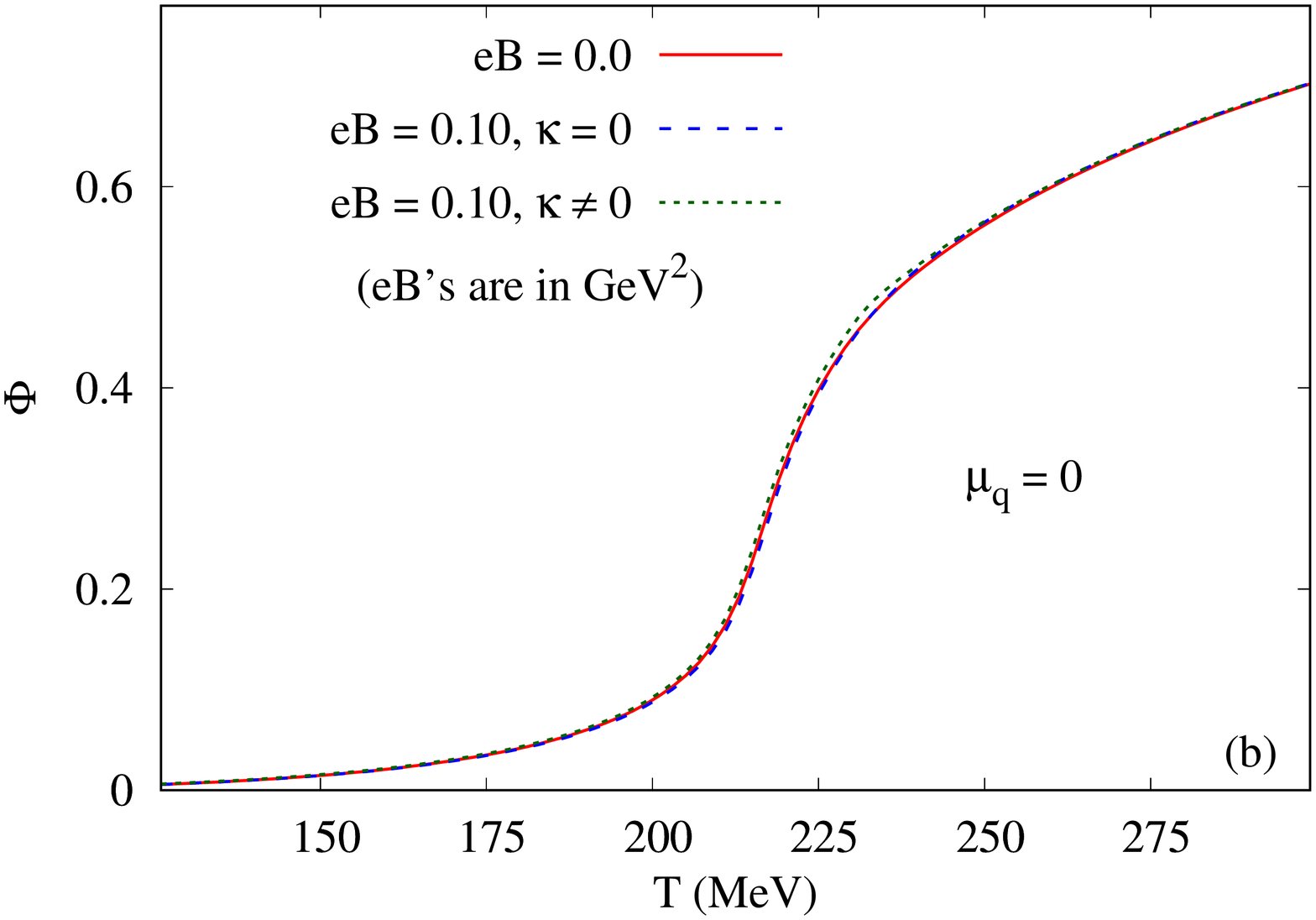}
		\includegraphics[angle=-90,scale=0.35]{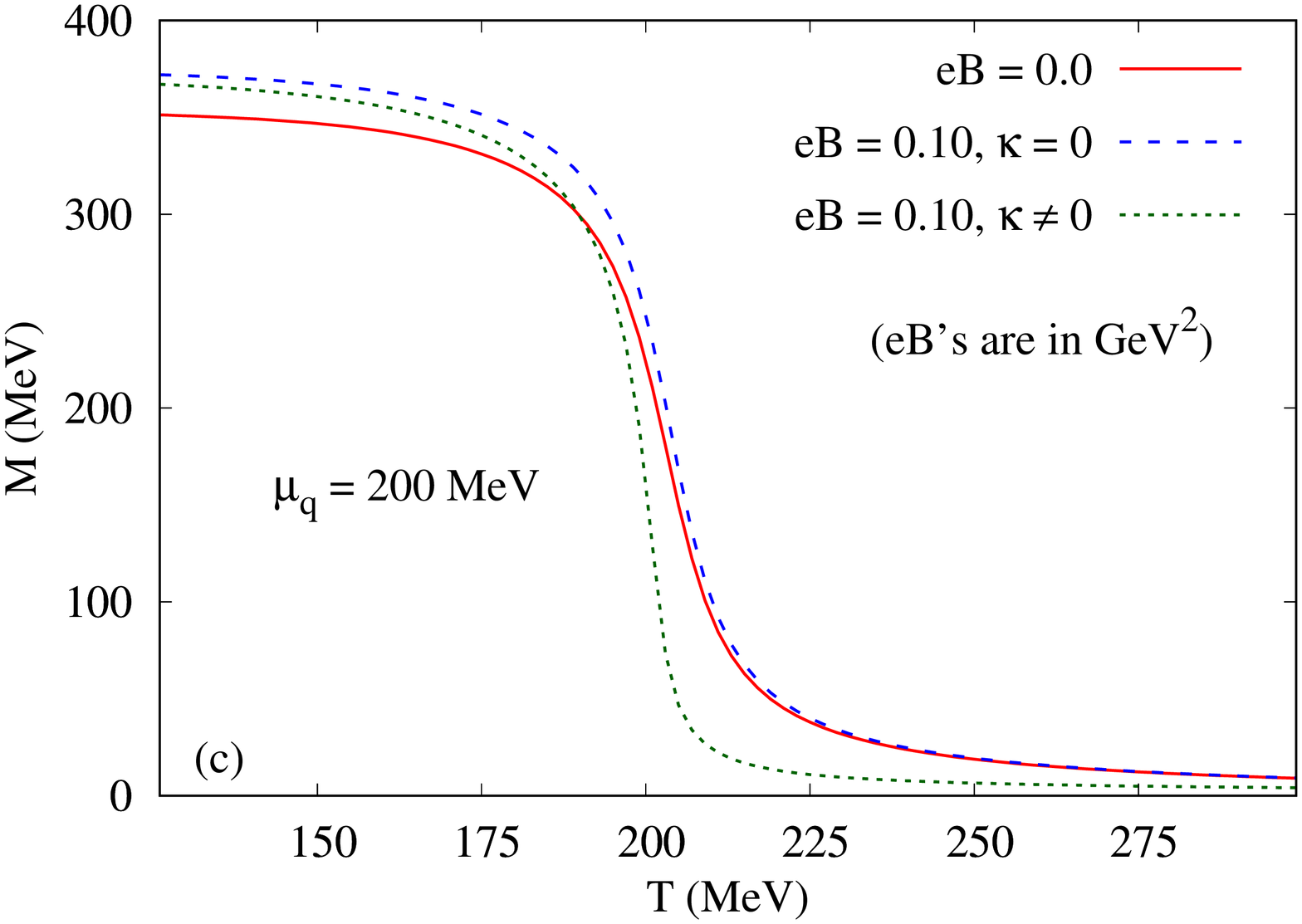} 
		\includegraphics[angle=-90,scale=0.35]{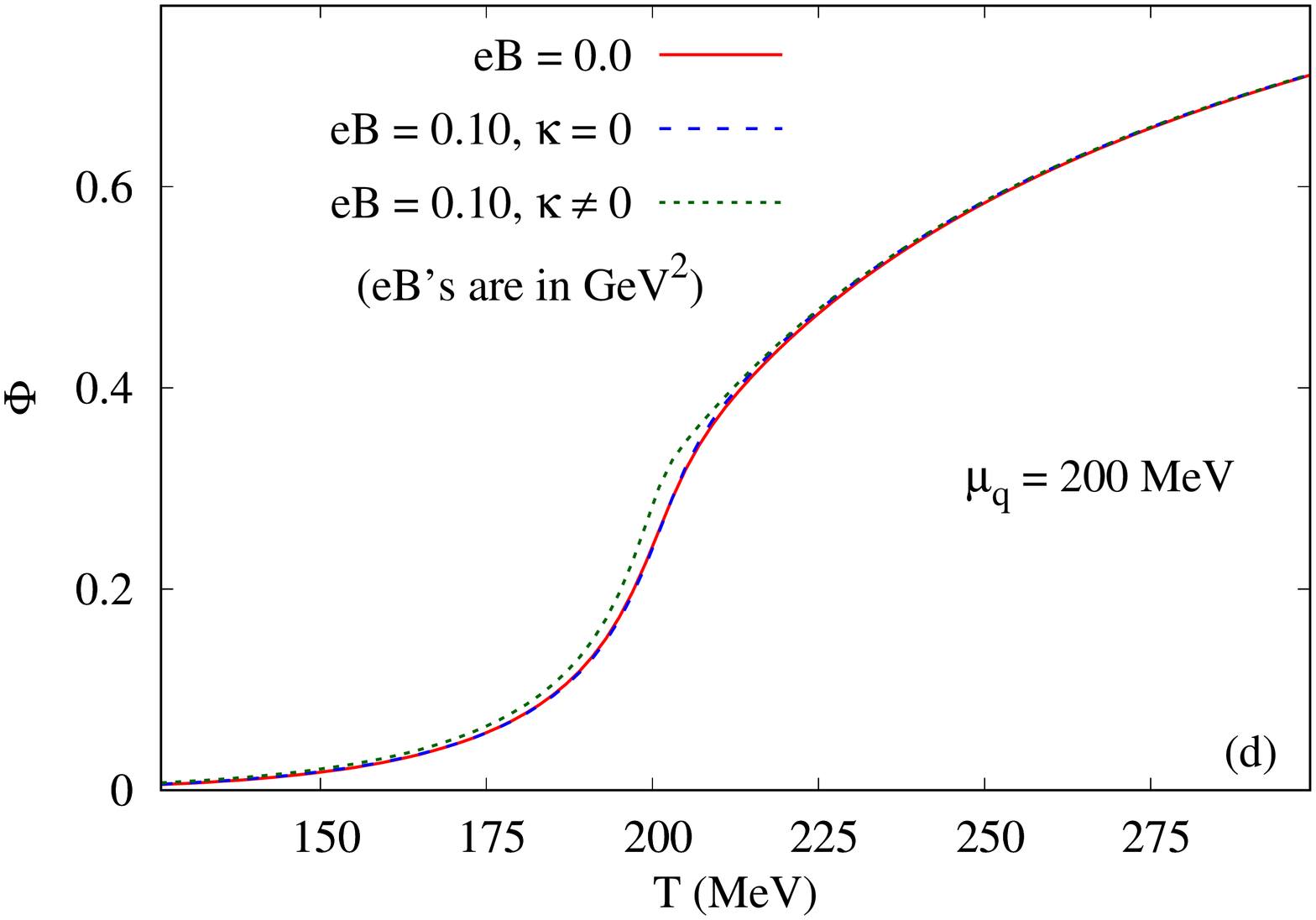}
	\caption{(Color Online) The variation of constituent quark mass ($ M $) as a function of temperature at (a) $ \mu_q =0$ and (c) $ \mu_q =200$ MeV. The variation of the expectation value of Polyakov loop ($ \Phi $) as a function of temperature at (b) $ \mu_q =0$ and (d) $ \mu_q =200$ MeV. Results are shown for different values background magnetic field ($ eB $) for both zero and non-zero values of the AMM of the quarks.}
	\label{Fig_MPhi_T}
\end{figure}

We first briefly discuss the numerical results for temperature dependence of the constituent quark masses ($ M $) and the expectation value of Polyakov loop ($ \Phi $) for different values of external parameters. In Figs.~\ref{Fig_MPhi_T}(a) and (b) we have shown the variation of $ M $ and $ \Phi $ as a function of temperature in the absence of $ \mu_q $ for both zero and nonzero values of the background magnetic field with and without considering the finite values of AMM of the quarks.
 The constituent mass starts from a high value at small $ T $, remains almost unaltered at the lower values of $ T $, drops smoothly in a small range of temperature and eventually,  at high $ T $ values, become nearly equal to the bare masses of the quarks illustrating the (partial) restoration of the chiral symmetry. In absence of background magnetic field this transition occurs at $ T_{0}^{ \rm~ch}\sim 229  $~MeV. On the contrary, the expectation values of Polyakov loop remains vanishingly small at lower values of $ T $, then it starts increasing around $T \approx 150 $ MeV and finally becomes close to unity at sufficiently high values of $ T $ indicating the transition from confined to deconfined phase of matter. At $ eB = 0 $, this transition occurs at $ T^{\rm~decon}_{0}\sim 217  $~MeV. Concentrating on Fig.~\ref{Fig_MPhi_T}(a) one can observe that, for non-zero values of the background magnetic field, in the absence of the AMM of the quarks, the magnitude of $ M $ increases at small values of $ T $ and the transitions from the chiral symmetry broken to the restored phase take place at higher values of temperature. This phenomenon is known as magnetic catalysis (MC)~\cite{Shovkovy:2012zn,Gusynin:1994re,Gusynin:1995nb,Gusynin:1999pq}, which explains that the magnetic field has a strong tendency to enhance (or catalyze) spin-zero fermion-antifermion condensates $ \langle\overline{\Psi}\Psi\rangle $. As we turn on finite values of the AMM of the quarks, the behaviour of $ M $ in low temperature region remains unchanged, but there is a slight decrease in the transition temperature implying inverse magnetic catalysis (IMC). The inclusion of finite values of the AMM of the quarks however does not bring any appreciable change in the $ T $-dependence of $ \Phi $ which is evident from Fig.~\ref{Fig_MPhi_T}(b). In Figs.~\ref{Fig_MPhi_T}(c) and (d), we have depicted the variation of $ M $ and $ \Phi $ as a function of temperature at $ \mu_q = 200 $~MeV keeping all the other parameters same as in Figs.~\ref{Fig_MPhi_T}(a) and (b). We observed further decrease in both chiral as well as deconfinement transition temperatures compared to the $ \mu_q =0 $ case but all the other qualitative features remaining same.
\begin{figure}[h]
	\includegraphics[angle=-90,scale=0.233]{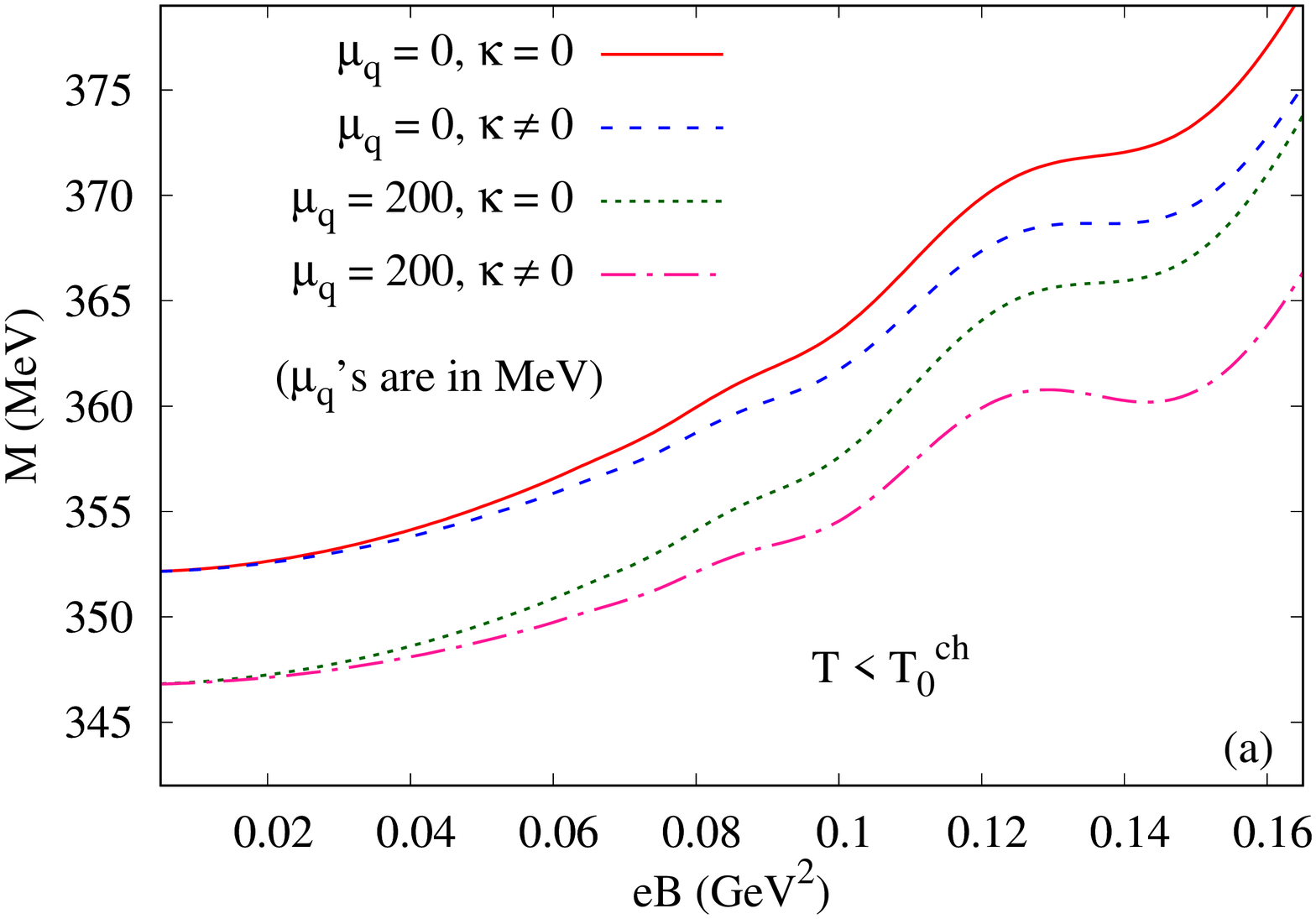} 
	\includegraphics[angle=-90,scale=0.233]{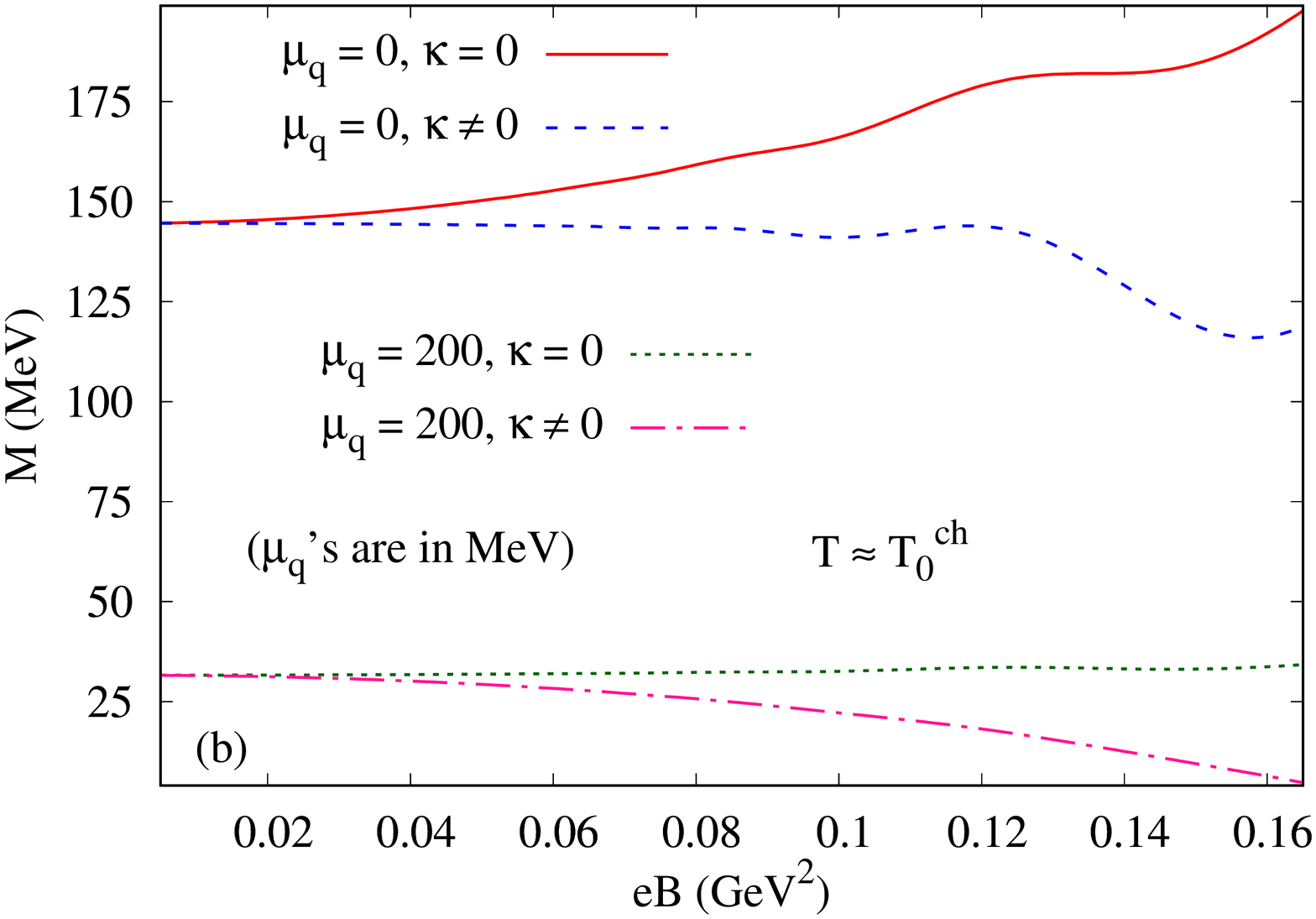}
	\includegraphics[angle=-90,scale=0.233]{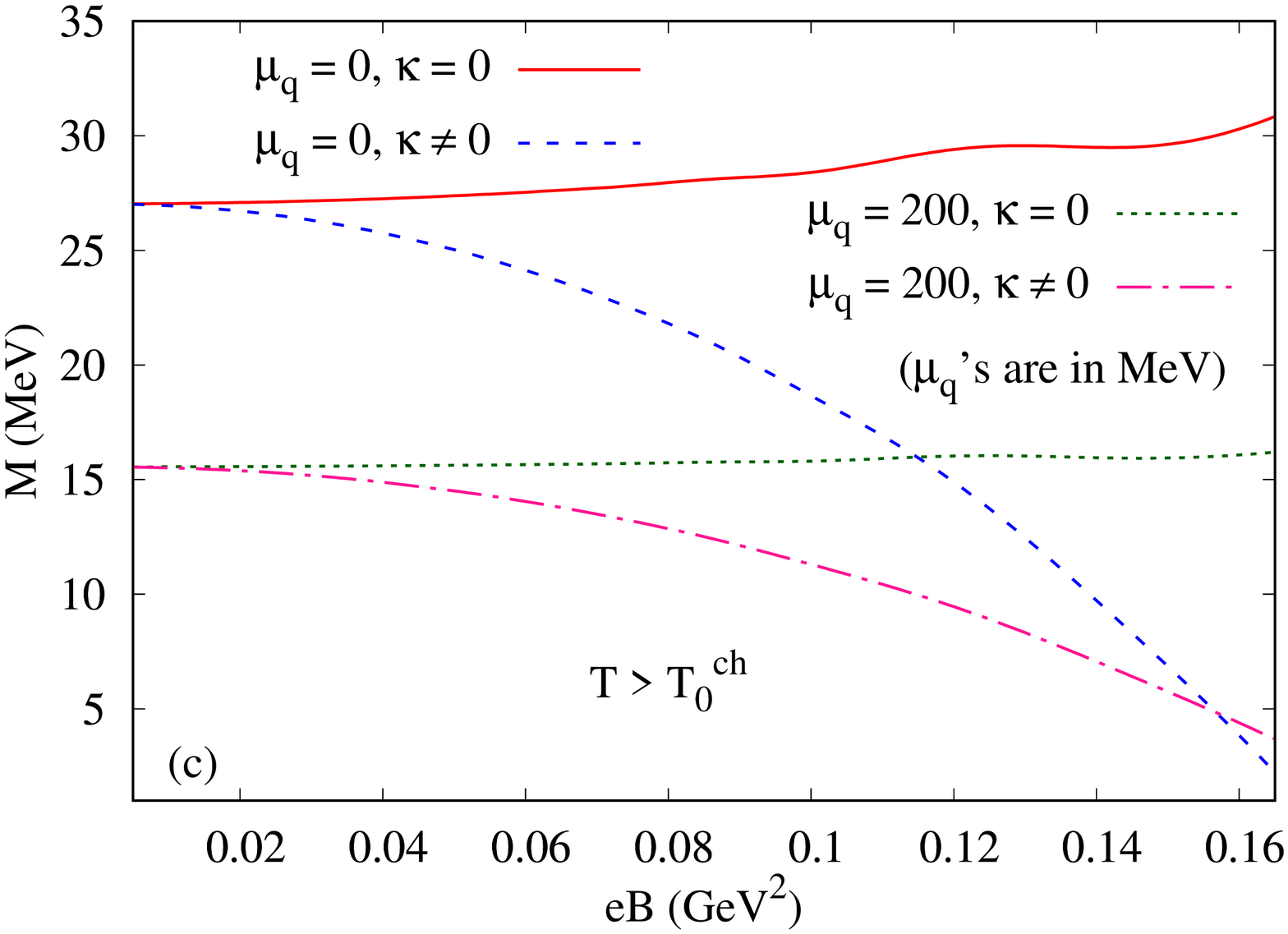}
	\caption{(Color Online) Constituent quark mass as a function of $ eB $ for different values of $ \mu_q $ with and without considering the AMM of the quarks at (a) $ T<  T_{0}^{ \rm~ch}$, (b) $ T\approx T_{0}^{ \rm~ch}$ and (c) $ T >  T_{0}^{ \rm~ch}$.}
	\label{Fig_Mass_eB}
\end{figure} 

In the remaining part of this section, we will study the background magnetic field dependence of constituent mass and several thermodynamic quantities in the presence as well as absence of quark chemical potential and the AMM of the quarks at three different values of temperature capturing different stages of chiral phase transition. These are as follows:
\begin{enumerate}
	\item $ T<  T_{0}^{ \rm~ch}$ : This will represent the scenario when chiral symmetry is broken. Analysing Figs.~\ref{Fig_MPhi_T}(a) to (d), we will choose $ T = 150 $~MeV to examine this situation.
	\item $ T\approx   T_{0}^{ \rm~ch}$ : Here we will choose $ T = 230 $ MeV to study the behaviour of strongly interacting matter in the vicinity of chiral phase transition.
	\item $ T >  T_{0}^{ \rm~ch}$ : We will consider $ T = 260 $ MeV which will correspond the (partial) restoration of the chiral symmetry.
\end{enumerate}

In Figs.~\ref{Fig_Mass_eB}(a), (b) and (c), we have plotted $ eB $-dependence of $ M $ at $ \mu_q = 0$ and  $ 200  $ MeV respectively with and without considering the AMM of quarks for three different values of temperature representing three different stages of the chiral phase transition mentioned above. From Fig.~\ref{Fig_Mass_eB}(a), it is evident that, in chiral symmetry broken phase, the constituent quark mass increases with the increasing values of $ eB $ for both zero and non-zero values of the AMM of the quarks although the magnitude  of $ M $ is marginally small in finite AMM case for both zero and non-zero values of $ \mu_q $. Now for a particular value of $ eB $, the magnitude of $ M $ is smaller in the presence of $ \mu_q $ compared to its absence which is understandable from Figs.~\ref{Fig_MPhi_T}(a) and (c). In the vicinity of transition temperature, at $ \mu_q =0$, the constituent quark mass increases with $ eB $ in the absence of the AMM of the quarks. However, as the finite values of the AMM of the quarks are taken into consideration, an opposite behaviour is observed which is evident from
 Fig.~\ref{Fig_Mass_eB}(b).  Now as we increase the temperature further so that the chiral symmetry is (partially) restored, inclusion of the AMM of the quarks results in further decrease in constituent quark mass as function of $ eB $, as can be seen from Fig.~\ref{Fig_Mass_eB}(c). Moreover, when finite values of quark chemical potential is considered  in  the vicinity as well as above the chiral transition temperature (see Figs.~\ref{Fig_Mass_eB}(b) and (c) respectively), similar qualitative behaviour of the $ eB $-dependence of the constituent quark mass is observed. Finally the difference between the magnitude of constituent quark mass for zero and nonzero values of the AMM of the quarks increases with increasing values of $ eB $ in all the cases. 
\begin{figure}[h]
	\begin{center}
		\includegraphics[angle=-90,scale=0.35]{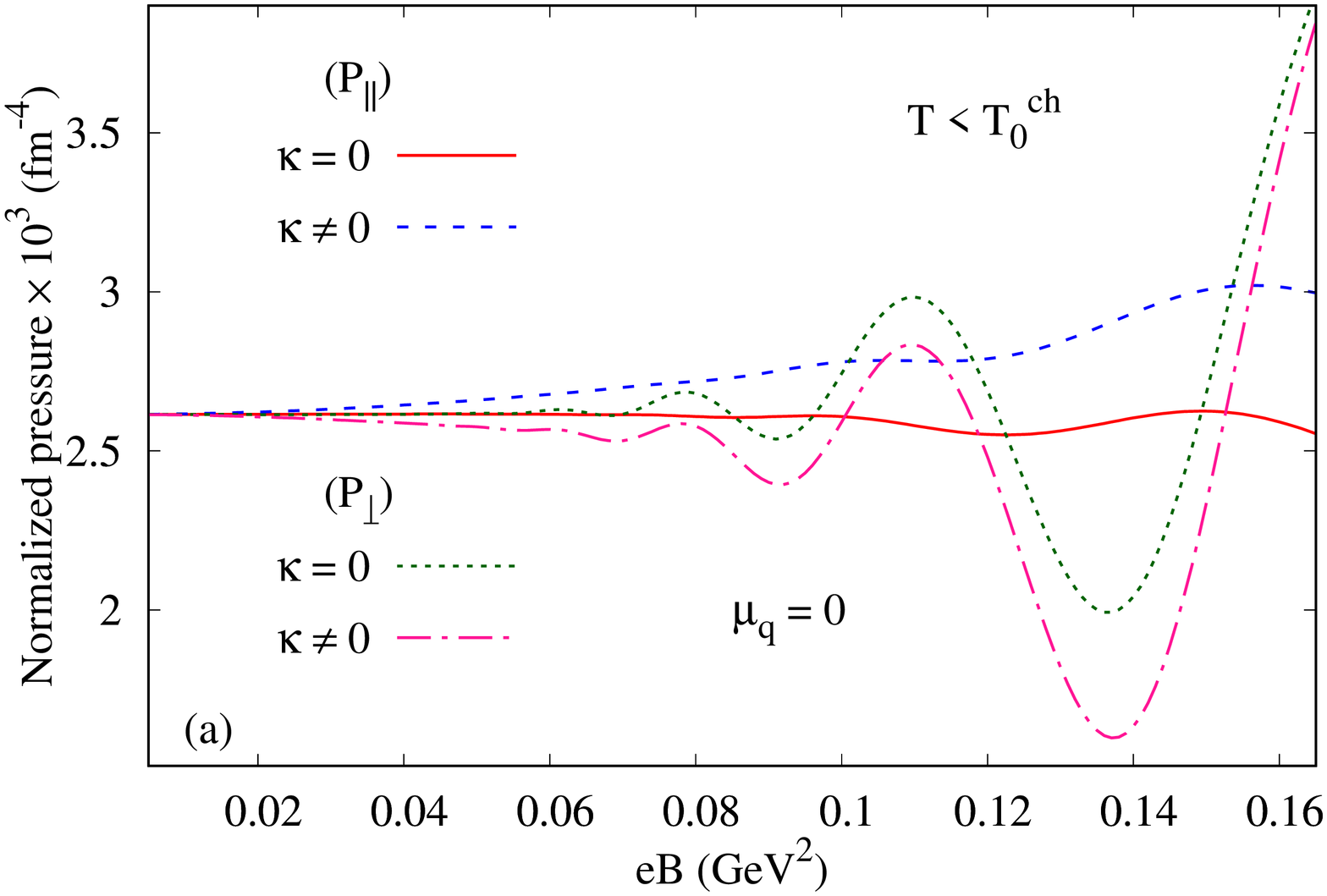} 
		\includegraphics[angle=-90,scale=0.35]{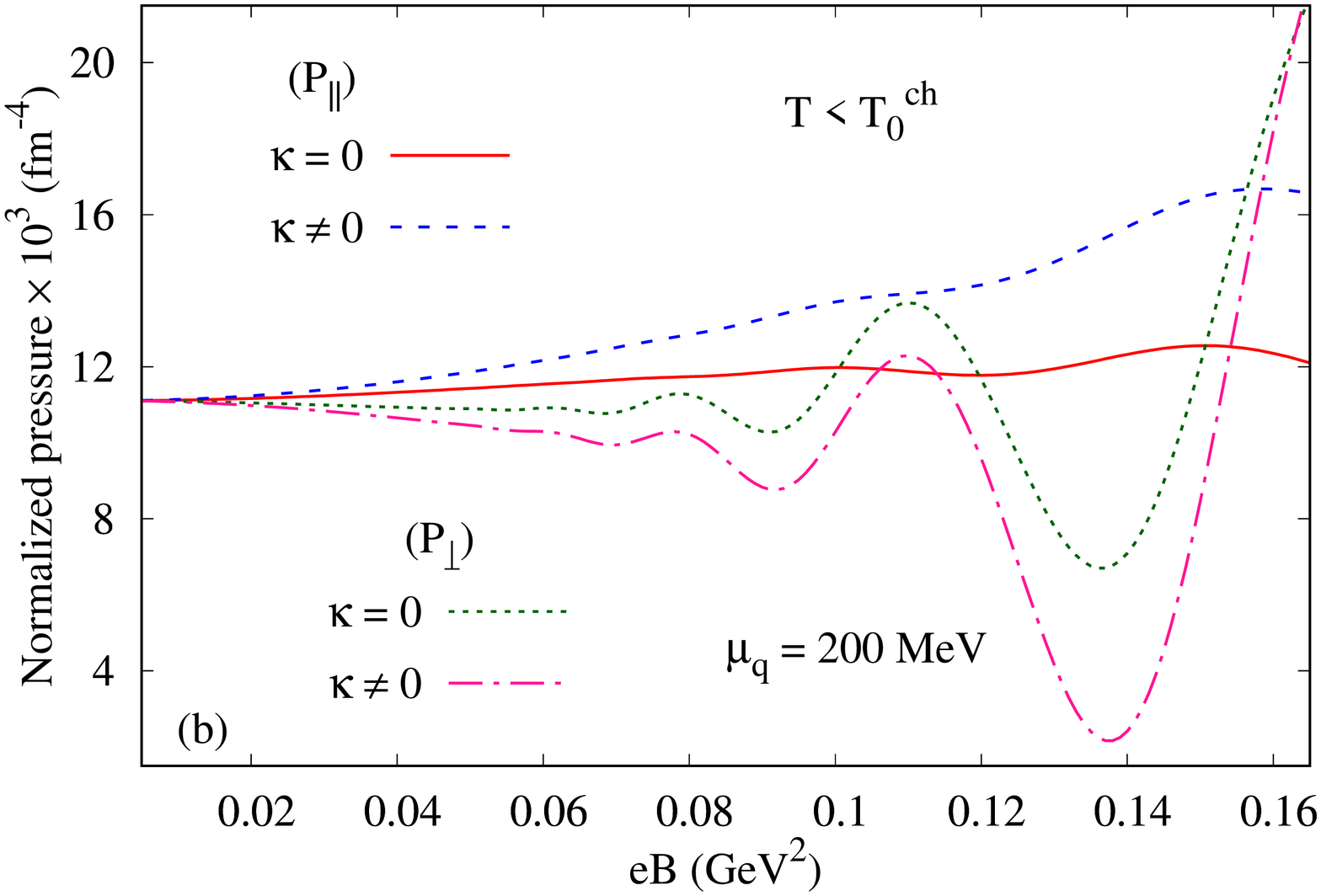}
	\end{center}
	\caption{(Color Online) Normalized longitudinal ($ P_\parallel $) and transverse ($ P_\perp $) pressure as a function of $ eB $ for $ T<  T_{0}^{ \rm~ch}$ with both vanishing and finite values of the AMM of the quarks at (a) $ \mu_q =0$ and (b) $ \mu_q =200$ MeV.}
	\label{Fig_Pressure1_eB}
\end{figure}

Next we focus on the $ eB $-dependence of several normalized thermodynamic quantities for different values of temperatures and quark chemical potentials. The scaling  of any thermodynamic quantity $ \Xi (T, \mu_q, eB) $ is done in the usual manner:
\begin{equation}
\Xi_N\FB{T,\mu_q,eB} \equiv \Xi\FB{T,\mu_q,eB} - \Xi \FB{T = 0, \mu_q, eB}.
\end{equation} 
For convenience, we will omit the subscript $ N $ as all the quantities discussed below are normalized in the above way unless otherwise specified. 

In Figs.~\ref{Fig_Pressure1_eB}(a) and (b) we have presented the variation of normalized longitudinal ($ P_\parallel $) and transverse ($ P_\perp $) pressures as a function of $ eB $ for $ \mu_q = 0 $ and 200 MeV respectively in the chiral symmetry broken phase considering both vanishing and finite values of the AMM of the quarks. Concentrating on Fig.~\ref{Fig_Pressure1_eB}(a), it can be observed that for small values of $ eB $, the longitudinal pressure almost coincides with transverse pressure independent of the consideration of the AMM of the quarks. But for $ eB > 0.03 $ Gev$ ^2 $, the pressures along the magnetic field and traverse to it, begin to be different. At higher values of magnetic field, $\PL  $ shows slight oscillation although its magnitude remains almost unchanged when the  AMM of the quarks is switched off. However, an overall increase in the magnitude of longitudinal pressure can be observed in finite AMM case. On the other hand, the transverse pressure, becomes highly oscillatory for large values of $ eB $ and the amplitude of oscillation is higher when AMM of the quarks are taken into consideration. From Fig.~\ref{Fig_Pressure1_eB}(b), it is evident that, for finite values of quark chemical potential, the variation of $ \PL $ and $ \PT $ are qualitatively similar to the case when $\mu_q$ is absent. However, the overall magnitudes of both the  longitudinal and transverse pressures increase.
\begin{figure}[h]
	\begin{center}
		\includegraphics[angle=-90,scale=0.35]{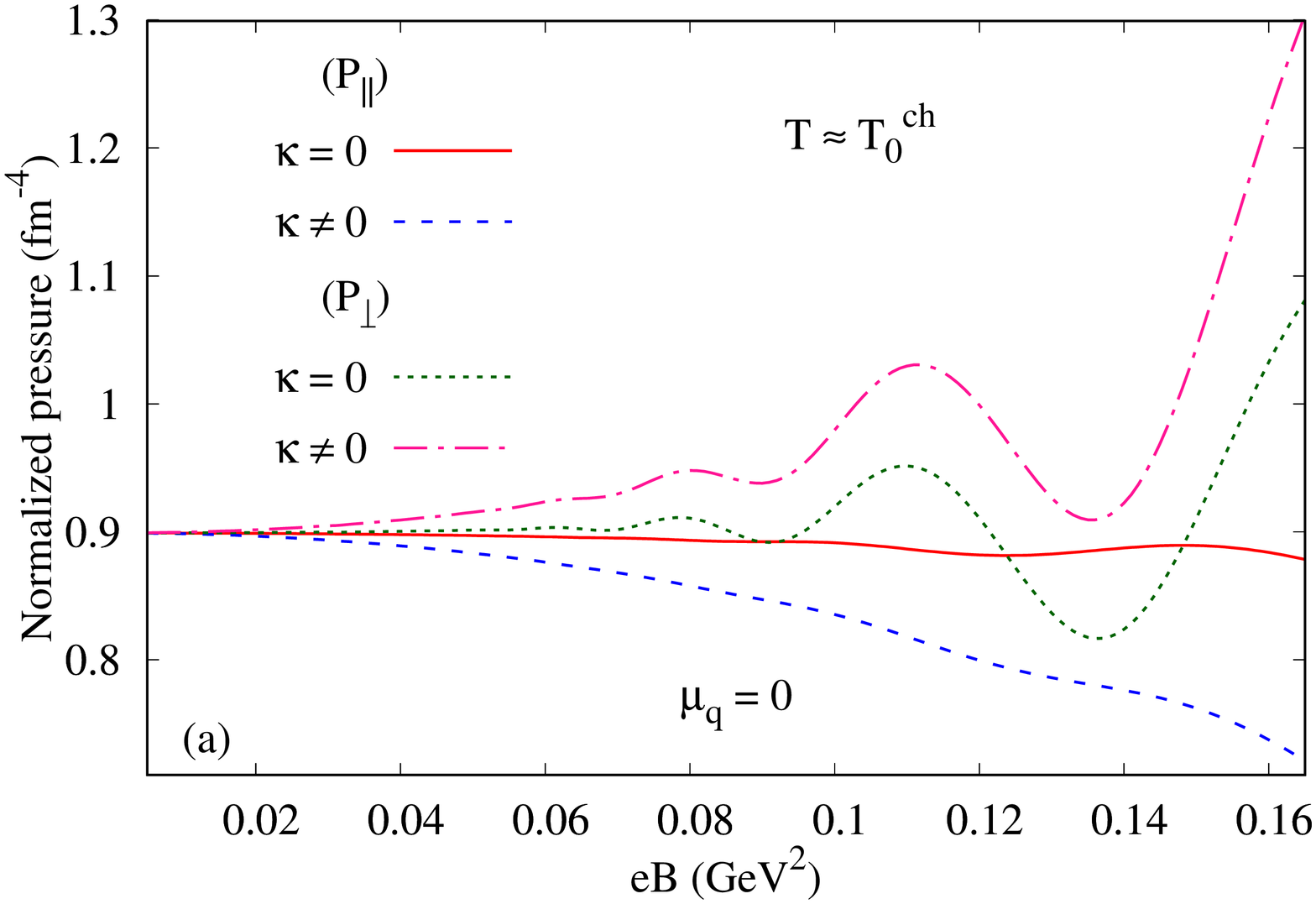} 
		\includegraphics[angle=-90,scale=0.35]{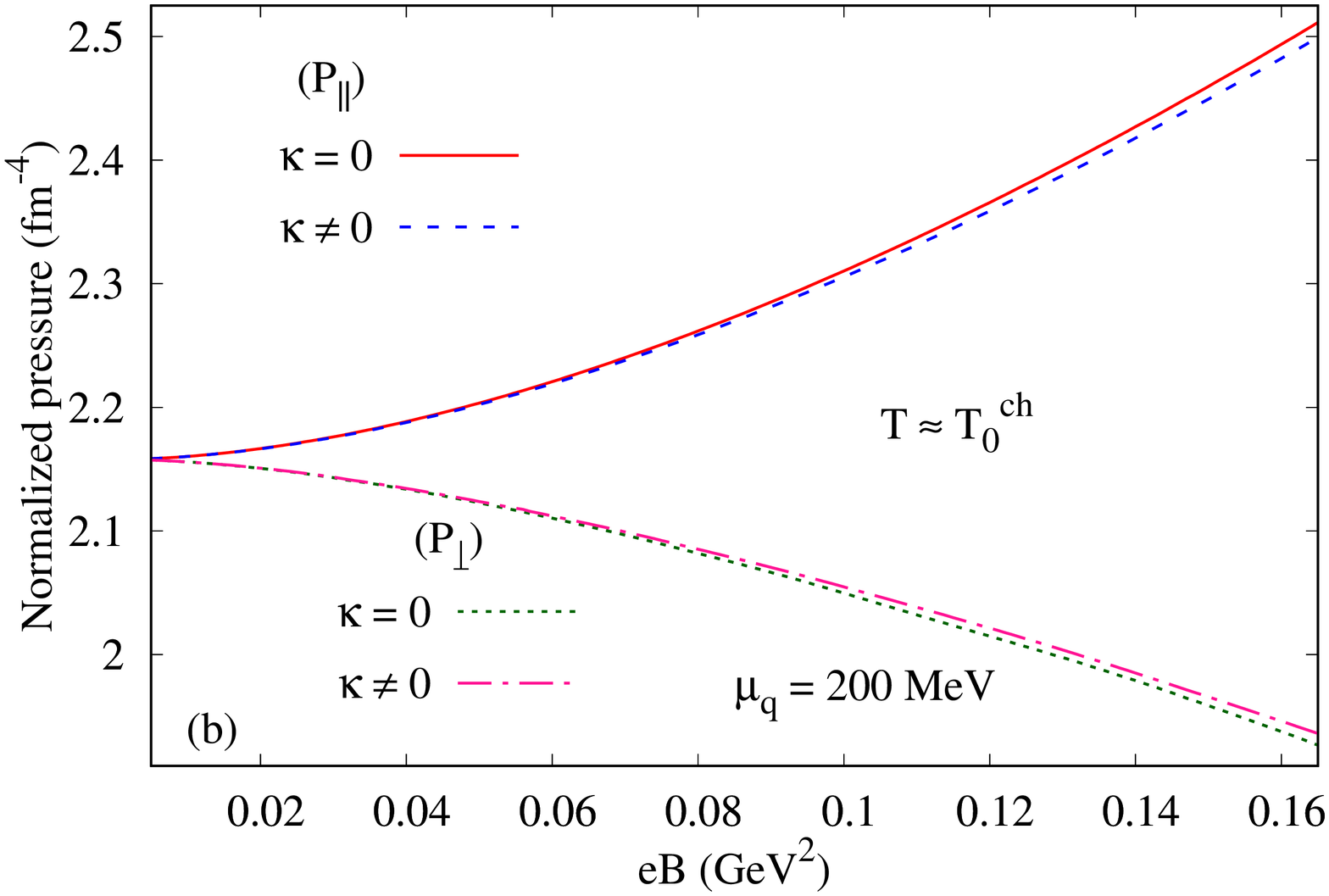}
	\end{center}
	\caption{(Color Online) Normalized longitudinal ($ P_\parallel $) and transverse ($ P_\perp $) pressure as a function of $ eB $ for $ T \approx T_{0}^{ \rm~ch}$ with both vanishing and finite values of the AMM of the quarks at (a) $ \mu_q =0$ and (b) $ \mu_q =200$ MeV.}
	\label{Fig_Pressure2_eB}
\end{figure} 

In Figs.~\ref{Fig_Pressure2_eB}(a) and (b) we have shown $ eB $ dependence of normalized longitudinal and transverse pressures for $ \mu_q = 0 $ and 200 MeV respectively in the vicinity of chiral transition temperature considering both zero and nonzero values of the AMM of the quarks. From Fig.~\ref{Fig_Pressure2_eB}(a), it is evident that, anisotropic   effects on $ \PL $ and $ \PT $ are noticeable for $ eB \gtrsim 0.04 $ GeV$ ^2 $. Moreover, it can be seen that, when finite values of the AMM of the quarks are taken into consideration, $ \PL $ starts to decrease with increasing values of magnetic field. However, in zero AMM case, the magnitude of the longitudinal pressure remains almost unaltered as a function of $ eB $ as observed in Fig.~\ref{Fig_Pressure1_eB}(a).  The transverse pressure exhibits oscillations at large $ eB $ values. Furthermore, the magnitude of $ \PT $ is higher when AMM of the quarks is taken into consideration. In this case, the difference between $ \PL $ and $ \PT $ never vanishes in the $ eB $ range considered in the plot which does not occur in the absence of the AMM. Now in Fig.~\ref{Fig_Pressure2_eB}(b), at $ \mu_q = 200 $ MeV, the pressure components in parallel and perpendicular directions to the magnetic field are different even at small values of $ eB $ and $ \PL $ ($ \PT $) increases (decreases) monotonically with the background field. Moreover, the effects of the inclusion of nonzero values of the AMM of the quarks are also negligible. Comparing Figs.~\ref{Fig_Pressure2_eB}(a) and (b), it can be seen that, the oscillatory behaviour of the transverse pressure for higher values of $ eB $ disappear at $ \mu_q = 200 $ MeV compared to the zero $ \mu_q $ case. This is due to the fact that, at $ \mu_q = 200 $ MeV, the constituent mass will go towards the bare mass limit at smaller values of temperature owing to decrease in chiral transition temperature at finite values of $ \mu_q $ (see Figs.~\ref{Fig_MPhi_T}(a) and (c)). As a result, the constituent mass increases (decreases) monotonically as function of $ eB $ in absence (presence) of AMM of the quarks as evident from Fig.~\ref{Fig_Mass_eB}(b). Consequently, the oscillations observed in $ \PT $ at higher values of $ eB $ disappears at $ T \approx  T_0^{\rm ch} $ in the case of $ \mu_q = 200 $ MeV. It should be noted that, the magnitudes of both $ \PL $ and $ \PT $ are $ \sim 10^3 $ times higher compared to that of the chiral symmetry broken phase.
\begin{figure}[h]
	\begin{center}
		\includegraphics[angle=-90,scale=0.35]{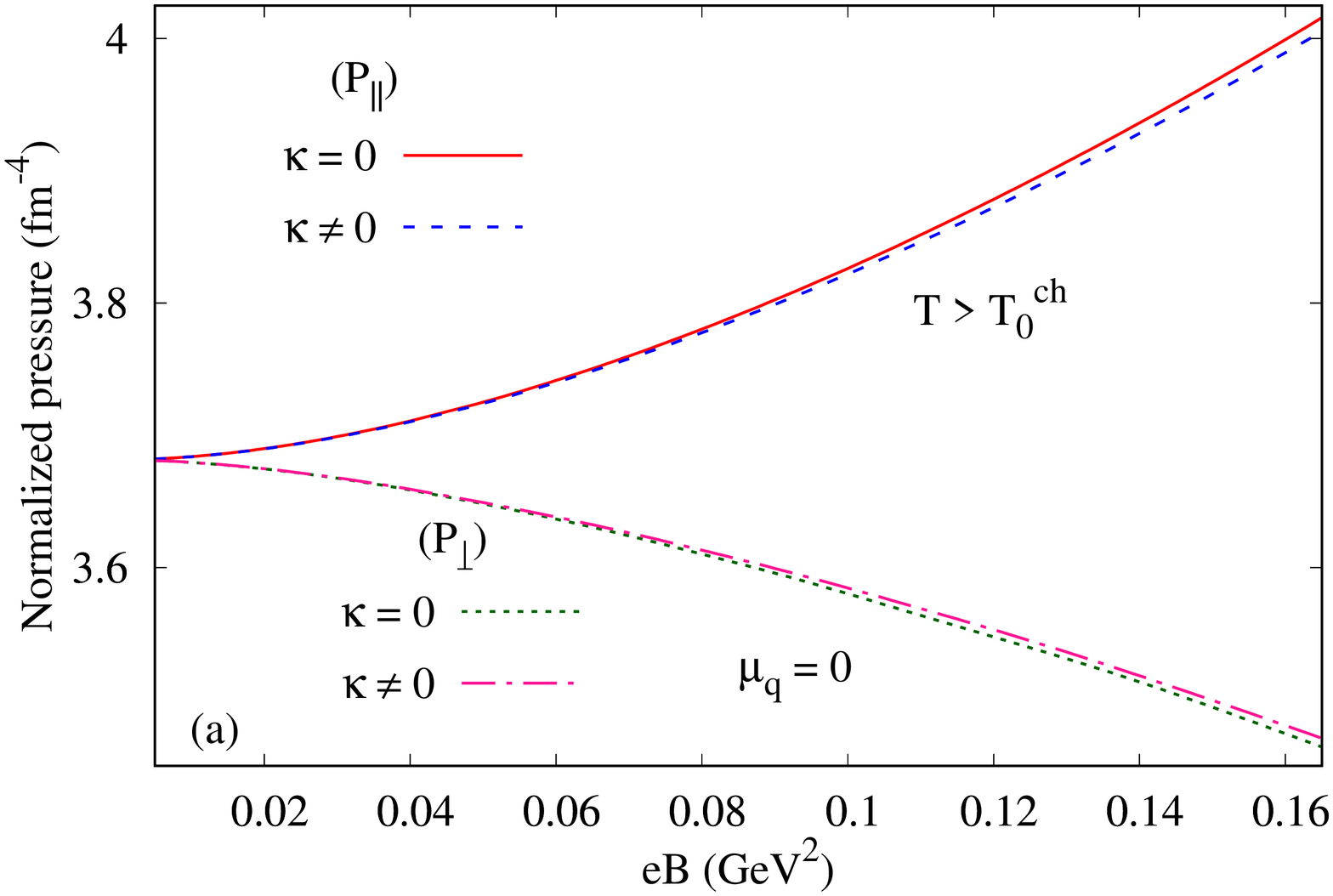} 
		\includegraphics[angle=-90,scale=0.35]{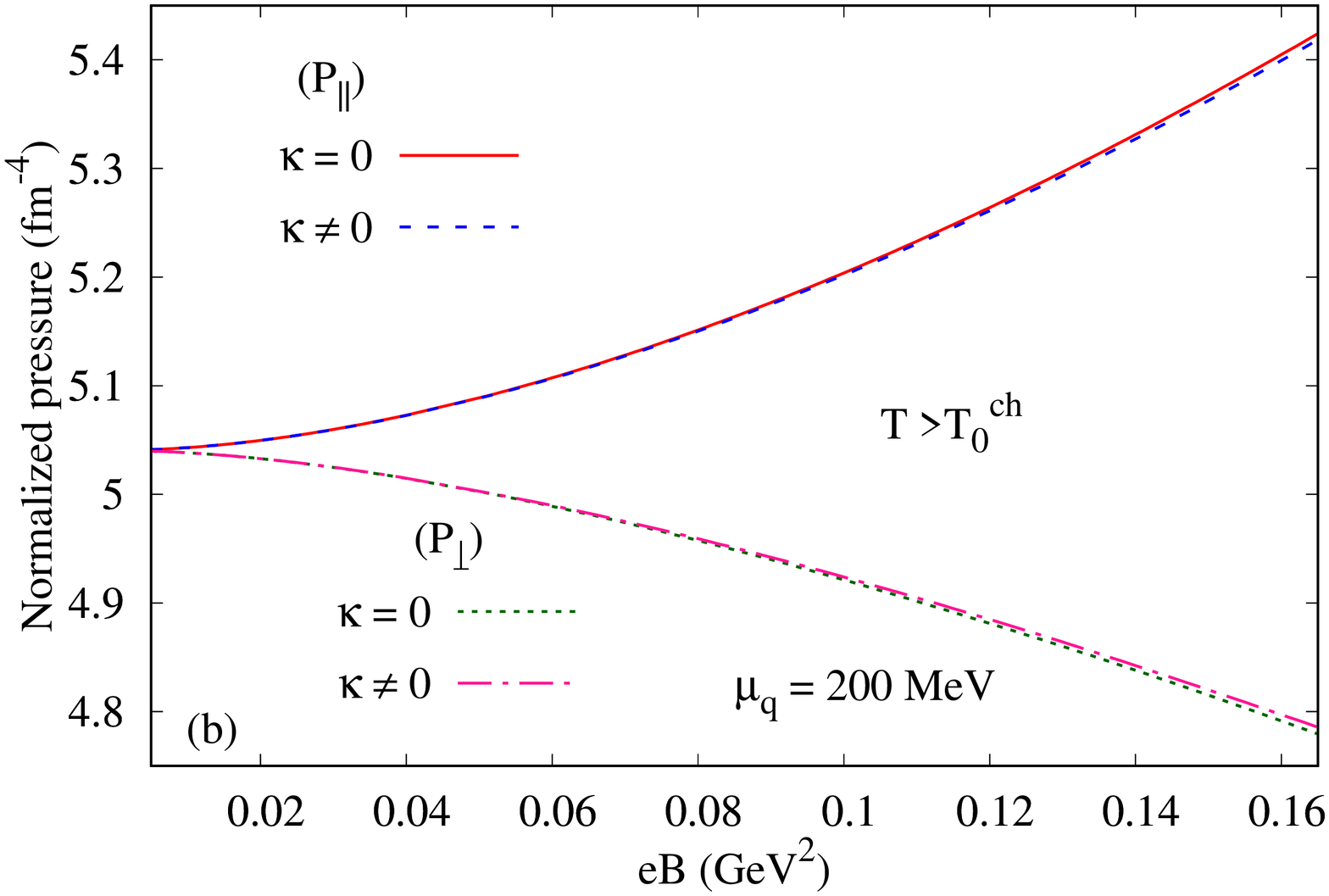}
	\end{center}
	\caption{(Color Online) Normalized longitudinal ($ P_\parallel $) and transverse ($ P_\perp $) pressure as a function of $ eB $ for $ T>  T_{0}^{ \rm~ch}$ with both vanishing and finite values of the AMM of the quarks at (a) $ \mu_q =0$ and (b) $ \mu_q =200$ MeV.}
	\label{Fig_Pressure3_eB}
\end{figure}

In Figs.~\ref{Fig_Pressure3_eB}(a) and (b) we have plotted the variation of normalized $ \PL $ and $ \PT $ for $ \mu_q = 0 $ and 200 MeV respectively  as a function of $ eB  $  considering both zero and nonzero values of the AMM of the quarks at a temperature where chiral symmetry is restored.  In both the figures the anisotropic effects in $ \PL $ and $ \PT $ are noticeable for all values of background magnetic field and the inclusion of AMM of the quarks bring negligible effects.  Moreover, the longitudinal (transverse) pressure increases (decreases) monotonically as a function of $ eB $ which is qualitatively same as we have seen in Fig.~\ref{Fig_Pressure2_eB}(b). 
\begin{figure}[h]
	\begin{center}
		\includegraphics[angle=-90,scale=0.233]{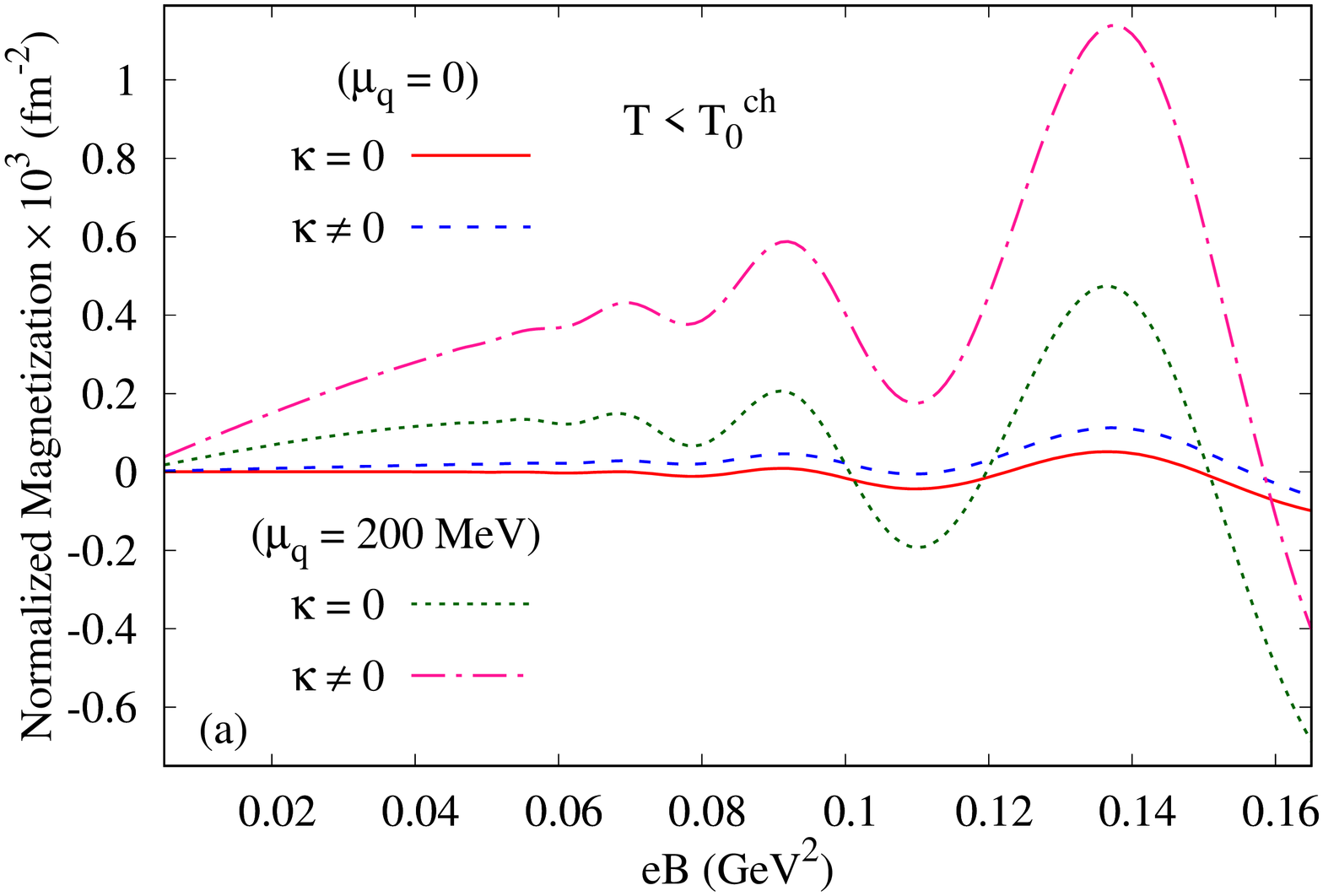} 
		\includegraphics[angle=-90,scale=0.233]{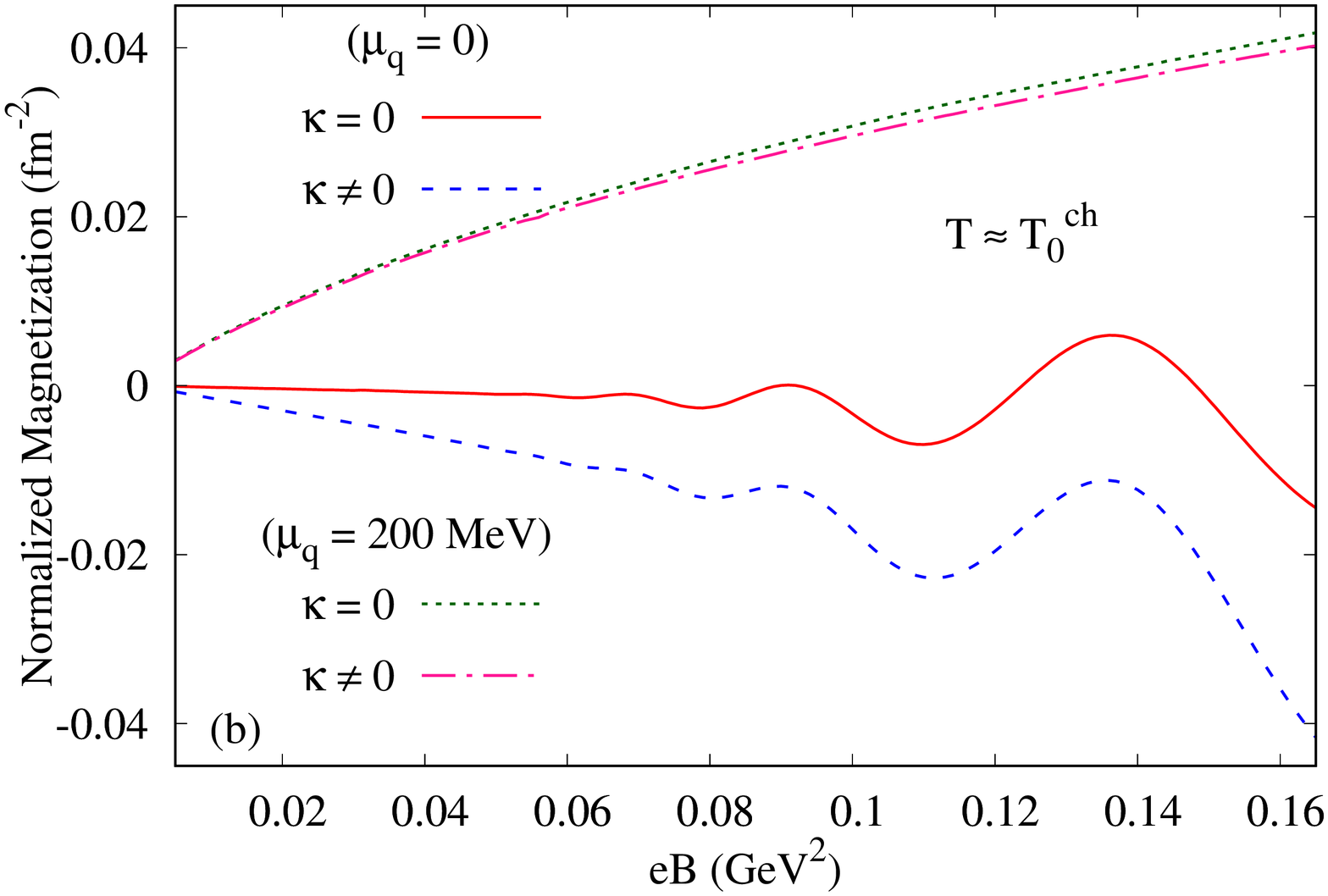}
		\includegraphics[angle=-90,scale=0.233]{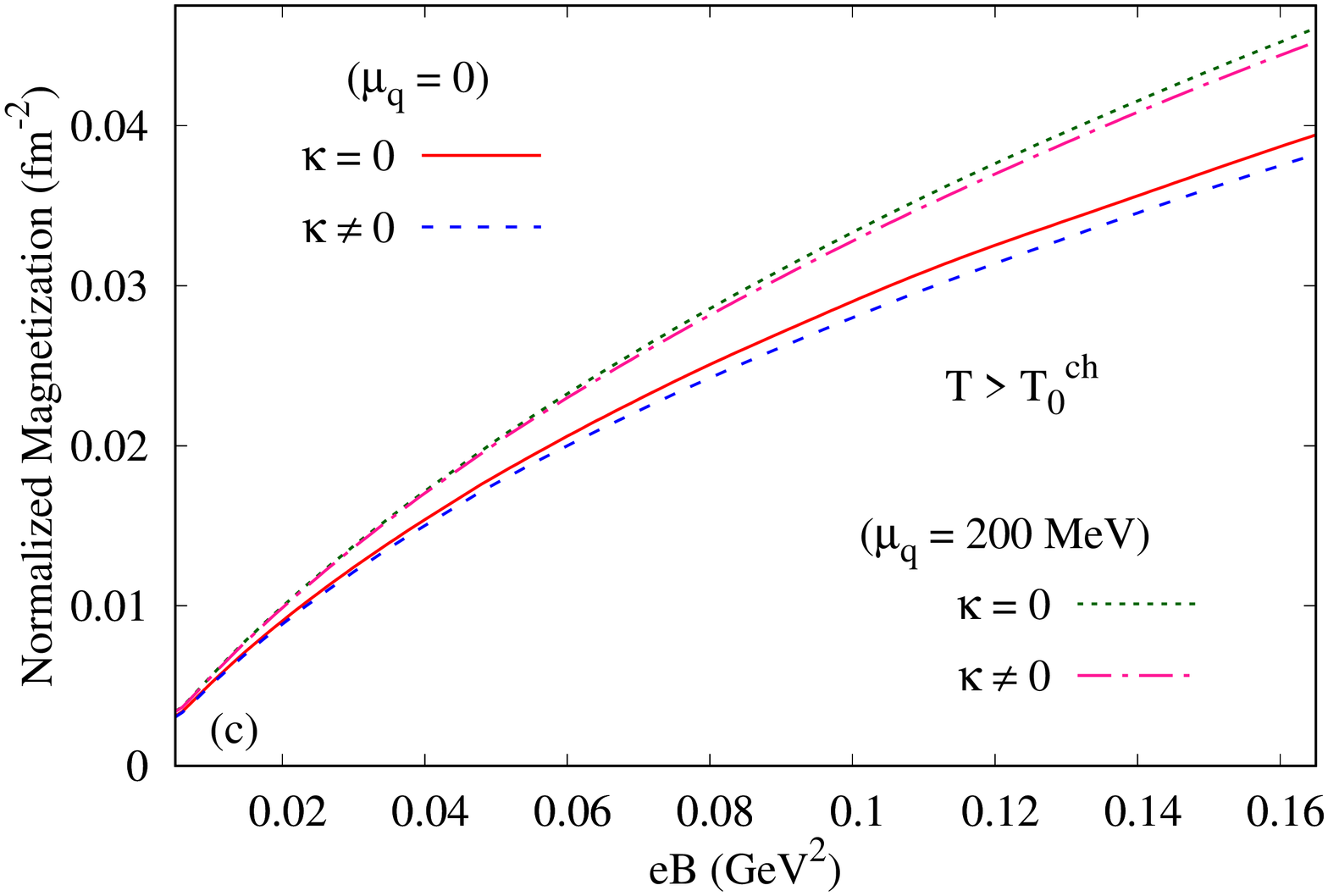}  
	\end{center}
	\caption{(Color Online) Normalized magnetization as a function of $ eB $ for different values of $ \mu_q $ with both vanishing and finite values of the AMM of the quarks at three different stages of the chiral phase transition namely at (a) $ T<  T_{0}^{ \rm~ch}$, (b) $ T\approx T_{0}^{ \rm~ch}$ and (c) $ T >  T_{0}^{ \rm~ch}$.}
	\label{Fig_Mag_eB}
\end{figure}

In Figs.~\ref{Fig_Mag_eB}(a), (b) and (c) we have depicted the variation of normalized magnetization ($\mathcal{M}$) as a function of the background magnetic field at $ \mu_q = 0$ and  $ 200  $ MeV  with and without considering the AMM of quarks for three different values of temperatures respectively, illustrating distinct phases of chiral symmetry breaking and its restoration. From Fig.~\ref{Fig_Mag_eB}(a) it is evident that, in the chiral symmetry broken phase, the magnetization highly oscillates around zero for large values of magnetic field for both zero and nonzero values of the AMM of the quarks and the amplitude of oscillation increases when finite quark chemical potential is considered. Moreover, for finite values of AMM of the quarks, the magnitude of $ \Mag $ is slightly higher for both zero and nonzero  values $ \mu_q $. One should note that, these oscillations in $ \Mag $ in the chiral symmetry broken phase has one to one correspondence with the oscillations seen in $ \PT $ in Figs.~\ref{Fig_Pressure1_eB}(a) and (b), as they are related by Eq.~\eqref{PT}. In Fig.~\ref{Fig_Mag_eB}(b) one can observe that, in the vicinity of chiral transition, at $ \mu_q  = 0$, the magnetization oscillates around zero when the AMM of the quarks are turned off for high $ eB $ values. However, when the finite values of the AMM of the quarks are taken into consideration, then $ \Mag $ remains negative throughout the whole range of background magnetic field  and displays oscillatory behaviour for high values of $ eB $. Now, when the finite values of quark chemical potential is taken into consideration, the magnetization monotonically increases for both zero and nonzero values of the AMM of the quarks and remains positive for all values of $ eB $. This nature of $ \Mag $ near the phase transition, is consistent with oscillations and the smooth decrease in $ \PT $ observed in Fig.~\ref{Fig_Pressure2_eB}(a)  and (b) respectively. Finally, when the chiral symmetry is restored (see Fig.~\ref{Fig_Mag_eB}(c)), $ \Mag $ increases smoothly with $ eB $ at $ \mu_q= 0 $ and 200 MeV respectively and remains positive throughout the whole range of $ eB $ for both zero and finite values of the AMM of the quarks. Notice that, the monotonically decreasing nature of the transverse pressure observed in Figs.~\ref{Fig_Pressure3_eB}(a)  and (b)  is compatible with this nature of $ \Mag $ at temperatures higher than the chiral transition temperature depicted in Fig.~\ref{Fig_Mag_eB}(c).
\begin{figure}[h]
	\begin{center}
		\includegraphics[angle=-90,scale=0.233]{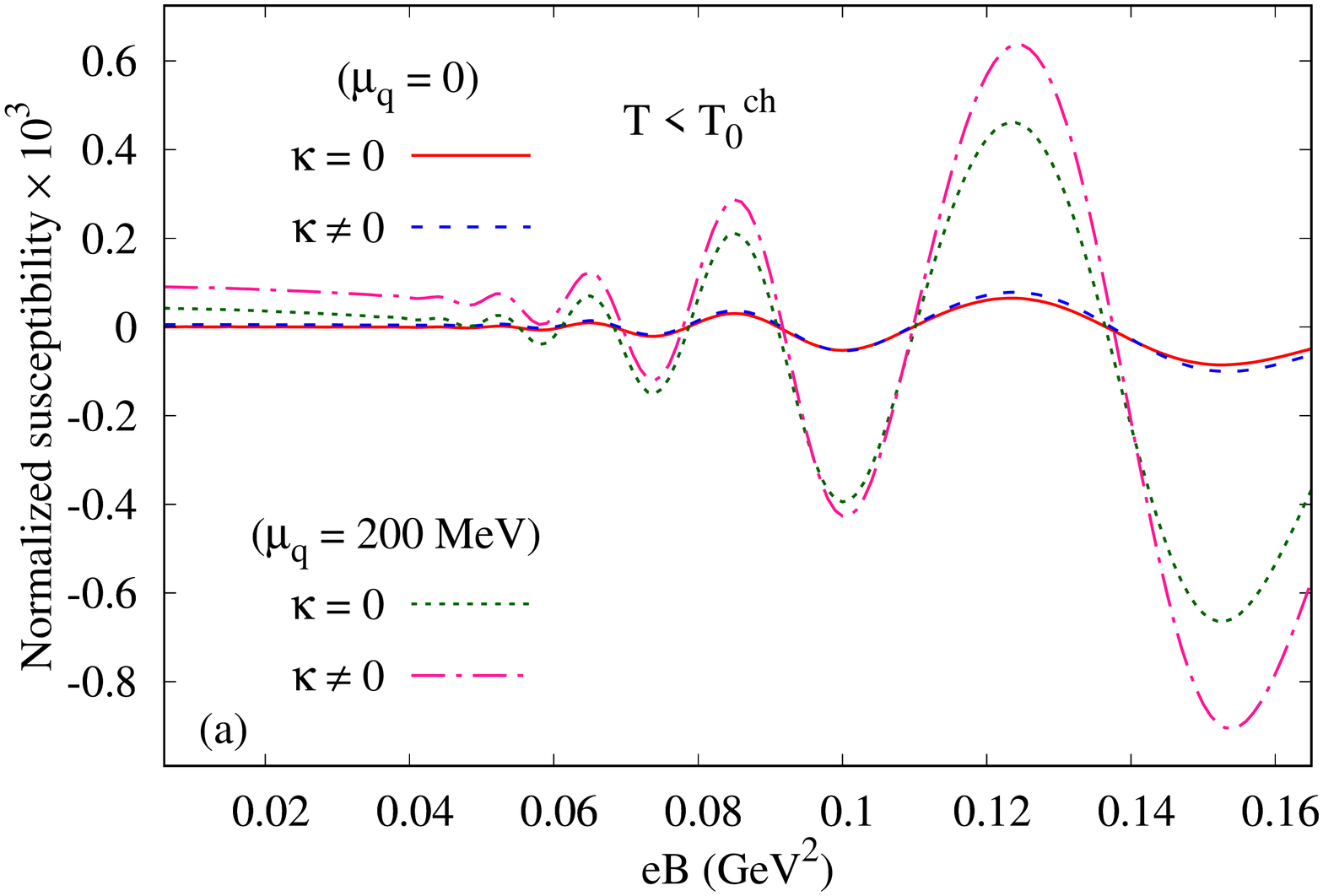} 
		\includegraphics[angle=-90,scale=0.223]{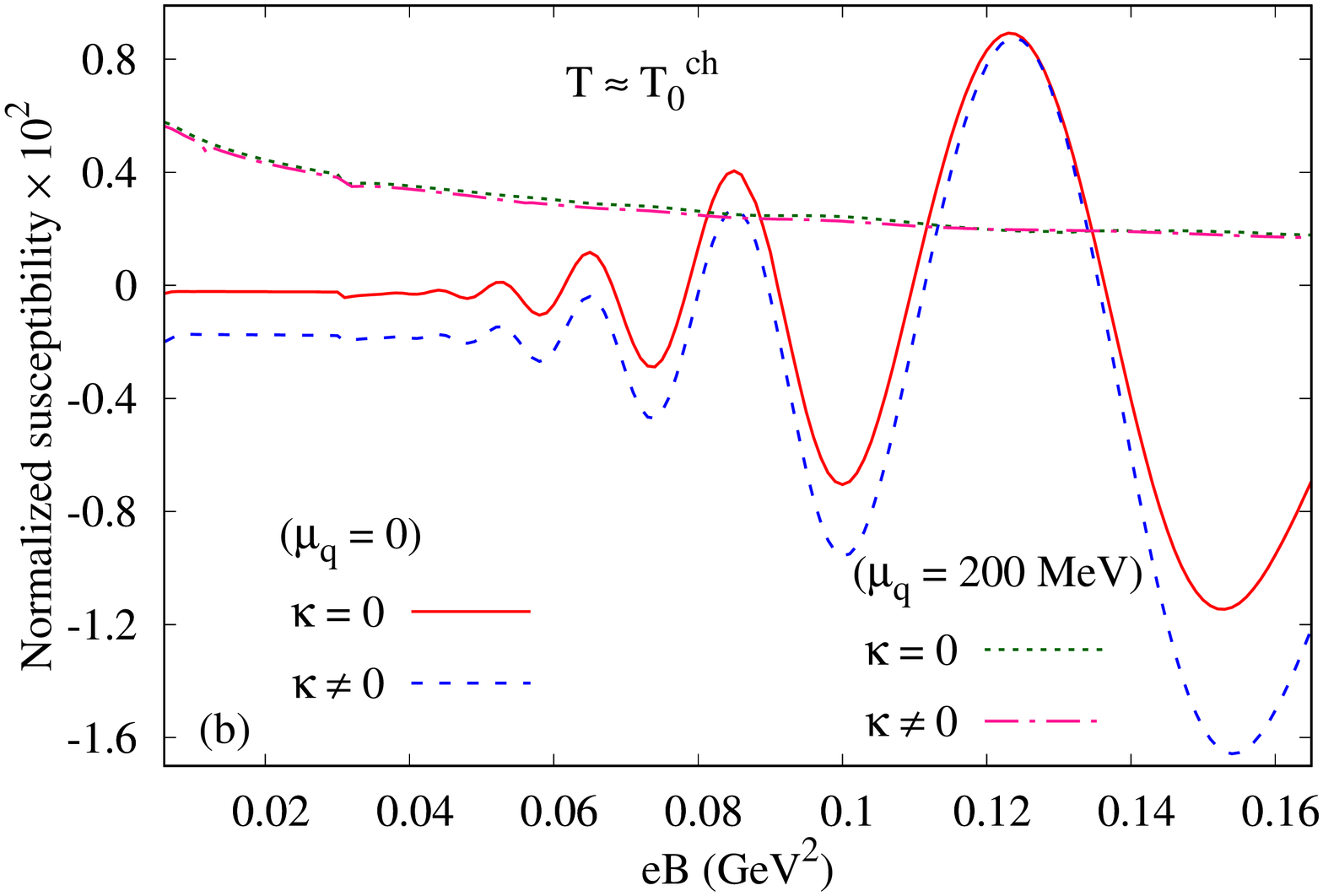}
		\includegraphics[angle=-90,scale=0.233]{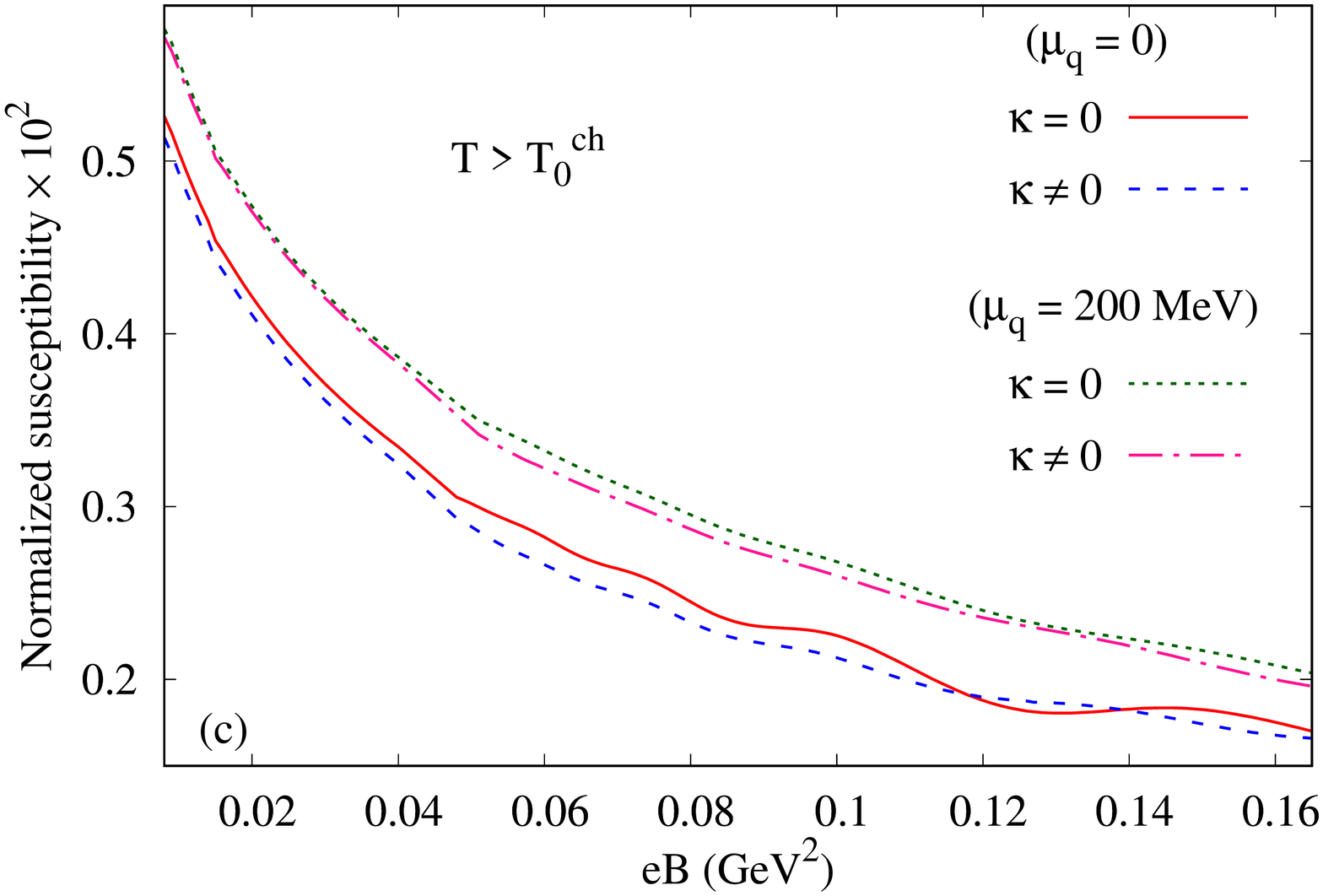}  
	\end{center}
	\caption{(Color Online) Normalized susceptibility as a function of $ eB $ for different values of $ \mu_q $ with both vanishing and finite values of the AMM of the quarks at three different stages of the chiral phase transition namely at (a) $ T<  T_{0}^{ \rm~ch}$, (b) $ T\approx T_{0}^{ \rm~ch}$ and (c) $ T >  T_{0}^{ \rm~ch}$.}
	\label{Fig_Sus_eB}
\end{figure}

In Figs.~\ref{Fig_Sus_eB}(a), (b) and (c) we have presented the variation of normalized susceptibility ($\chi_B$) as a function of $ eB $ at $ \mu_q = 0$ and  $ 200  $ MeV  with and without considering the AMM of quarks for temperatures characterizing the three distinct stages of chiral phase transition. Concentrating on Fig.~\ref{Fig_Sus_eB}(a), it is evident that, in the chiral symmetry broken phase at $ \mu_q = 0 $, $ \chi_B $ remains vanishingly small at small values of $ eB $ and fluctuates around zero for $ eB \gtrsim 0.06  $~GeV$ ^2 $ for both zero and nonzero values of the AMM of the quarks. At finite quark chemical potential, at small $ eB $ values susceptibility becomes slightly positive  and  exhibits highly oscillatory nature at large values of background magnetic field. Thus it can be inferred that, in the case when chiral symmetry is broken, the strongly interacting matter oscillates between diamagnetism ($ \chi_B <0 $) and paramagnetism ($ \chi_B>0 $) for different values of $ eB $. Now from Fig.~\ref{Fig_Sus_eB}(b), it can be seen that, in the vicinity of chiral transition temperature at vanishing quark chemical potential, the susceptibility remains close to zero at samll $ eB $ values and undergoes large oscillations for high values of $ eB $ when the AMM of the quarks are not taken into consideration. However, when the AMM of the quarks are turned on, $ \chi_B $ picks up small negative values indicating diamagnetism at small $ eB $ values and for $eB \gtrsim 0.05  $~GeV$ ^2 $ keeps fluctuating around zero.  Moreover, at finite values of $\mu_q $, $ \chi_B $ remains positive showing the paramagnetic nature throughout the whole range of $ eB $ and the effects of finite values of the AMM of the quarks becomes negligible. Furthermore, as we increase the temperature beyond the chiral transition temperature (see Fig.~\ref{Fig_Sus_eB}(c)), we observe that, $ \chi_B $ remains positive for whole range of $ eB $, although the overall magnitude decreases with increasing values of $ eB $ resulting a paramagnetic character of the QCD matter. Here also the inclusion of the AMM of the quarks does not bring any substantial change in the $ eB $-dependence of $ \chi_B $.
\begin{figure}[h]
	\begin{center}
		\includegraphics[angle=-90,scale=0.233]{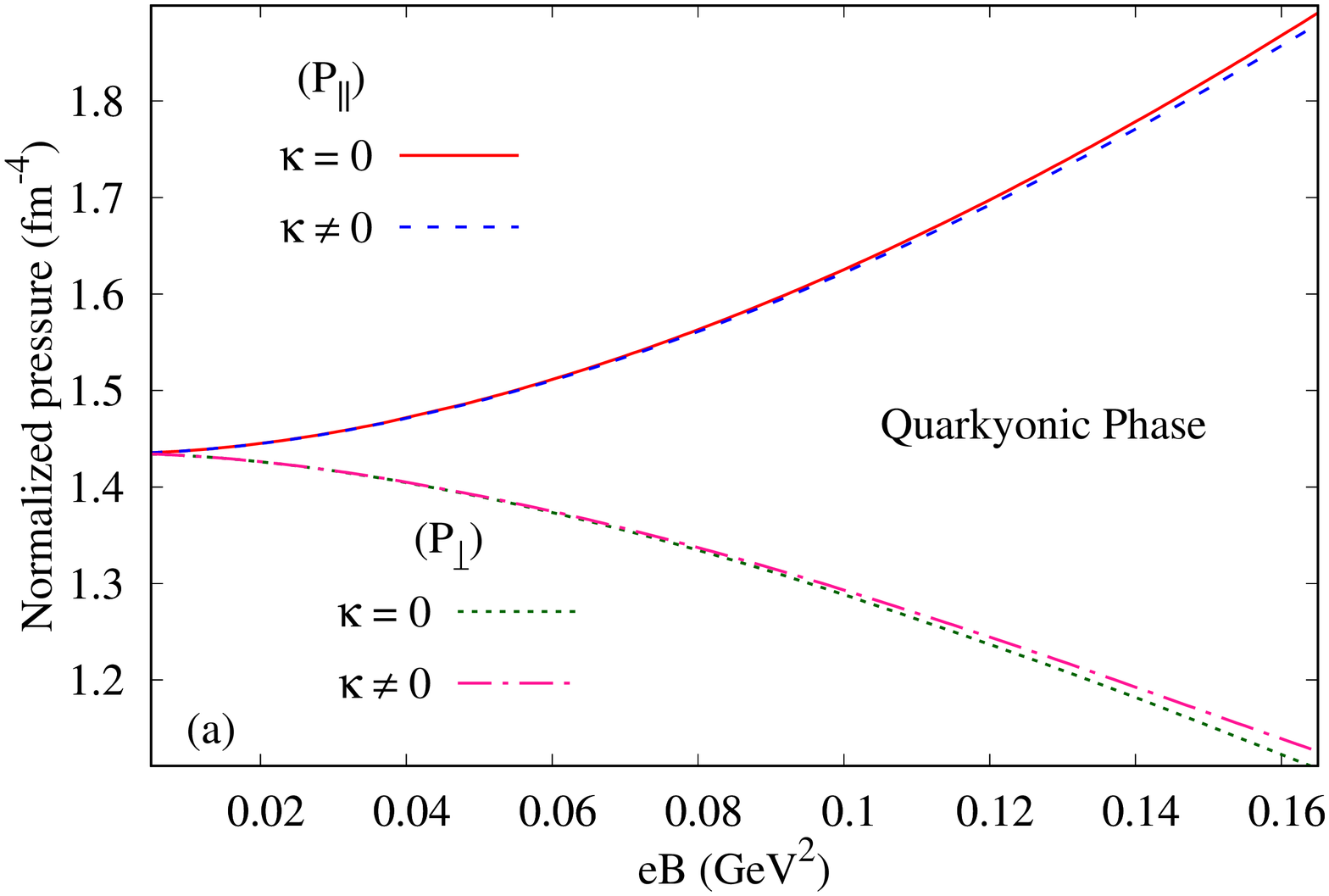} 
		\includegraphics[angle=-90,scale=0.233]{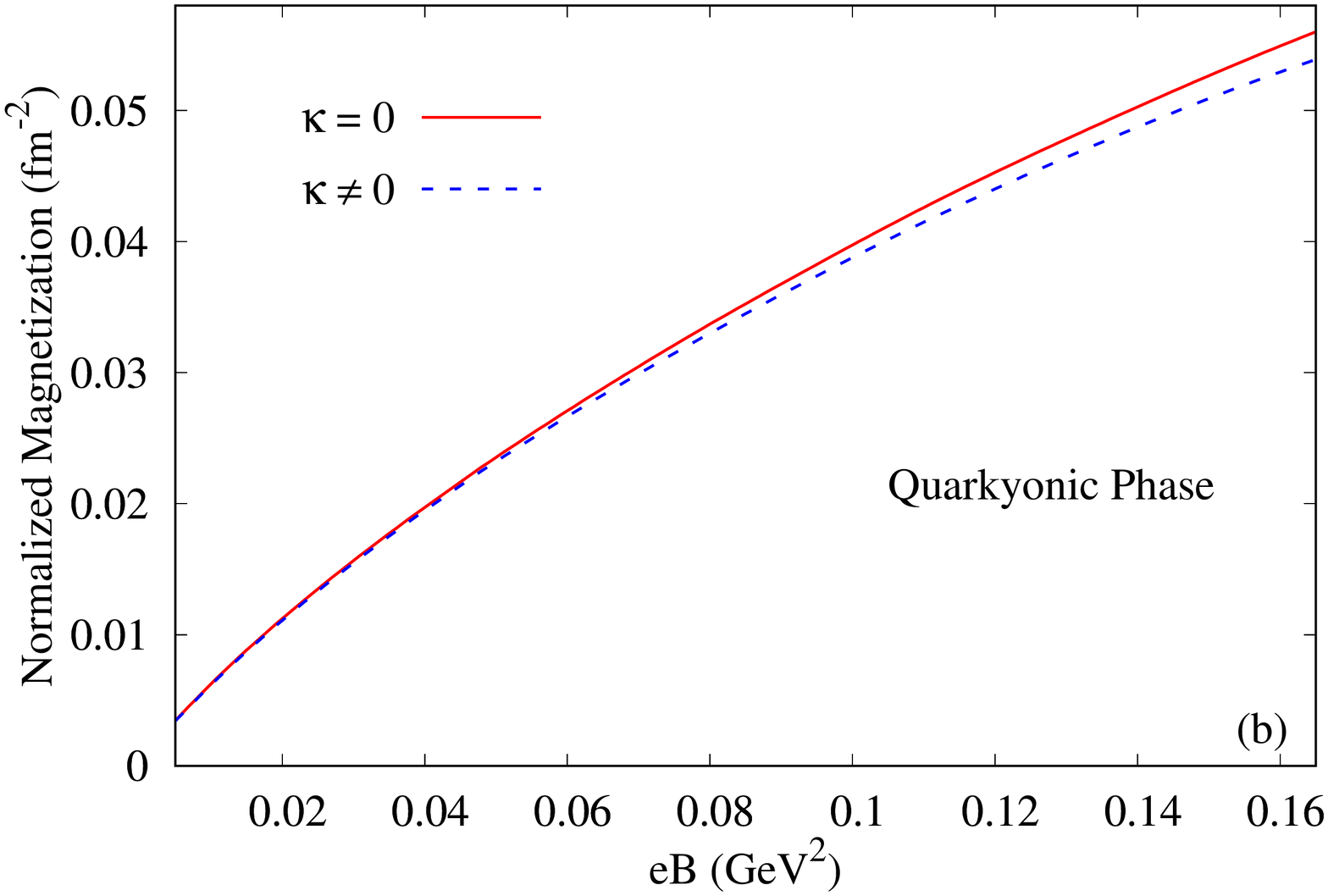}
		\includegraphics[angle=-90,scale=0.233]{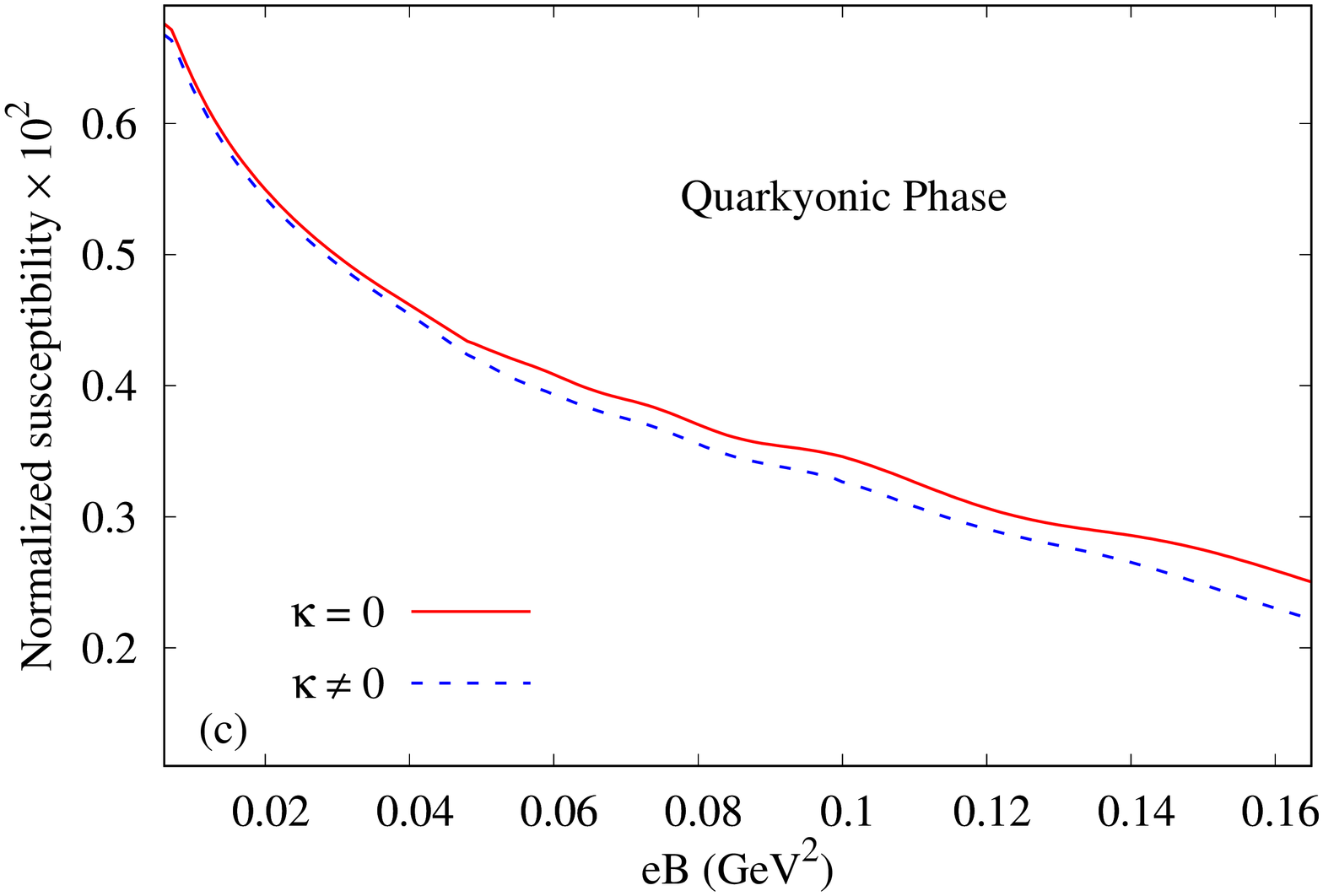}  
	\end{center}
	\caption{(Color Online) Normalized (a) pressure, (b) magnetization and (c) susceptibility as a function of $ eB $ at the quarkyonic phase with both vanishing and finite values of the AMM of the quarks.}
	\label{Fig_Quarky}
\end{figure}

The quarkyonic phase is proposed as a new phase of QCD~\cite{McLerran:2007qj,McLerran:2008ua,Abuki:2008nm,Fukushima:2008wg,Hidaka:2008yy,Buisseret:2011ms,Chaudhuri:2020lga} which is expected to exist at large chemical potentials where the chiral symmetry has been (partially) restored though the matter is still in the confined phase. This basically indicates that, as the chemical potential increases,  the restoration of chiral symmetry occurs earlier than the deconfinement transition. For the current set of parameters of the PNJL model (see Table~\ref{Table_parameters}), we find that at $ T = 175 $~MeV and $ \mu_q  = 350$~MeV, the constituent quark mass goes to bare mass limit, however $ \Phi \lesssim 0.3 $ for both zero and nonzero values of the AMM of the quarks which we will refer as quarkyonic phase~\cite{Abuki:2008nm} or confined chirally restored phase (CCS)~\cite{Hansen:2019lnf}. In Figs~\ref{Fig_Quarky}(a), (b) and (c), we have taken the previously mentioned numerical values of $ T $ and $ \mu_q $ to study the $ eB $-dependence of longitudinal and transverse pressure, magnetization and susceptibility respectively in the quarkyonic or CCS phase. It is evident that, the inclusion of the AMM of the quarks leads to negligible change in the variation of all these thermodynamic quantities as a function of $ eB $. Concentrating on Fig.~\ref{Fig_Quarky}(a), it can be seen that, $ \PL $ ($ \PT $) is a monotonically increasing (decreasing) function of the background field and the  anisotropic effects in the pressure components in parallel and perpendicular direction to the magnetic field can be observed even at small values of $ eB $. The magnetization of the quarkyonic matter is found to be positive throughout the whole range of $ eB $ and is a smoothly growing function. The susceptibility of the strongly interacting matter also remains positive for the whole range of $ eB $ indicating a paramagnetic nature of the matter in this exotic phase of QCD.

\begin{figure}[h]
	\includegraphics[angle=-90,scale=0.34] {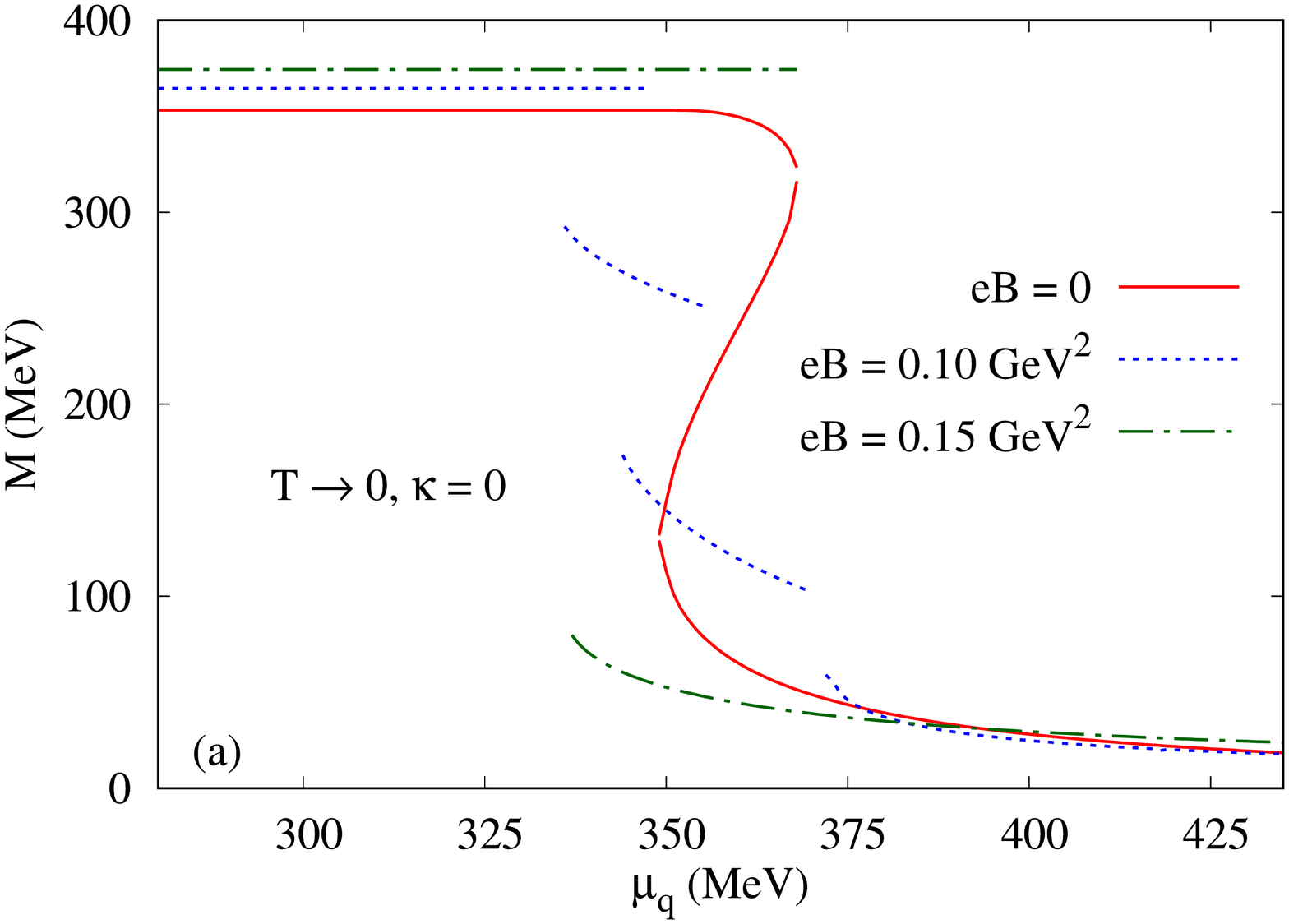}
	\includegraphics[angle=-90,scale=0.34] {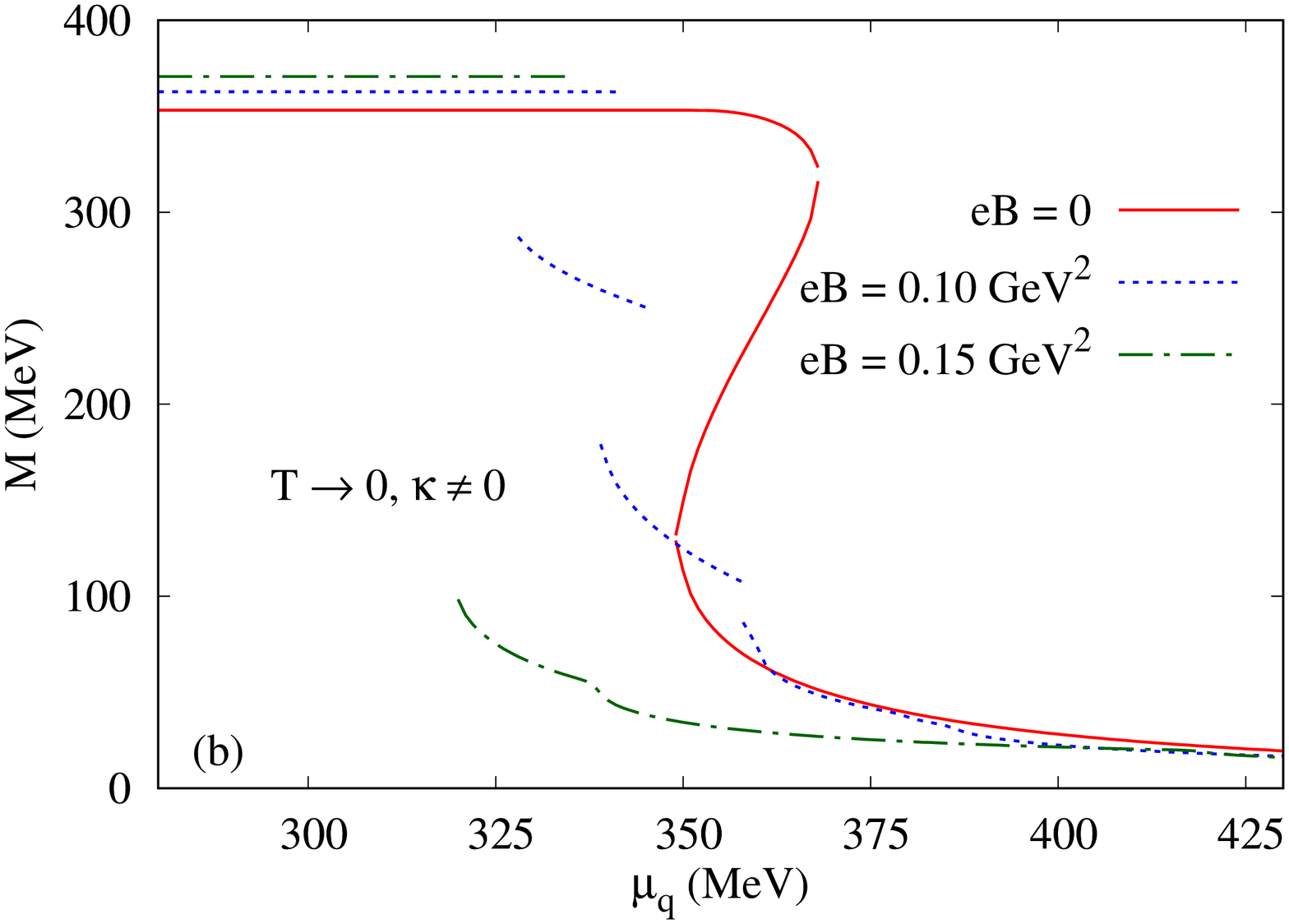}
	\caption{(Color Online) Constituent quark mass as a function of $ \mu_q $ at $ T\rightarrow 0 $ for different values of $ eB $ (a) in absence of AMM of the quarks and (b) considering finite values of the AMM of the quarks.}
	\label{Fig_1st_Mmu}
\end{figure}
Now it has been conjectured that, at lower values of $ T $, the transition from chiral symmetry broken to the restored phase will be first order~\cite{Yagi:2005yb,Letessier:2002gp}. So in Figs.~\ref{Fig_1st_Mmu}(a) and (b) we have shown the variation of constituent quark mass as a function of $ \mu_q $ at $ T\rightarrow 0 $ for different values of $ eB $ (a) in absence of AMM of the quarks and (b) considering finite values of the AMM of the quarks. In both the plots, the $ M $ behaves qualitatively in a similar manner as it remains constant at low values of $ \mu_q $, becomes multiple-valued around $ \mu_q \sim 350$ MeV which is a signature of first order transition and finally goes to the bare mass limit~\cite{Chaudhuri:2019lbw}. The presence of finite background field increases the constituent mass at low $ \mu_q $ values for both zero and nonzero values of AMM of the quarks indicating MC~\cite{Wang:2022xxp} effect. However, when the AMM of the quarks are taken into consideration, the transition occurs at smaller values of $ \mu_q $ implying IMC (see Fig.~\ref{Fig_1st_Mmu}(b)). In both the figures, for $ eB = 0.10 $ GeV$ ^2 $, multiple branch structure of the constituent mass is observed due to the presence of a mismatch in the maximum Landau level for $ u $ and $ d $ quarks which is consistent with the observations made in~\cite{Wang:2022xxp}. Due to the multi-valued nature of $ M $, which goes as input in the calculations of  longituidinal and transverse pressure, magnetization and susceptibility, in the following, we will study the $ eB $-dependence of constituent mass and several thermodynamic quantities in the limit $ T\rightarrow 0 $ at $ \mu_q =300  $ and $ 425 $ MeV which will correspond to chiral symmetry broken and restored phase respectively.
\begin{figure}[h]
	\includegraphics[angle=-90,scale=0.34] {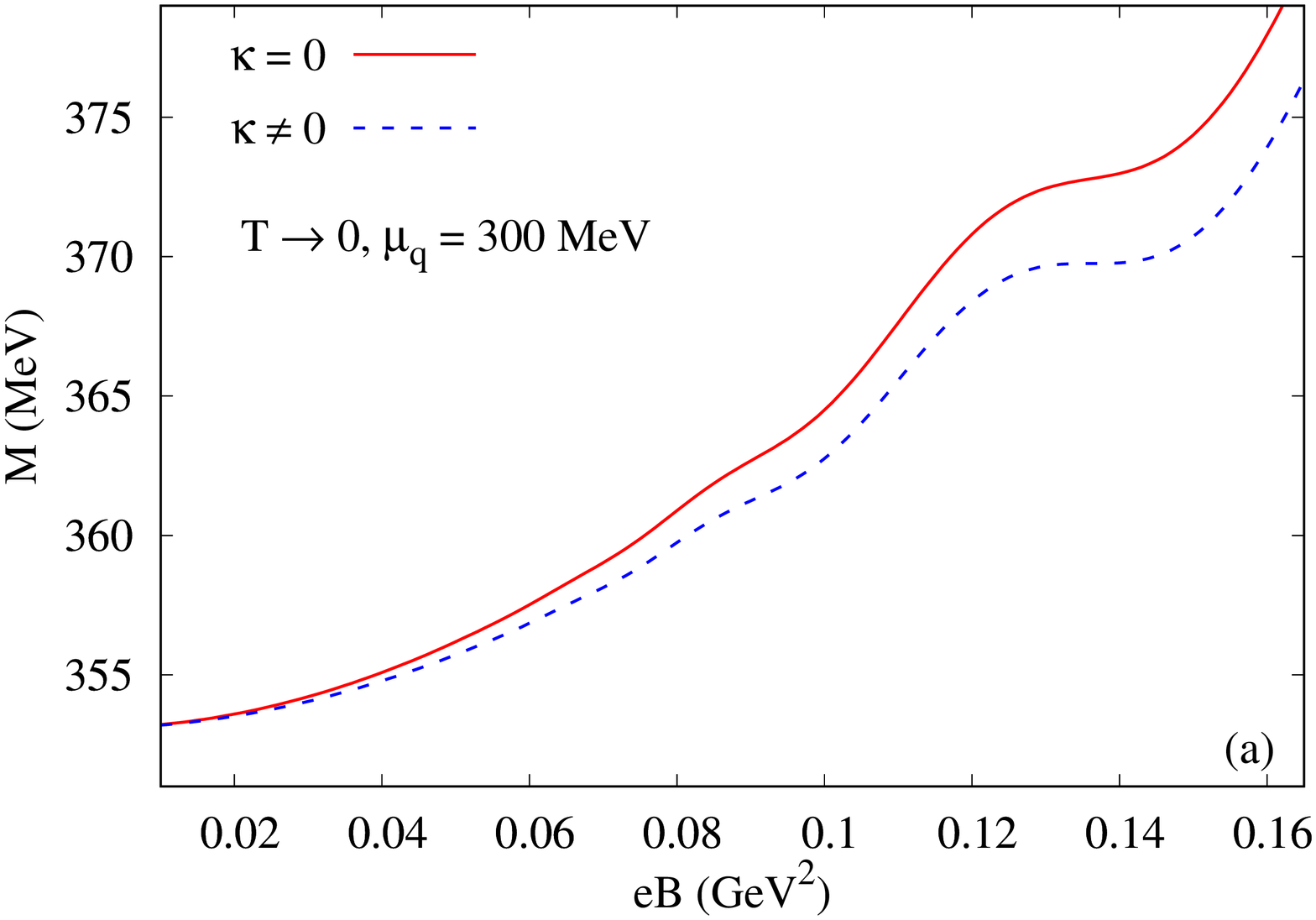}
	\includegraphics[angle=-90,scale=0.34] {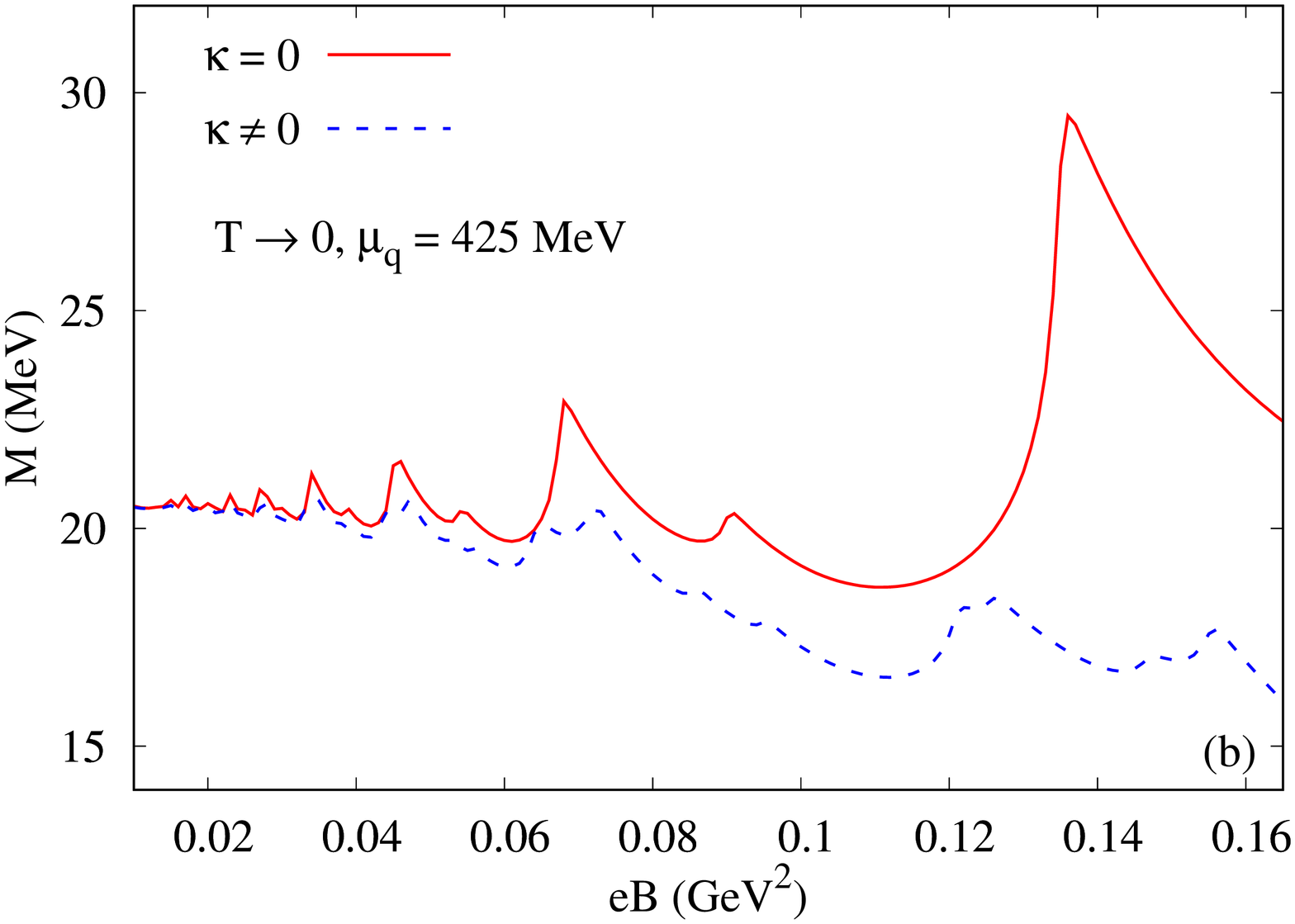}
	\caption{(Color Online) $ eB $-dependence of the constituent quark mass at (a) $ \mu_q $ = 300 MeV and (b) $ \mu_q $ = 425 MeV at $ T\rightarrow 0 $ with both vanishing and finite values of the AMM of the quarks.}
	\label{Fig_1st_MeB}
\end{figure}

In Figs.~\ref{Fig_1st_MeB}(a) and (b), we have depicted the $ eB $- dependence of the constituent quark mass at (a) $ \mu_q $ = 300 MeV and (b) $ \mu_q $ = 425 MeV respectively in the limit $ T\rightarrow 0 $ for both zero and finite values of the AMM of the quarks. From, Fig.~\ref{Fig_1st_MeB}(a), one can observe that, owing to the MC effect, $ M $ increases with $ eB $ and shows slight oscillations at high $ eB $ values independent of the consideration of AMM of the quarks. However, in case of finite values of AMM of the quarks, the overall magnitude of the constituent quark mass is slightly lower. At $ \mu_q  = 425$ MeV, an oscillatory nature of the constituent mass is seen in Fig.~\ref{Fig_1st_MeB}(b), with an overall increase(decrease) in absence(presence) of the AMM of the quarks.
\begin{figure}[h]
	\includegraphics[angle=-90,scale=0.225] {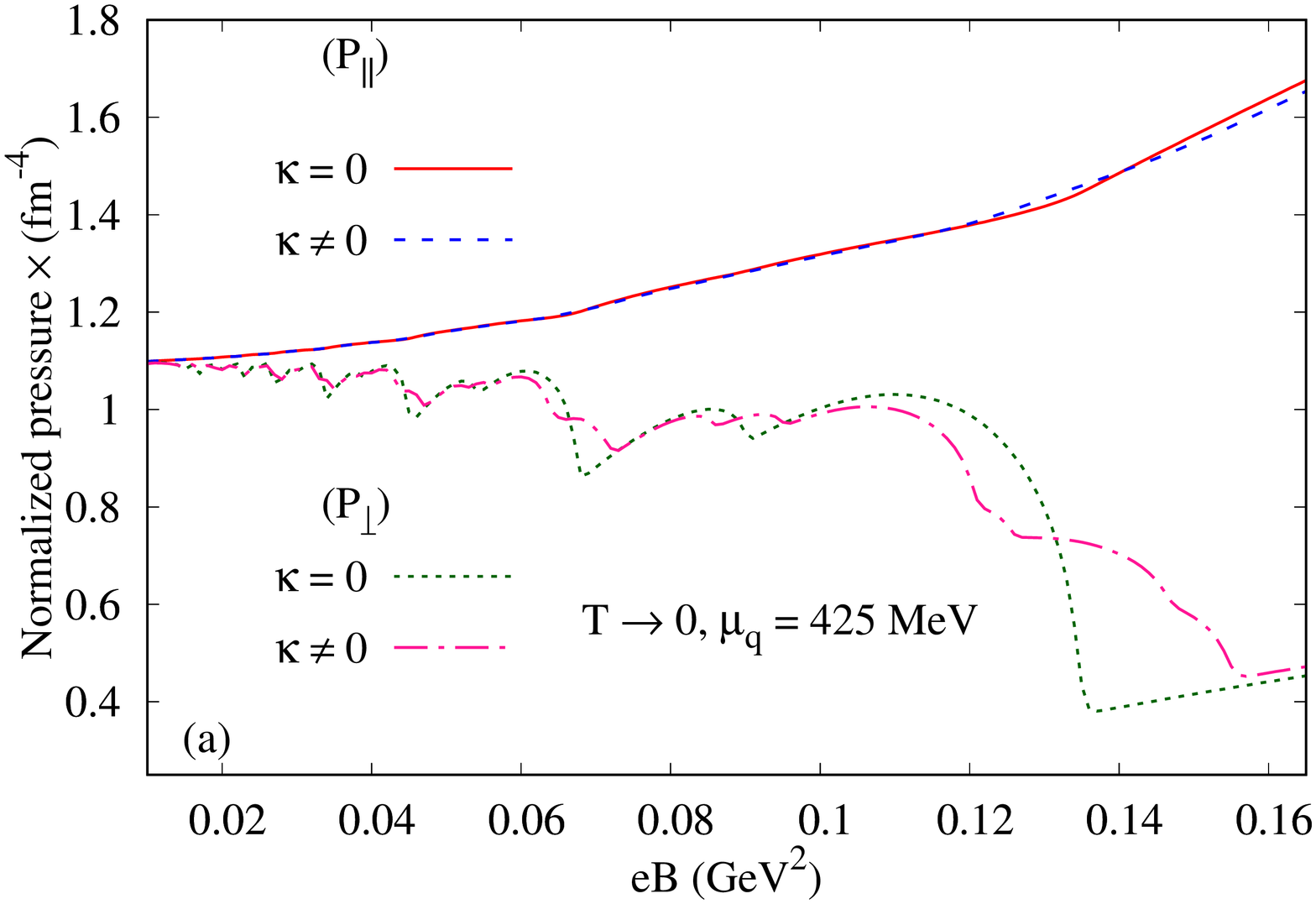}
	\includegraphics[angle=-90,scale=0.225] {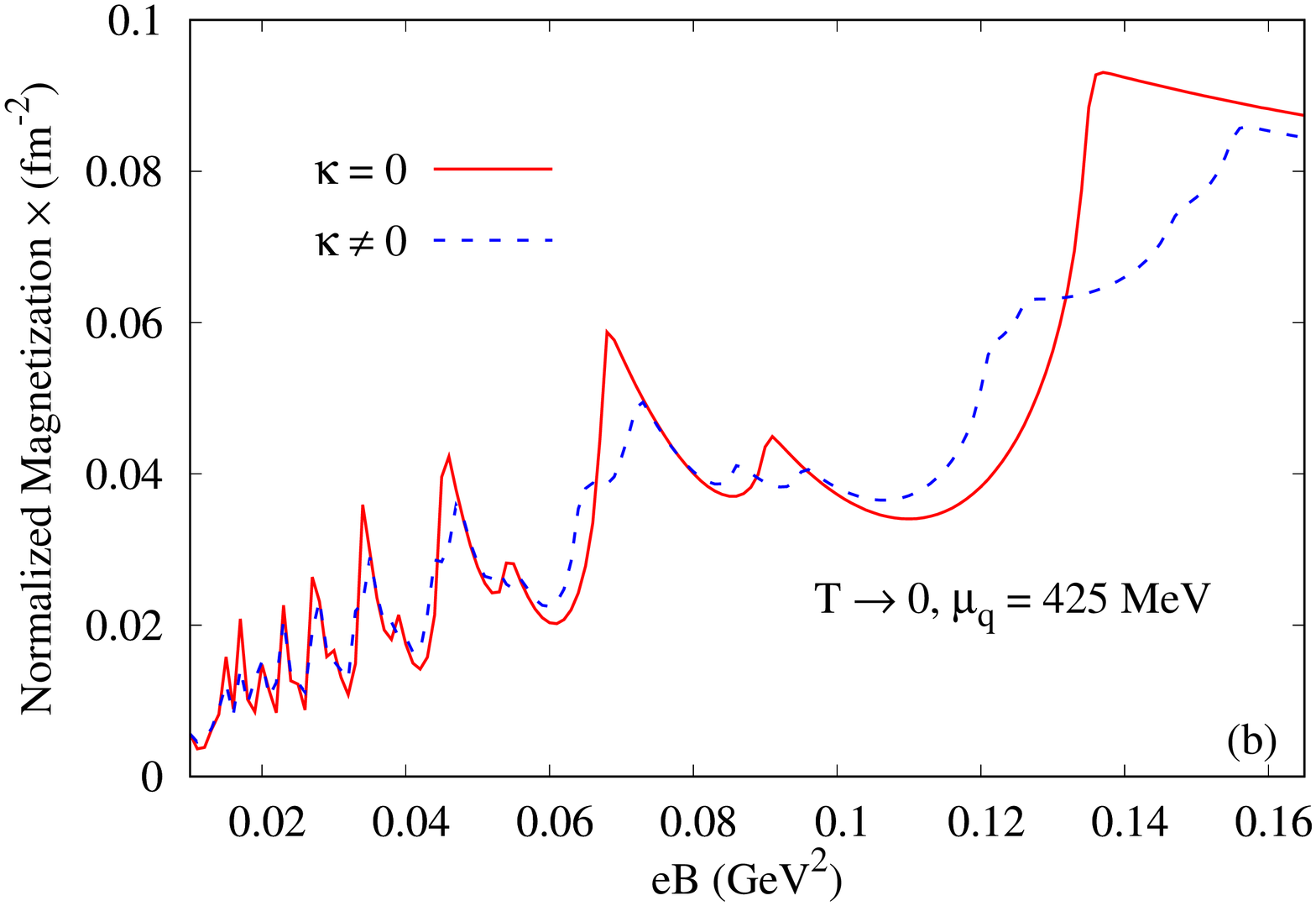}
	\includegraphics[angle=-90,scale=0.225] {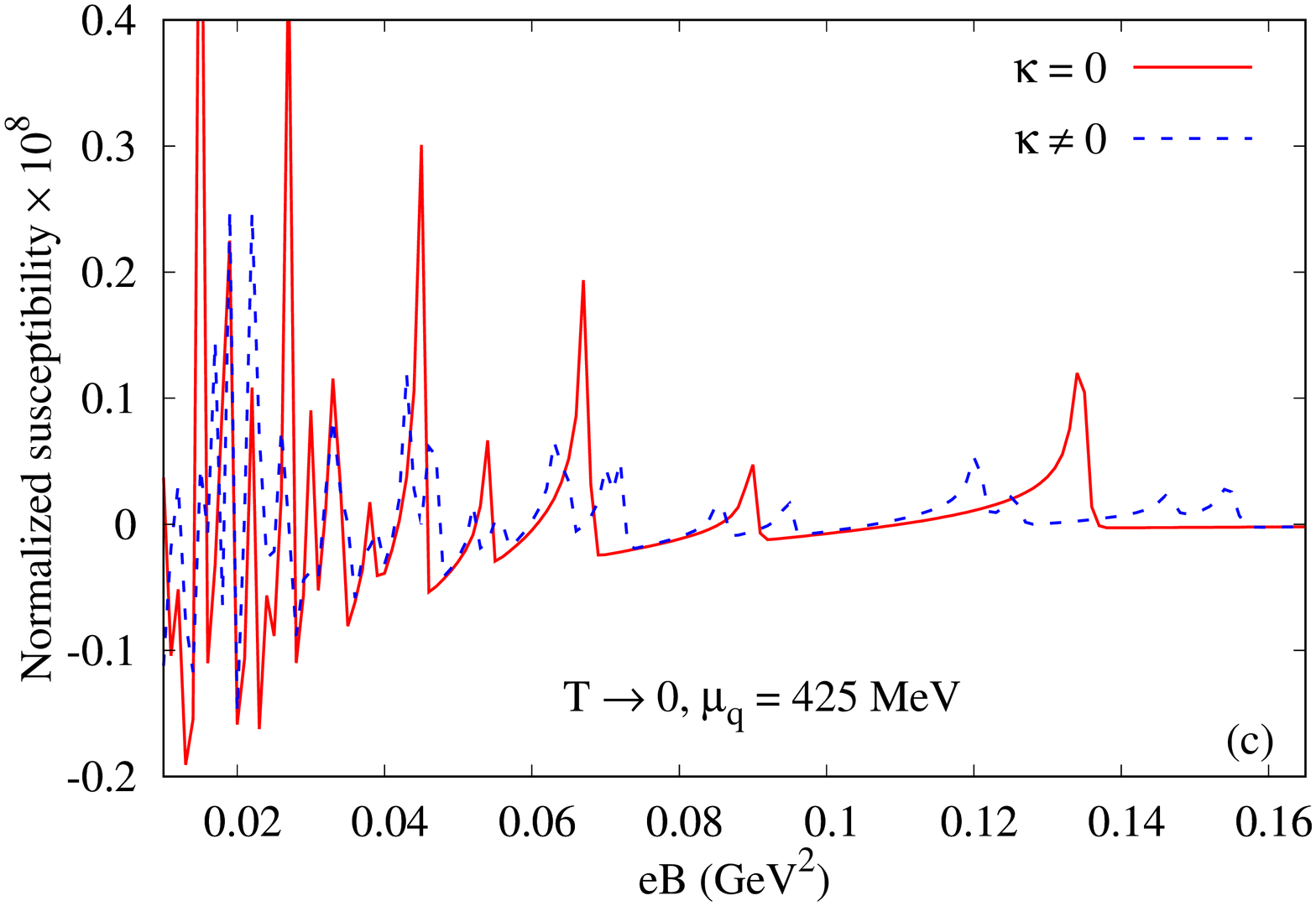}
	\caption{(Color Online) Normalized (a) pressure, (b) magnetization and (c) susceptibility as a function of $ eB $ at $ T\rightarrow 0 $ and $ \mu_q  = 300$ MeV with both vanishing and finite values of the AMM of the quarks.}
	\label{Fig_1st_PMS2}
\end{figure}

In Figs~\ref{Fig_1st_PMS2}(a), (b) and (c), we have shown the variation of normalized longitudinal and transverse pressure, magnetization and susceptibility respectively as a function of $ eB $  at $ \mu_q  = 425 $ MeV in the limit $ T \rightarrow 0  $ for both zero and nonzero values of AMM of the quarks. From Fig.~\ref{Fig_1st_PMS2}(a), it can be seen that, the anisotropic behaviour of $ \PL $ and $ \PT $ is present for all values of $ eB $ and $ \PT $ shows a non-monotonic nature and decreases as a function of $ eB $. As $ \PT $ and magnetization are related by Eq.~(20), a non-monotonic increase in $ \Mag$ is evident from Fig.~\ref{Fig_1st_PMS2}(b). The nornalized susceptibility also oscillates around zero (see Fig.~\ref{Fig_1st_PMS2}(c)) for for smaller values of $ eB $, however for high $ eB $-values, it is showing a tendency to remain positive indicating a paramagnetic nature. Unlike at $ \mu_q = 425 $ MeV, where the constituent mass $ M \rightarrow m_0 $ (see Figs.~\ref{Fig_1st_Mmu}(a) and (b)), the value of $ M $ at $ \mu_q = 300$ MeV is around $ 350 $ MeV. This leads to a thermal suppression of the pressure in this case leading to vanishingly small values of the normalized pressure, magnetization and susceptibility. These observations at small and high values of $ \mu_q $ in the limit $ T \rightarrow 0  $ are consistent with the results shown in ~\cite{Huang:2009ue,Menezes:2015fla}.

\section{Summary and Conclusion}
\label{sec.summary}
We have employed the PNJL model with the inclusion of AMM of the quarks to numerically calculate the canonical expressions for the longitudinal and transverse pressure, magnetization and susceptibility at finite temperature and density in presence of background magnetic field. To this end, we start with the $eB$ and $T$ dependence of constituent quark mass and the expectation values of the Polyakov loop which go as input to the evaluation  of the other thermodynamical variables. It is found that, in the chiral symmetry broken phase, for small values of $ eB $, the longitudinal pressure almost coincides with the transverse pressure independent of the consideration of AMM of the quarks.  However, for higher values of $eB$ the results tend to differ revealing the anisotropic nature and the AMM dependence begins to show up. Both $ \PL $ and $ \PT $ exhibits small and high oscillatory behaviour respectively at higher values of $ eB $. Moreover, for finite values of $ \mu_q $, we observe a qualitatively similar nature of $ \PL $  and $ \PT $ with an increase in the overall magnitudes.	The oscillations in both the pressures disappear for higher values of $ T $ as the chiral symmetry is partially restored for both zero and nonzero values of $ \mu_q $ and $ \PL $ ($ \PT $) becomes a smoothly increasing (decreasing) function of $ eB $ independent of the consideration of the AMM of the quarks. Consequently, the anisotropic nature in $ \PL $ and $ \PT $ is manifested for all values of background field. We then calculate the magnetization  ${\cal M}$ which also shows similar behaviour as $ \PT $ as evident from Eq.~\eqref{PT} i.e. it is highly oscillatory for $ T < T_{0}^{ \rm~ch} $ and becomes a smoothly increasing function of $ eB $ after restoration of chiral symmetry. Next we study the variation of magnetic susceptibility as a function of $ eB $ and observe that the strongly interacting matter oscillates between diamagnetic and paramagnetic phases in the chiral symmetry broken phase for both zero and nonzero values of the AMM of the quarks and $ \mu_q $ respectively. However, at large values fo $ T $, $ \chi_B $ remains positive for whole range of $ eB $ resulting a paramagnetic character of the QCD matter as observed earlier~\cite{Karmakar:2019tdp,Lin:2022ied}. We have also investigated the magnetic property of the quarkyonic phase and found it to be paramagntic in nature. Both the pressures as well as magnetization vary smoothly as a function of $ eB $ in this exotic phase of QCD. Finally we have also explored, $ eB $-dependence of constituent mass and several 	thermodynamic quantities in the limit $ T\rightarrow 0 $ where the transition from chiral symmetry broken to the restored phase is found to be first order by studying the constituent quark mass as a function of $ \mu_q $. The variation of the normalized susceptibility with $ eB $ is showing a strong tendency to remain positive indicating paramagnetic nature  of strongly interacting matter at high $ \mu_q $ values in absence of temperature. We end with the comment that inclusion of strange quark is essential for a more realistic approach. This will be addressed in a future work.

\section*{Acknowledgments}
N.C., P.R. and S.S. are funded by the Department of Atomic Energy (DAE), Government of India. S.G. is funded by the Department of Higher Education, Government of West Bengal, India. 

\appendix
\section{FEW IMPORTANT RESULTS TO EVALUATE THE MAGNETIZATION} 
\label{Appendix1}
Here we note down few important results which are required to evaluate the magnetization of the system. Differentiating Eq.~\eqref{eq.wnfs}, with respect to $B$ we get, 
\begin{eqnarray}
\DerivativePartialOne{\omega_{nfs}}{B} &=& \frac{1}{\omega_{nfs} }~ \FB{ 1 - \frac{s \kappa_f e_f B}{M_{nfs}} }\TB{ M \DerivativePartialOne{M}{B} + \SB{ \FB{\frac{2n + 1 -s}{2}} \MB{e_f} - s \kappa_f e_f M_{nfs}  }}.
\end{eqnarray}

The $B$-derivative of Eqs.~\eqref{eq.gp} and \eqref{eq.gm} comes out to be 
\begin{eqnarray}
\DerivativePartialOne{\ln g^{(+)}}{B} &=& 3\TB{\frac{e^{-(\omega_{nfs} - \mu_q)/T}}{g^{(+)}} \DerivativePartialOne{\Phi}{B} + \frac{e^{-2(\omega_{nfs} - \mu_q)/T}}{g^{(+)}} \DerivativePartialOne{\Phibar}{B} - \frac{1}{T} \DerivativePartialOne{\omega_{nfs}}{B} },
\end{eqnarray}
and,
\begin{eqnarray}
\DerivativePartialOne{\ln g^{(-)}}{B} &=& 3\TB{\frac{e^{-(\omega_{nfs} + \mu_q)/T}}{g^{(-)}} \DerivativePartialOne{\Phibar}{B} + \frac{e^{-2(\omega_{nfs} + \mu_q)/T}}{g^{(-)}} \DerivativePartialOne{\Phi}{B} - \frac{1}{T} \DerivativePartialOne{\omega_{nfs}}{B} }.
\end{eqnarray}
respectively. Finally, we have from Eq.~\eqref{polyakov_potential},
\begin{eqnarray}
\DerivativePartialOne{\mathcal{U}}{B} &=& -T^4 \TB{ \SB{  \frac{1}{2}A ( T )\Fb + 6B(T)\frac{\Fb  - 2 \F^2 + \fb{ \Fb \F } \F}{  1 - 6 \Fb \F + 4 \fb{\F^3 + \Fb^3} + 3\fb{\Fb \F}^2 }  }  \DerivativePartialOne{\Phi}{B} \nn \right. \\ &&  \left. + \SB{ \frac{1}{2}A ( T )\F + 6B(T)\frac{ \F - 2 \Fb^2 + \fb{ \Fb \F } \Fb}{  1 - 6 \Fb \F + 4 \fb{\F^3 + \Fb^3} - 3\fb{\Fb \F}^2 } } \DerivativePartialOne{\Fb}{B} }.
\end{eqnarray}
\section{EXPLICIT EXPRESSIONS OF THE COEFFICIENTS $\chi_B^{(0)}, \mathcal{A}_{M,B}, \mathcal{A}_{\Phi,B}$ AND $\mathcal{A}_{\Fb,B}$ }
\label{Appendix2}
In this appendix, we provide the explicit expressions of the coefficients $\chi_B^{(0)}$, $\mathcal{A}_{M,B}$, $\mathcal{A}_{\Phi,B}$ and $\mathcal{A}_{\Phibar,B}$ appearing in Eq.~\eqref{eq.sus}:
\begin{eqnarray}
	\chi_B^{(0)}  &=& - 1 + 6 \sum_{nfs} \frac{\mb{e_f}}{2\pi} \kzint{p} \frac{1}{\omega_{nfs}} \FB{1 - \frac{s\kappa_f e_f B}{M_{nfs}}} \FB{ (2n + 1 -s)\frac{\mb{e_f}}{2} - s\kappa_f e_f M_{nfs} } \SB{f_\Lambda - f^{(+)} - f^{(-)}}  \nn \\ 
	&& + 3 \sum_{nfs} \frac{\mb{e_fB}}{2\pi} \kzint{p} \Bigg[ - \frac{1}{\omega_{nfs}^3} \FB{1 - \frac{s\kappa_f e_f B}{M_{nfs}}}^2 \FB{ (2n + 1 -s)\frac{\mb{e_f}}{2} - s\kappa_f e_f M_{nfs} }^2 \nn  \\ 
	&& + \frac{1}{\omega_{nfs}} \FB{ - \frac{s\kappa_f e_f}{M_{nfs}  } + \frac{s\kappa_fe_f B}{M_{nfs}^3} (2n + 1 -s)\frac{\mb{e_f}}{2}  } \FB{ (2n + 1 -s)\frac{\mb{e_f}}{2} - s\kappa_f e_f M_{nfs} } \nn  \\ 
	&&  +\frac{1}{\omega_{nfs}} \FB{1 - \frac{s\kappa_f e_f B}{M_{nfs}}}  \FB{  - \frac{s\kappa_fe_f B}{M_{nfs}} (2n + 1 -s)\frac{\mb{e_f}}{2}} \Bigg] \SB{f_\Lambda - f^{(+)}  - f^{(-)} } \nn \\
	&& - \frac{3}{T} \sum_{nfs} \frac{\mb{e_fB}}{2\pi} \kzint{p} \frac{1}{\omega_{nfs}^2} \FB{1 - \frac{s\kappa_f e_f B}{M_{nfs}}}^2 \FB{ (2n + 1 -s)\frac{\mb{e_f}}{2} - s\kappa_f e_f M_{nfs} }^2 \nn \\ 
	&& \times \Big[ \frac{1}{g^{(+)}} \SB{  \{3 f^{(+)}-1\} \F  e^{-( \omega_{nfs}-\mu_q )/T} + 2\{3f^{(+)}-2\} \Fb e^{-2( \omega_{nfs}-\mu_q ) /T} + 3 \{ f^{(+)}-1\} e^{-3( \omega_{nfs}  -\mu_q )/T } }  \nn \\ 
	&& + \frac{1}{g^{(-)}} \SB{  \{3 f^{(-)}-1\} \Fb  e^{-( \omega_{nfs}+\mu_q )/T} + 2\{3f^{(-)}-2\} \F e^{-2( \omega_{nfs}+\mu_q ) /T} + 3 \{ f^{(-)}-1\} e^{-3( \omega_{nfs} + \mu_q )/T } } \Big]  \nn \\ 
	&&- 6\sum_{nfs} \frac{\mb{e_f}}{2\pi } \kzint{p}\omega_{nfs} \dfrac{2 n N \mb{e_f}\SB{p_z^2 + (2n + 1 -s )\mb{e_f B} }^{N-1} \Lambda^{2N} }{ \SB{\FB{p_z^2 + (2n + 1 -s )\mb{e_f B} }^{N} + \Lambda^{2N}}^2}\nn \\ 
	&& \hspace{-0.5 IN}
	+ 3\sum_{nfs} \frac{\mb{e_f B}}{2\pi } \kzint{p}\omega_{nfs} \frac{4 e_f^2 n^2 \SB{p_z^2 + (2n + 1 -s )\mb{e_f B} }^{N-2} \Lambda^{2N} 
		\TB{ N (N-1) \Lambda^N - N ( N+1) \SB{p_z^2 + (2n + 1 -s )\mb{e_f B} }^{N} } }
	{ \SB{\FB{p_z^2 + (2n + 1 -s )\mb{e_f B} }^{N} + \Lambda^{2N}}^3} \nn \\ 
	&& \hspace{-0.5IN}
	- 6\sum_{nfs} \frac{\mb{e_fB}}{2\pi } \kzint{p}\frac{1}{\omega_{nfs}} \FB{1 - \frac{s\kappa_f e_f B}{M_{nfs}}}  \FB{ (2n + 1 -s)\frac{\mb{e_f}}{2} - s\kappa_f e_f M_{nfs} } 
	\frac{2 n N \mb{e_f}\SB{p_z^2 + (2n + 1 -s )\mb{e_f B} }^{N-1} \Lambda^{2N} }
	{ \SB{\FB{p_z^2 + (2n + 1 -s )\mb{e_f B} }^{N} + \Lambda^{2N}}^2}, \nn \\
\end{eqnarray}
\begin{eqnarray}
	\mathcal{A}_{M,B}  &=& 3 \sum_{nfs} \frac{\mb{e_f}}{2 \pi} \kzint{p}~\frac{M}{\omega_{nfs}} \FB{ 1- \frac{s\kappa_f e_f B}{M_{nfs}} } \SB{f_\Lambda - f^{(+)} - f^{(-)} } \nn \\ 
	&& + 3 \sum_{nfs} \frac{\mb{e_fB}}{2 \pi} \kzint{p}~ \bigg[  - \frac{M}{\omega_{nfs}^3} \FB{  1- \frac{s\kappa_f e_f B}{M_{nfs}} }^2 \FB{ (2n + 1 -s)\frac{\mb{e_f}}{2} - s\kappa_f e_f M_{nfs} } \nn  \\
	&&  + \frac{M}{M^3_{nfs}} \frac{s \kappa_f e_f B}{\omega_{nfs}}(2n + 1 -s)\frac{\mb{e_f}}{2} - \frac{M}{\omega_{nfs}} \frac{s\kappa_f e_f }{M_{nfs}}    \bigg]
	\SB{f_\Lambda - f^{(+)}  - f^{(-)} } \nn \\ 
	&& -  3 \sum_{nfs} \frac{\mb{e_fB}}{2 \pi} \kzint{p}~ \bigg[  \frac{ M}{T \omega_{nfs}^2}  \FB{  1- \frac{s\kappa_f e_f B}{M_{nfs}} }^2    \FB{ (2n + 1 -s)\frac{\mb{e_f}}{2} - s\kappa_f e_f M_{nfs} } \nn  \\ 
	&&  \times \SB{  \{3 f^{(+)}-1\} \F  e^{-( \omega_{nfs}-\mu_q )/T} + 2\{3f^{(+)}-2\} \Fb e^{-2( \omega_{nfs}-\mu_q ) /T} + 3 \{ f^{(+)}-1\} e^{-3( \omega_{nfs}  -\mu_q )/T } \right. \nn \\
		&&  + \left.  \{3 f^{(-)}-1\} \Fb  e^{-( \omega_{nfs}+\mu_q )/T} + 2\{3f^{(-)}-2\} \F e^{-2( \omega_{nfs}+\mu_q ) /T} + 3 \{ f^{(-)}-1\} e^{-3( \omega_{nfs} + \mu_q )/T } }  \nn  \\
	&&  - 3\sum_{nfs} \frac{\mb{e_fB}}{2\pi } \kzint{p}\frac{M}{\omega_{nfs}} \FB{1 - \frac{s\kappa_f e_f B}{M_{nfs}}}  \frac{2 n N \mb{e_f}\SB{p_z^2 + (2n + 1 -s )\mb{e_f B} }^{N-1} \Lambda^{2N} }{ \SB{\fb{p_z^2 + (2n + 1 -s )\mb{e_f B} }^{N} + \Lambda^{2N}}^2}  \bigg].
\end{eqnarray}
	\begin{eqnarray}
	\mathcal{A}_{\Phi,B} &=& 3 T \sum_{nfs}\frac{\mb{e_f}}{2\pi}  \kzint{p} ~ \TB{ \frac{1}{g^{(+)}}e^{-( \omega_{nfs}  -\mu_q ) /T} + \frac{1}{g^{(-)}}e^{-2( \omega_{nfs}  + \mu_q ) /T} } \nn \\ 
	&& - 3\sum_{nfs} \frac{\mb{e_fB}}{2\pi } \kzint{p}\frac{M}{\omega_{nfs}} \FB{1 - \frac{s\kappa_f e_f B}{M_{nfs}}  } \FB{ (2n + 1 -s)\frac{\mb{e_f}}{2} - s\kappa_f e_f M_{nfs} } \nn \\ 
	&& \times\TB{ \frac{1}{g^{(+)}} \{1 - 3 f^{(+)}\}e^{-( \omega_{nfs}  -\mu_q ) /T} + \frac{1}{g^{(-)}} \{1 - 3 f^{(-)}\}e^{-2( \omega_{nfs}  + \mu_q ) /T}  }, 
\end{eqnarray}
\begin{eqnarray}
	\mathcal{A}_{\Fb,B} &=& 3 T \sum_{nfs}\frac{\mb{e_f}}{2\pi}  \kzint{p} ~ \TB{ \frac{1}{g^{(-)}}e^{-( \omega_{nfs}  +\mu_q ) /T} + \frac{1}{g^{(+)}}e^{-2( \omega_{nfs}  - \mu_q ) /T} } \nn \\ 
	&& - 3\sum_{nfs} \frac{\mb{e_fB}}{2\pi } \kzint{p}\frac{M}{\omega_{nfs}} \FB{1 - \frac{s\kappa_f e_f B}{M_{nfs}}  } \FB{ (2n + 1 -s)\frac{\mb{e_f}}{2} - s\kappa_f e_f M_{nfs} } \nn \\ 
	&& \times\TB{ \frac{1}{g^{(-)}} \{1 - 3 f^{(-)}\}e^{-( \omega_{nfs}  +\mu_q ) /T} + \frac{1}{g^{(+)}} \{1 - 3 f^{(+)}\}e^{-2( \omega_{nfs}  - \mu_q ) /T}  }.
\end{eqnarray}

\bibliography{Z-Pressure}

\begin{thebibliography}{124}%
\makeatletter
\providecommand \@ifxundefined [1]{%
 \@ifx{#1\undefined}
}%
\providecommand \@ifnum [1]{%
 \ifnum #1\expandafter \@firstoftwo
 \else \expandafter \@secondoftwo
 \fi
}%
\providecommand \@ifx [1]{%
 \ifx #1\expandafter \@firstoftwo
 \else \expandafter \@secondoftwo
 \fi
}%
\providecommand \natexlab [1]{#1}%
\providecommand \enquote  [1]{``#1''}%
\providecommand \bibnamefont  [1]{#1}%
\providecommand \bibfnamefont [1]{#1}%
\providecommand \citenamefont [1]{#1}%
\providecommand \href@noop [0]{\@secondoftwo}%
\providecommand \href [0]{\begingroup \@sanitize@url \@href}%
\providecommand \@href[1]{\@@startlink{#1}\@@href}%
\providecommand \@@href[1]{\endgroup#1\@@endlink}%
\providecommand \@sanitize@url [0]{\catcode `\\12\catcode `\$12\catcode
  `\&12\catcode `\#12\catcode `\^12\catcode `\_12\catcode `\%12\relax}%
\providecommand \@@startlink[1]{}%
\providecommand \@@endlink[0]{}%
\providecommand \url  [0]{\begingroup\@sanitize@url \@url }%
\providecommand \@url [1]{\endgroup\@href {#1}{\urlprefix }}%
\providecommand \urlprefix  [0]{URL }%
\providecommand \Eprint [0]{\href }%
\providecommand \doibase [0]{http://dx.doi.org/}%
\providecommand \selectlanguage [0]{\@gobble}%
\providecommand \bibinfo  [0]{\@secondoftwo}%
\providecommand \bibfield  [0]{\@secondoftwo}%
\providecommand \translation [1]{[#1]}%
\providecommand \BibitemOpen [0]{}%
\providecommand \bibitemStop [0]{}%
\providecommand \bibitemNoStop [0]{.\EOS\space}%
\providecommand \EOS [0]{\spacefactor3000\relax}%
\providecommand \BibitemShut  [1]{\csname bibitem#1\endcsname}%
\let\auto@bib@innerbib\@empty
\bibitem [{\citenamefont {Kharzeev}\ \emph
  {et~al.}(2013{\natexlab{a}})\citenamefont {Kharzeev}, \citenamefont
  {Landsteiner}, \citenamefont {Schmitt},\ and\ \citenamefont
  {Yee}}]{Kharzeev:2013jha}%
  \BibitemOpen
  \bibinfo {editor} {\bibfnamefont {D.}~\bibnamefont {Kharzeev}}, \bibinfo
  {editor} {\bibfnamefont {K.}~\bibnamefont {Landsteiner}}, \bibinfo {editor}
  {\bibfnamefont {A.}~\bibnamefont {Schmitt}}, \ and\ \bibinfo {editor}
  {\bibfnamefont {H.-U.}\ \bibnamefont {Yee}},\ eds.,\ \href {\doibase
  10.1007/978-3-642-37305-3} {\emph {\bibinfo {title} {{Strongly Interacting
  Matter in Magnetic Fields}}}},\ Vol.\ \bibinfo {volume} {871}\ (\bibinfo
  {year} {2013})\BibitemShut {NoStop}%
\bibitem [{\citenamefont {Miransky}\ and\ \citenamefont
  {Shovkovy}(2015)}]{Miransky:2015ava}%
  \BibitemOpen
  \bibfield  {author} {\bibinfo {author} {\bibfnamefont {V.~A.}\ \bibnamefont
  {Miransky}}\ and\ \bibinfo {author} {\bibfnamefont {I.~A.}\ \bibnamefont
  {Shovkovy}},\ }\href {\doibase 10.1016/j.physrep.2015.02.003} {\bibfield
  {journal} {\bibinfo  {journal} {Phys. Rept.}\ }\textbf {\bibinfo {volume}
  {576}},\ \bibinfo {pages} {1} (\bibinfo {year} {2015})},\ \Eprint
  {http://arxiv.org/abs/1503.00732} {arXiv:1503.00732 [hep-ph]} \BibitemShut
  {NoStop}%
\bibitem [{\citenamefont {Andersen}\ \emph {et~al.}(2016)\citenamefont
  {Andersen}, \citenamefont {Naylor},\ and\ \citenamefont
  {Tranberg}}]{Andersen:2014xxa}%
  \BibitemOpen
  \bibfield  {author} {\bibinfo {author} {\bibfnamefont {J.~O.}\ \bibnamefont
  {Andersen}}, \bibinfo {author} {\bibfnamefont {W.~R.}\ \bibnamefont
  {Naylor}}, \ and\ \bibinfo {author} {\bibfnamefont {A.}~\bibnamefont
  {Tranberg}},\ }\href {\doibase 10.1103/RevModPhys.88.025001} {\bibfield
  {journal} {\bibinfo  {journal} {Rev. Mod. Phys.}\ }\textbf {\bibinfo {volume}
  {88}},\ \bibinfo {pages} {025001} (\bibinfo {year} {2016})},\ \Eprint
  {http://arxiv.org/abs/1411.7176} {arXiv:1411.7176 [hep-ph]} \BibitemShut
  {NoStop}%
\bibitem [{\citenamefont {Friman}\ \emph {et~al.}(2011)\citenamefont {Friman},
  \citenamefont {Hohne}, \citenamefont {Knoll}, \citenamefont {Leupold},
  \citenamefont {Randrup}, \citenamefont {Rapp},\ and\ \citenamefont
  {Senger}}]{Friman:2011zz}%
  \BibitemOpen
  \bibinfo {editor} {\bibfnamefont {B.}~\bibnamefont {Friman}}, \bibinfo
  {editor} {\bibfnamefont {C.}~\bibnamefont {Hohne}}, \bibinfo {editor}
  {\bibfnamefont {J.}~\bibnamefont {Knoll}}, \bibinfo {editor} {\bibfnamefont
  {S.}~\bibnamefont {Leupold}}, \bibinfo {editor} {\bibfnamefont
  {J.}~\bibnamefont {Randrup}}, \bibinfo {editor} {\bibfnamefont
  {R.}~\bibnamefont {Rapp}}, \ and\ \bibinfo {editor} {\bibfnamefont
  {P.}~\bibnamefont {Senger}},\ eds.,\ \href {\doibase
  10.1007/978-3-642-13293-3} {\emph {\bibinfo {title} {{The CBM physics book:
  Compressed baryonic matter in laboratory experiments}}}},\ Vol.\ \bibinfo
  {volume} {814}\ (\bibinfo {year} {2011})\BibitemShut {NoStop}%
\bibitem [{\citenamefont {Bali}\ \emph
  {et~al.}(2012{\natexlab{a}})\citenamefont {Bali}, \citenamefont {Bruckmann},
  \citenamefont {Endrodi}, \citenamefont {Fodor}, \citenamefont {Katz},
  \citenamefont {Krieg}, \citenamefont {Schafer},\ and\ \citenamefont
  {Szabo}}]{Bali:2011qj}%
  \BibitemOpen
  \bibfield  {author} {\bibinfo {author} {\bibfnamefont {G.~S.}\ \bibnamefont
  {Bali}}, \bibinfo {author} {\bibfnamefont {F.}~\bibnamefont {Bruckmann}},
  \bibinfo {author} {\bibfnamefont {G.}~\bibnamefont {Endrodi}}, \bibinfo
  {author} {\bibfnamefont {Z.}~\bibnamefont {Fodor}}, \bibinfo {author}
  {\bibfnamefont {S.~D.}\ \bibnamefont {Katz}}, \bibinfo {author}
  {\bibfnamefont {S.}~\bibnamefont {Krieg}}, \bibinfo {author} {\bibfnamefont
  {A.}~\bibnamefont {Schafer}}, \ and\ \bibinfo {author} {\bibfnamefont
  {K.~K.}\ \bibnamefont {Szabo}},\ }\href {\doibase 10.1007/JHEP02(2012)044}
  {\bibfield  {journal} {\bibinfo  {journal} {JHEP}\ }\textbf {\bibinfo
  {volume} {02}},\ \bibinfo {pages} {044} (\bibinfo {year}
  {2012}{\natexlab{a}})},\ \Eprint {http://arxiv.org/abs/1111.4956}
  {arXiv:1111.4956 [hep-lat]} \BibitemShut {NoStop}%
\bibitem [{\citenamefont {Abdallah}\ \emph {et~al.}(2022)\citenamefont
  {Abdallah} \emph {et~al.}}]{STAR:2021mii}%
  \BibitemOpen
  \bibfield  {author} {\bibinfo {author} {\bibfnamefont {M.}~\bibnamefont
  {Abdallah}} \emph {et~al.} (\bibinfo {collaboration} {STAR}),\ }\href
  {\doibase 10.1103/PhysRevC.105.014901} {\bibfield  {journal} {\bibinfo
  {journal} {Phys. Rev. C}\ }\textbf {\bibinfo {volume} {105}},\ \bibinfo
  {pages} {014901} (\bibinfo {year} {2022})},\ \Eprint
  {http://arxiv.org/abs/2109.00131} {arXiv:2109.00131 [nucl-ex]} \BibitemShut
  {NoStop}%
\bibitem [{\citenamefont {An}\ \emph {et~al.}(2022)\citenamefont {An} \emph
  {et~al.}}]{An:2021wof}%
  \BibitemOpen
  \bibfield  {author} {\bibinfo {author} {\bibfnamefont {X.}~\bibnamefont {An}}
  \emph {et~al.},\ }\href {\doibase 10.1016/j.nuclphysa.2021.122343} {\bibfield
   {journal} {\bibinfo  {journal} {Nucl. Phys. A}\ }\textbf {\bibinfo {volume}
  {1017}},\ \bibinfo {pages} {122343} (\bibinfo {year} {2022})},\ \Eprint
  {http://arxiv.org/abs/2108.13867} {arXiv:2108.13867 [nucl-th]} \BibitemShut
  {NoStop}%
\bibitem [{\citenamefont {Milton}\ \emph {et~al.}(2021)\citenamefont {Milton},
  \citenamefont {Wang}, \citenamefont {Sergeeva}, \citenamefont {Shi},
  \citenamefont {Liao},\ and\ \citenamefont {Huang}}]{Milton:2021wku}%
  \BibitemOpen
  \bibfield  {author} {\bibinfo {author} {\bibfnamefont {R.}~\bibnamefont
  {Milton}}, \bibinfo {author} {\bibfnamefont {G.}~\bibnamefont {Wang}},
  \bibinfo {author} {\bibfnamefont {M.}~\bibnamefont {Sergeeva}}, \bibinfo
  {author} {\bibfnamefont {S.}~\bibnamefont {Shi}}, \bibinfo {author}
  {\bibfnamefont {J.}~\bibnamefont {Liao}}, \ and\ \bibinfo {author}
  {\bibfnamefont {H.~Z.}\ \bibnamefont {Huang}},\ }\href {\doibase
  10.1103/PhysRevC.104.064906} {\bibfield  {journal} {\bibinfo  {journal}
  {Phys. Rev. C}\ }\textbf {\bibinfo {volume} {104}},\ \bibinfo {pages}
  {064906} (\bibinfo {year} {2021})},\ \Eprint
  {http://arxiv.org/abs/2110.01435} {arXiv:2110.01435 [nucl-th]} \BibitemShut
  {NoStop}%
\bibitem [{\citenamefont {Kharzeev}\ \emph {et~al.}(2022)\citenamefont
  {Kharzeev}, \citenamefont {Liao},\ and\ \citenamefont
  {Shi}}]{Kharzeev:2022hqz}%
  \BibitemOpen
  \bibfield  {author} {\bibinfo {author} {\bibfnamefont {D.~E.}\ \bibnamefont
  {Kharzeev}}, \bibinfo {author} {\bibfnamefont {J.}~\bibnamefont {Liao}}, \
  and\ \bibinfo {author} {\bibfnamefont {S.}~\bibnamefont {Shi}},\ }\href@noop
  {} {\  (\bibinfo {year} {2022})},\ \Eprint {http://arxiv.org/abs/2205.00120}
  {arXiv:2205.00120 [nucl-th]} \BibitemShut {NoStop}%
\bibitem [{\citenamefont {Kharzeev}\ \emph {et~al.}(2008)\citenamefont
  {Kharzeev}, \citenamefont {McLerran},\ and\ \citenamefont
  {Warringa}}]{Kharzeev:2007jp}%
  \BibitemOpen
  \bibfield  {author} {\bibinfo {author} {\bibfnamefont {D.~E.}\ \bibnamefont
  {Kharzeev}}, \bibinfo {author} {\bibfnamefont {L.~D.}\ \bibnamefont
  {McLerran}}, \ and\ \bibinfo {author} {\bibfnamefont {H.~J.}\ \bibnamefont
  {Warringa}},\ }\href {\doibase 10.1016/j.nuclphysa.2008.02.298} {\bibfield
  {journal} {\bibinfo  {journal} {Nucl. Phys.}\ }\textbf {\bibinfo {volume}
  {A803}},\ \bibinfo {pages} {227} (\bibinfo {year} {2008})},\ \Eprint
  {http://arxiv.org/abs/0711.0950} {arXiv:0711.0950 [hep-ph]} \BibitemShut
  {NoStop}%
\bibitem [{\citenamefont {Skokov}\ \emph {et~al.}(2009)\citenamefont {Skokov},
  \citenamefont {Illarionov},\ and\ \citenamefont {Toneev}}]{Skokov:2009qp}%
  \BibitemOpen
  \bibfield  {author} {\bibinfo {author} {\bibfnamefont {V.}~\bibnamefont
  {Skokov}}, \bibinfo {author} {\bibfnamefont {A.~{\relax Yu}.}\ \bibnamefont
  {Illarionov}}, \ and\ \bibinfo {author} {\bibfnamefont {V.}~\bibnamefont
  {Toneev}},\ }\href {\doibase 10.1142/S0217751X09047570} {\bibfield  {journal}
  {\bibinfo  {journal} {Int. J. Mod. Phys.}\ }\textbf {\bibinfo {volume}
  {A24}},\ \bibinfo {pages} {5925} (\bibinfo {year} {2009})},\ \Eprint
  {http://arxiv.org/abs/0907.1396} {arXiv:0907.1396 [nucl-th]} \BibitemShut
  {NoStop}%
\bibitem [{\citenamefont {Tuchin}(2013{\natexlab{a}})}]{Tuchin:2013apa}%
  \BibitemOpen
  \bibfield  {author} {\bibinfo {author} {\bibfnamefont {K.}~\bibnamefont
  {Tuchin}},\ }\href {\doibase 10.1103/PhysRevC.88.024911} {\bibfield
  {journal} {\bibinfo  {journal} {Phys. Rev. C}\ }\textbf {\bibinfo {volume}
  {88}},\ \bibinfo {pages} {024911} (\bibinfo {year} {2013}{\natexlab{a}})},\
  \Eprint {http://arxiv.org/abs/1305.5806} {arXiv:1305.5806 [hep-ph]}
  \BibitemShut {NoStop}%
\bibitem [{\citenamefont {Tuchin}(2016)}]{Tuchin:2015oka}%
  \BibitemOpen
  \bibfield  {author} {\bibinfo {author} {\bibfnamefont {K.}~\bibnamefont
  {Tuchin}},\ }\href {\doibase 10.1103/PhysRevC.93.014905} {\bibfield
  {journal} {\bibinfo  {journal} {Phys. Rev. C}\ }\textbf {\bibinfo {volume}
  {93}},\ \bibinfo {pages} {014905} (\bibinfo {year} {2016})},\ \Eprint
  {http://arxiv.org/abs/1508.06925} {arXiv:1508.06925 [hep-ph]} \BibitemShut
  {NoStop}%
\bibitem [{\citenamefont {Tuchin}(2013{\natexlab{b}})}]{Tuchin:2013ie}%
  \BibitemOpen
  \bibfield  {author} {\bibinfo {author} {\bibfnamefont {K.}~\bibnamefont
  {Tuchin}},\ }\href {\doibase 10.1155/2013/490495} {\bibfield  {journal}
  {\bibinfo  {journal} {Adv. High Energy Phys.}\ }\textbf {\bibinfo {volume}
  {2013}},\ \bibinfo {pages} {490495} (\bibinfo {year} {2013}{\natexlab{b}})},\
  \Eprint {http://arxiv.org/abs/1301.0099} {arXiv:1301.0099 [hep-ph]}
  \BibitemShut {NoStop}%
\bibitem [{\citenamefont {Gursoy}\ \emph {et~al.}(2014)\citenamefont {Gursoy},
  \citenamefont {Kharzeev},\ and\ \citenamefont {Rajagopal}}]{Gursoy:2014aka}%
  \BibitemOpen
  \bibfield  {author} {\bibinfo {author} {\bibfnamefont {U.}~\bibnamefont
  {Gursoy}}, \bibinfo {author} {\bibfnamefont {D.}~\bibnamefont {Kharzeev}}, \
  and\ \bibinfo {author} {\bibfnamefont {K.}~\bibnamefont {Rajagopal}},\ }\href
  {\doibase 10.1103/PhysRevC.89.054905} {\bibfield  {journal} {\bibinfo
  {journal} {Phys. Rev.}\ }\textbf {\bibinfo {volume} {C89}},\ \bibinfo {pages}
  {054905} (\bibinfo {year} {2014})},\ \Eprint {http://arxiv.org/abs/1401.3805}
  {arXiv:1401.3805 [hep-ph]} \BibitemShut {NoStop}%
\bibitem [{\citenamefont {Duncan}\ and\ \citenamefont
  {Thompson}(1992)}]{Duncan:1992hi}%
  \BibitemOpen
  \bibfield  {author} {\bibinfo {author} {\bibfnamefont {R.~C.}\ \bibnamefont
  {Duncan}}\ and\ \bibinfo {author} {\bibfnamefont {C.}~\bibnamefont
  {Thompson}},\ }\href {\doibase 10.1086/186413} {\bibfield  {journal}
  {\bibinfo  {journal} {Astrophys. J.}\ }\textbf {\bibinfo {volume} {392}},\
  \bibinfo {pages} {L9} (\bibinfo {year} {1992})}\BibitemShut {NoStop}%
\bibitem [{\citenamefont {Thompson}\ and\ \citenamefont
  {Duncan}(1993)}]{Thompson:1993hn}%
  \BibitemOpen
  \bibfield  {author} {\bibinfo {author} {\bibfnamefont {C.}~\bibnamefont
  {Thompson}}\ and\ \bibinfo {author} {\bibfnamefont {R.~C.}\ \bibnamefont
  {Duncan}},\ }\href {\doibase 10.1086/172580} {\bibfield  {journal} {\bibinfo
  {journal} {Astrophys. J.}\ }\textbf {\bibinfo {volume} {408}},\ \bibinfo
  {pages} {194} (\bibinfo {year} {1993})}\BibitemShut {NoStop}%
\bibitem [{\citenamefont {Vachaspati}(1991)}]{Vachaspati:1991nm}%
  \BibitemOpen
  \bibfield  {author} {\bibinfo {author} {\bibfnamefont {T.}~\bibnamefont
  {Vachaspati}},\ }\href {\doibase 10.1016/0370-2693(91)90051-Q} {\bibfield
  {journal} {\bibinfo  {journal} {Phys. Lett.}\ }\textbf {\bibinfo {volume}
  {B265}},\ \bibinfo {pages} {258} (\bibinfo {year} {1991})}\BibitemShut
  {NoStop}%
\bibitem [{\citenamefont {Campanelli}(2013)}]{Campanelli:2013mea}%
  \BibitemOpen
  \bibfield  {author} {\bibinfo {author} {\bibfnamefont {L.}~\bibnamefont
  {Campanelli}},\ }\href {\doibase 10.1103/PhysRevLett.111.061301} {\bibfield
  {journal} {\bibinfo  {journal} {Phys. Rev. Lett.}\ }\textbf {\bibinfo
  {volume} {111}},\ \bibinfo {pages} {061301} (\bibinfo {year} {2013})},\
  \Eprint {http://arxiv.org/abs/1304.6534} {arXiv:1304.6534 [astro-ph.CO]}
  \BibitemShut {NoStop}%
\bibitem [{\citenamefont {Kharzeev}\ \emph
  {et~al.}(2013{\natexlab{b}})\citenamefont {Kharzeev}, \citenamefont
  {Landsteiner}, \citenamefont {Schmitt},\ and\ \citenamefont
  {Yee}}]{Kharzeev:2012ph}%
  \BibitemOpen
  \bibfield  {author} {\bibinfo {author} {\bibfnamefont {D.~E.}\ \bibnamefont
  {Kharzeev}}, \bibinfo {author} {\bibfnamefont {K.}~\bibnamefont
  {Landsteiner}}, \bibinfo {author} {\bibfnamefont {A.}~\bibnamefont
  {Schmitt}}, \ and\ \bibinfo {author} {\bibfnamefont {H.-U.}\ \bibnamefont
  {Yee}},\ }\href {\doibase 10.1007/978-3-642-37305-3_1} {\bibfield  {journal}
  {\bibinfo  {journal} {Lect. Notes Phys.}\ }\textbf {\bibinfo {volume}
  {871}},\ \bibinfo {pages} {1} (\bibinfo {year} {2013}{\natexlab{b}})},\
  \Eprint {http://arxiv.org/abs/1211.6245} {arXiv:1211.6245 [hep-ph]}
  \BibitemShut {NoStop}%
\bibitem [{\citenamefont {Kharzeev}\ and\ \citenamefont
  {Zhitnitsky}(2007)}]{Kharzeev:2007tn}%
  \BibitemOpen
  \bibfield  {author} {\bibinfo {author} {\bibfnamefont {D.}~\bibnamefont
  {Kharzeev}}\ and\ \bibinfo {author} {\bibfnamefont {A.}~\bibnamefont
  {Zhitnitsky}},\ }\href {\doibase 10.1016/j.nuclphysa.2007.10.001} {\bibfield
  {journal} {\bibinfo  {journal} {Nucl. Phys.}\ }\textbf {\bibinfo {volume}
  {A797}},\ \bibinfo {pages} {67} (\bibinfo {year} {2007})},\ \Eprint
  {http://arxiv.org/abs/0706.1026} {arXiv:0706.1026 [hep-ph]} \BibitemShut
  {NoStop}%
\bibitem [{\citenamefont {Chernodub}(2010)}]{Chernodub:2010qx}%
  \BibitemOpen
  \bibfield  {author} {\bibinfo {author} {\bibfnamefont {M.~N.}\ \bibnamefont
  {Chernodub}},\ }\href {\doibase 10.1103/PhysRevD.82.085011} {\bibfield
  {journal} {\bibinfo  {journal} {Phys. Rev.}\ }\textbf {\bibinfo {volume}
  {D82}},\ \bibinfo {pages} {085011} (\bibinfo {year} {2010})},\ \Eprint
  {http://arxiv.org/abs/1008.1055} {arXiv:1008.1055 [hep-ph]} \BibitemShut
  {NoStop}%
\bibitem [{\citenamefont {Chernodub}(2013)}]{Chernodub:2012tf}%
  \BibitemOpen
  \bibfield  {author} {\bibinfo {author} {\bibfnamefont {M.~N.}\ \bibnamefont
  {Chernodub}},\ }\href {\doibase 10.1007/978-3-642-37305-3_6} {\bibfield
  {journal} {\bibinfo  {journal} {Lect. Notes Phys.}\ }\textbf {\bibinfo
  {volume} {871}},\ \bibinfo {pages} {143} (\bibinfo {year} {2013})},\ \Eprint
  {http://arxiv.org/abs/1208.5025} {arXiv:1208.5025 [hep-ph]} \BibitemShut
  {NoStop}%
\bibitem [{\citenamefont {Fukushima}\ \emph {et~al.}(2008)\citenamefont
  {Fukushima}, \citenamefont {Kharzeev},\ and\ \citenamefont
  {Warringa}}]{Fukushima:2008xe}%
  \BibitemOpen
  \bibfield  {author} {\bibinfo {author} {\bibfnamefont {K.}~\bibnamefont
  {Fukushima}}, \bibinfo {author} {\bibfnamefont {D.~E.}\ \bibnamefont
  {Kharzeev}}, \ and\ \bibinfo {author} {\bibfnamefont {H.~J.}\ \bibnamefont
  {Warringa}},\ }\href {\doibase 10.1103/PhysRevD.78.074033} {\bibfield
  {journal} {\bibinfo  {journal} {Phys. Rev.}\ }\textbf {\bibinfo {volume}
  {D78}},\ \bibinfo {pages} {074033} (\bibinfo {year} {2008})},\ \Eprint
  {http://arxiv.org/abs/0808.3382} {arXiv:0808.3382 [hep-ph]} \BibitemShut
  {NoStop}%
\bibitem [{\citenamefont {Kharzeev}\ and\ \citenamefont
  {Warringa}(2009)}]{Kharzeev:2009pj}%
  \BibitemOpen
  \bibfield  {author} {\bibinfo {author} {\bibfnamefont {D.~E.}\ \bibnamefont
  {Kharzeev}}\ and\ \bibinfo {author} {\bibfnamefont {H.~J.}\ \bibnamefont
  {Warringa}},\ }\href {\doibase 10.1103/PhysRevD.80.034028} {\bibfield
  {journal} {\bibinfo  {journal} {Phys. Rev.}\ }\textbf {\bibinfo {volume}
  {D80}},\ \bibinfo {pages} {034028} (\bibinfo {year} {2009})},\ \Eprint
  {http://arxiv.org/abs/0907.5007} {arXiv:0907.5007 [hep-ph]} \BibitemShut
  {NoStop}%
\bibitem [{\citenamefont {Shovkovy}(2013)}]{Shovkovy:2012zn}%
  \BibitemOpen
  \bibfield  {author} {\bibinfo {author} {\bibfnamefont {I.~A.}\ \bibnamefont
  {Shovkovy}},\ }\href {\doibase 10.1007/978-3-642-37305-3_2} {\bibfield
  {journal} {\bibinfo  {journal} {Lect. Notes Phys.}\ }\textbf {\bibinfo
  {volume} {871}},\ \bibinfo {pages} {13} (\bibinfo {year} {2013})},\ \Eprint
  {http://arxiv.org/abs/1207.5081} {arXiv:1207.5081 [hep-ph]} \BibitemShut
  {NoStop}%
\bibitem [{\citenamefont {Gusynin}\ \emph {et~al.}(1994)\citenamefont
  {Gusynin}, \citenamefont {Miransky},\ and\ \citenamefont
  {Shovkovy}}]{Gusynin:1994re}%
  \BibitemOpen
  \bibfield  {author} {\bibinfo {author} {\bibfnamefont {V.~P.}\ \bibnamefont
  {Gusynin}}, \bibinfo {author} {\bibfnamefont {V.~A.}\ \bibnamefont
  {Miransky}}, \ and\ \bibinfo {author} {\bibfnamefont {I.~A.}\ \bibnamefont
  {Shovkovy}},\ }\href {\doibase 10.1103/PhysRevLett.76.1005,
  10.1103/PhysRevLett.73.3499} {\bibfield  {journal} {\bibinfo  {journal}
  {Phys. Rev. Lett.}\ }\textbf {\bibinfo {volume} {73}},\ \bibinfo {pages}
  {3499} (\bibinfo {year} {1994})},\ \bibinfo {note} {[Erratum: Phys. Rev.
  Lett.76,1005(1996)]},\ \Eprint {http://arxiv.org/abs/hep-ph/9405262}
  {arXiv:hep-ph/9405262 [hep-ph]} \BibitemShut {NoStop}%
\bibitem [{\citenamefont {Gusynin}\ \emph {et~al.}(1996)\citenamefont
  {Gusynin}, \citenamefont {Miransky},\ and\ \citenamefont
  {Shovkovy}}]{Gusynin:1995nb}%
  \BibitemOpen
  \bibfield  {author} {\bibinfo {author} {\bibfnamefont {V.~P.}\ \bibnamefont
  {Gusynin}}, \bibinfo {author} {\bibfnamefont {V.~A.}\ \bibnamefont
  {Miransky}}, \ and\ \bibinfo {author} {\bibfnamefont {I.~A.}\ \bibnamefont
  {Shovkovy}},\ }\href {\doibase 10.1016/0550-3213(96)00021-1} {\bibfield
  {journal} {\bibinfo  {journal} {Nucl. Phys.}\ }\textbf {\bibinfo {volume}
  {B462}},\ \bibinfo {pages} {249} (\bibinfo {year} {1996})},\ \Eprint
  {http://arxiv.org/abs/hep-ph/9509320} {arXiv:hep-ph/9509320 [hep-ph]}
  \BibitemShut {NoStop}%
\bibitem [{\citenamefont {Gusynin}\ \emph {et~al.}(1999)\citenamefont
  {Gusynin}, \citenamefont {Miransky},\ and\ \citenamefont
  {Shovkovy}}]{Gusynin:1999pq}%
  \BibitemOpen
  \bibfield  {author} {\bibinfo {author} {\bibfnamefont {V.~P.}\ \bibnamefont
  {Gusynin}}, \bibinfo {author} {\bibfnamefont {V.~A.}\ \bibnamefont
  {Miransky}}, \ and\ \bibinfo {author} {\bibfnamefont {I.~A.}\ \bibnamefont
  {Shovkovy}},\ }\href {\doibase 10.1016/S0550-3213(99)00573-8} {\bibfield
  {journal} {\bibinfo  {journal} {Nucl. Phys.}\ }\textbf {\bibinfo {volume}
  {B563}},\ \bibinfo {pages} {361} (\bibinfo {year} {1999})},\ \Eprint
  {http://arxiv.org/abs/hep-ph/9908320} {arXiv:hep-ph/9908320 [hep-ph]}
  \BibitemShut {NoStop}%
\bibitem [{\citenamefont {Preis}\ \emph {et~al.}(2011)\citenamefont {Preis},
  \citenamefont {Rebhan},\ and\ \citenamefont {Schmitt}}]{Preis:2010cq}%
  \BibitemOpen
  \bibfield  {author} {\bibinfo {author} {\bibfnamefont {F.}~\bibnamefont
  {Preis}}, \bibinfo {author} {\bibfnamefont {A.}~\bibnamefont {Rebhan}}, \
  and\ \bibinfo {author} {\bibfnamefont {A.}~\bibnamefont {Schmitt}},\ }\href
  {\doibase 10.1007/JHEP03(2011)033} {\bibfield  {journal} {\bibinfo  {journal}
  {JHEP}\ }\textbf {\bibinfo {volume} {03}},\ \bibinfo {pages} {033} (\bibinfo
  {year} {2011})},\ \Eprint {http://arxiv.org/abs/1012.4785} {arXiv:1012.4785
  [hep-th]} \BibitemShut {NoStop}%
\bibitem [{\citenamefont {Preis}\ \emph {et~al.}(2013)\citenamefont {Preis},
  \citenamefont {Rebhan},\ and\ \citenamefont {Schmitt}}]{Preis:2012fh}%
  \BibitemOpen
  \bibfield  {author} {\bibinfo {author} {\bibfnamefont {F.}~\bibnamefont
  {Preis}}, \bibinfo {author} {\bibfnamefont {A.}~\bibnamefont {Rebhan}}, \
  and\ \bibinfo {author} {\bibfnamefont {A.}~\bibnamefont {Schmitt}},\ }\href
  {\doibase 10.1007/978-3-642-37305-3_3} {\bibfield  {journal} {\bibinfo
  {journal} {Lect. Notes Phys.}\ }\textbf {\bibinfo {volume} {871}},\ \bibinfo
  {pages} {51} (\bibinfo {year} {2013})},\ \Eprint
  {http://arxiv.org/abs/1208.0536} {arXiv:1208.0536 [hep-ph]} \BibitemShut
  {NoStop}%
\bibitem [{\citenamefont {Chernodub}\ \emph {et~al.}(2012)\citenamefont
  {Chernodub}, \citenamefont {Van~Doorsselaere},\ and\ \citenamefont
  {Verschelde}}]{Chernodub:2011gs}%
  \BibitemOpen
  \bibfield  {author} {\bibinfo {author} {\bibfnamefont {M.~N.}\ \bibnamefont
  {Chernodub}}, \bibinfo {author} {\bibfnamefont {J.}~\bibnamefont
  {Van~Doorsselaere}}, \ and\ \bibinfo {author} {\bibfnamefont
  {H.}~\bibnamefont {Verschelde}},\ }\href {\doibase
  10.1103/PhysRevD.85.045002} {\bibfield  {journal} {\bibinfo  {journal} {Phys.
  Rev.}\ }\textbf {\bibinfo {volume} {D85}},\ \bibinfo {pages} {045002}
  (\bibinfo {year} {2012})},\ \Eprint {http://arxiv.org/abs/1111.4401}
  {arXiv:1111.4401 [hep-ph]} \BibitemShut {NoStop}%
\bibitem [{\citenamefont {Chernodub}(2011)}]{Chernodub:2011mc}%
  \BibitemOpen
  \bibfield  {author} {\bibinfo {author} {\bibfnamefont {M.~N.}\ \bibnamefont
  {Chernodub}},\ }\href {\doibase 10.1103/PhysRevLett.106.142003} {\bibfield
  {journal} {\bibinfo  {journal} {Phys. Rev. Lett.}\ }\textbf {\bibinfo
  {volume} {106}},\ \bibinfo {pages} {142003} (\bibinfo {year} {2011})},\
  \Eprint {http://arxiv.org/abs/1101.0117} {arXiv:1101.0117 [hep-ph]}
  \BibitemShut {NoStop}%
\bibitem [{\citenamefont {de~Forcrand}\ and\ \citenamefont
  {Philipsen}(2007)}]{deForcrand:2006pv}%
  \BibitemOpen
  \bibfield  {author} {\bibinfo {author} {\bibfnamefont {P.}~\bibnamefont
  {de~Forcrand}}\ and\ \bibinfo {author} {\bibfnamefont {O.}~\bibnamefont
  {Philipsen}},\ }\href {\doibase 10.1088/1126-6708/2007/01/077} {\bibfield
  {journal} {\bibinfo  {journal} {JHEP}\ }\textbf {\bibinfo {volume} {01}},\
  \bibinfo {pages} {077} (\bibinfo {year} {2007})},\ \Eprint
  {http://arxiv.org/abs/hep-lat/0607017} {arXiv:hep-lat/0607017} \BibitemShut
  {NoStop}%
\bibitem [{\citenamefont {Aoki}\ \emph {et~al.}(2006)\citenamefont {Aoki},
  \citenamefont {Fodor}, \citenamefont {Katz},\ and\ \citenamefont
  {Szabo}}]{Aoki:2006br}%
  \BibitemOpen
  \bibfield  {author} {\bibinfo {author} {\bibfnamefont {Y.}~\bibnamefont
  {Aoki}}, \bibinfo {author} {\bibfnamefont {Z.}~\bibnamefont {Fodor}},
  \bibinfo {author} {\bibfnamefont {S.~D.}\ \bibnamefont {Katz}}, \ and\
  \bibinfo {author} {\bibfnamefont {K.~K.}\ \bibnamefont {Szabo}},\ }\href
  {\doibase 10.1016/j.physletb.2006.10.021} {\bibfield  {journal} {\bibinfo
  {journal} {Phys. Lett. B}\ }\textbf {\bibinfo {volume} {643}},\ \bibinfo
  {pages} {46} (\bibinfo {year} {2006})},\ \Eprint
  {http://arxiv.org/abs/hep-lat/0609068} {arXiv:hep-lat/0609068} \BibitemShut
  {NoStop}%
\bibitem [{\citenamefont {Aoki}\ \emph {et~al.}(2009)\citenamefont {Aoki},
  \citenamefont {Borsanyi}, \citenamefont {Durr}, \citenamefont {Fodor},
  \citenamefont {Katz}, \citenamefont {Krieg},\ and\ \citenamefont
  {Szabo}}]{Aoki:2009sc}%
  \BibitemOpen
  \bibfield  {author} {\bibinfo {author} {\bibfnamefont {Y.}~\bibnamefont
  {Aoki}}, \bibinfo {author} {\bibfnamefont {S.}~\bibnamefont {Borsanyi}},
  \bibinfo {author} {\bibfnamefont {S.}~\bibnamefont {Durr}}, \bibinfo {author}
  {\bibfnamefont {Z.}~\bibnamefont {Fodor}}, \bibinfo {author} {\bibfnamefont
  {S.~D.}\ \bibnamefont {Katz}}, \bibinfo {author} {\bibfnamefont
  {S.}~\bibnamefont {Krieg}}, \ and\ \bibinfo {author} {\bibfnamefont {K.~K.}\
  \bibnamefont {Szabo}},\ }\href {\doibase 10.1088/1126-6708/2009/06/088}
  {\bibfield  {journal} {\bibinfo  {journal} {JHEP}\ }\textbf {\bibinfo
  {volume} {06}},\ \bibinfo {pages} {088} (\bibinfo {year} {2009})},\ \Eprint
  {http://arxiv.org/abs/0903.4155} {arXiv:0903.4155 [hep-lat]} \BibitemShut
  {NoStop}%
\bibitem [{\citenamefont {Bazavov}\ \emph {et~al.}(2009)\citenamefont {Bazavov}
  \emph {et~al.}}]{Bazavov:2009zn}%
  \BibitemOpen
  \bibfield  {author} {\bibinfo {author} {\bibfnamefont {A.}~\bibnamefont
  {Bazavov}} \emph {et~al.},\ }\href {\doibase 10.1103/PhysRevD.80.014504}
  {\bibfield  {journal} {\bibinfo  {journal} {Phys. Rev. D}\ }\textbf {\bibinfo
  {volume} {80}},\ \bibinfo {pages} {014504} (\bibinfo {year} {2009})},\
  \Eprint {http://arxiv.org/abs/0903.4379} {arXiv:0903.4379 [hep-lat]}
  \BibitemShut {NoStop}%
\bibitem [{\citenamefont {Cheng}\ \emph {et~al.}(2008)\citenamefont {Cheng}
  \emph {et~al.}}]{Cheng:2007jq}%
  \BibitemOpen
  \bibfield  {author} {\bibinfo {author} {\bibfnamefont {M.}~\bibnamefont
  {Cheng}} \emph {et~al.},\ }\href {\doibase 10.1103/PhysRevD.77.014511}
  {\bibfield  {journal} {\bibinfo  {journal} {Phys. Rev. D}\ }\textbf {\bibinfo
  {volume} {77}},\ \bibinfo {pages} {014511} (\bibinfo {year} {2008})},\
  \Eprint {http://arxiv.org/abs/0710.0354} {arXiv:0710.0354 [hep-lat]}
  \BibitemShut {NoStop}%
\bibitem [{\citenamefont {Muroya}\ \emph {et~al.}(2003)\citenamefont {Muroya},
  \citenamefont {Nakamura}, \citenamefont {Nonaka},\ and\ \citenamefont
  {Takaishi}}]{Muroya:2003qs}%
  \BibitemOpen
  \bibfield  {author} {\bibinfo {author} {\bibfnamefont {S.}~\bibnamefont
  {Muroya}}, \bibinfo {author} {\bibfnamefont {A.}~\bibnamefont {Nakamura}},
  \bibinfo {author} {\bibfnamefont {C.}~\bibnamefont {Nonaka}}, \ and\ \bibinfo
  {author} {\bibfnamefont {T.}~\bibnamefont {Takaishi}},\ }\href {\doibase
  10.1143/PTP.110.615} {\bibfield  {journal} {\bibinfo  {journal} {Prog. Theor.
  Phys.}\ }\textbf {\bibinfo {volume} {110}},\ \bibinfo {pages} {615} (\bibinfo
  {year} {2003})},\ \Eprint {http://arxiv.org/abs/hep-lat/0306031}
  {arXiv:hep-lat/0306031} \BibitemShut {NoStop}%
\bibitem [{\citenamefont {Splittorff}\ and\ \citenamefont
  {Verbaarschot}(2007)}]{Splittorff:2006fu}%
  \BibitemOpen
  \bibfield  {author} {\bibinfo {author} {\bibfnamefont {K.}~\bibnamefont
  {Splittorff}}\ and\ \bibinfo {author} {\bibfnamefont {J.~J.~M.}\ \bibnamefont
  {Verbaarschot}},\ }\href {\doibase 10.1103/PhysRevLett.98.031601} {\bibfield
  {journal} {\bibinfo  {journal} {Phys. Rev. Lett.}\ }\textbf {\bibinfo
  {volume} {98}},\ \bibinfo {pages} {031601} (\bibinfo {year} {2007})},\
  \Eprint {http://arxiv.org/abs/hep-lat/0609076} {arXiv:hep-lat/0609076}
  \BibitemShut {NoStop}%
\bibitem [{\citenamefont {Fukushima}\ and\ \citenamefont
  {Hidaka}(2007)}]{Fukushima:2006uv}%
  \BibitemOpen
  \bibfield  {author} {\bibinfo {author} {\bibfnamefont {K.}~\bibnamefont
  {Fukushima}}\ and\ \bibinfo {author} {\bibfnamefont {Y.}~\bibnamefont
  {Hidaka}},\ }\href {\doibase 10.1103/PhysRevD.75.036002} {\bibfield
  {journal} {\bibinfo  {journal} {Phys. Rev. D}\ }\textbf {\bibinfo {volume}
  {75}},\ \bibinfo {pages} {036002} (\bibinfo {year} {2007})},\ \Eprint
  {http://arxiv.org/abs/hep-ph/0610323} {arXiv:hep-ph/0610323} \BibitemShut
  {NoStop}%
\bibitem [{\citenamefont {Nambu}\ and\ \citenamefont
  {Jona-Lasinio}(1961{\natexlab{a}})}]{Nambu:1961fr}%
  \BibitemOpen
  \bibfield  {author} {\bibinfo {author} {\bibfnamefont {Y.}~\bibnamefont
  {Nambu}}\ and\ \bibinfo {author} {\bibfnamefont {G.}~\bibnamefont
  {Jona-Lasinio}},\ }\href {\doibase 10.1103/PhysRev.124.246} {\bibfield
  {journal} {\bibinfo  {journal} {Phys. Rev.}\ }\textbf {\bibinfo {volume}
  {124}},\ \bibinfo {pages} {246} (\bibinfo {year}
  {1961}{\natexlab{a}})}\BibitemShut {NoStop}%
\bibitem [{\citenamefont {Nambu}\ and\ \citenamefont
  {Jona-Lasinio}(1961{\natexlab{b}})}]{Nambu:1961tp}%
  \BibitemOpen
  \bibfield  {author} {\bibinfo {author} {\bibfnamefont {Y.}~\bibnamefont
  {Nambu}}\ and\ \bibinfo {author} {\bibfnamefont {G.}~\bibnamefont
  {Jona-Lasinio}},\ }\href {\doibase 10.1103/PhysRev.122.345} {\bibfield
  {journal} {\bibinfo  {journal} {Phys. Rev.}\ }\textbf {\bibinfo {volume}
  {122}},\ \bibinfo {pages} {345} (\bibinfo {year}
  {1961}{\natexlab{b}})}\BibitemShut {NoStop}%
\bibitem [{\citenamefont {Klevansky}(1992)}]{Klevansky:1992qe}%
  \BibitemOpen
  \bibfield  {author} {\bibinfo {author} {\bibfnamefont {S.~P.}\ \bibnamefont
  {Klevansky}},\ }\href {\doibase 10.1103/RevModPhys.64.649} {\bibfield
  {journal} {\bibinfo  {journal} {Rev. Mod. Phys.}\ }\textbf {\bibinfo {volume}
  {64}},\ \bibinfo {pages} {649} (\bibinfo {year} {1992})}\BibitemShut
  {NoStop}%
\bibitem [{\citenamefont {Vogl}\ and\ \citenamefont
  {Weise}(1991)}]{Vogl:1991qt}%
  \BibitemOpen
  \bibfield  {author} {\bibinfo {author} {\bibfnamefont {U.}~\bibnamefont
  {Vogl}}\ and\ \bibinfo {author} {\bibfnamefont {W.}~\bibnamefont {Weise}},\
  }\href {\doibase 10.1016/0146-6410(91)90005-9} {\bibfield  {journal}
  {\bibinfo  {journal} {Prog. Part. Nucl. Phys.}\ }\textbf {\bibinfo {volume}
  {27}},\ \bibinfo {pages} {195} (\bibinfo {year} {1991})}\BibitemShut
  {NoStop}%
\bibitem [{\citenamefont {Buballa}(2005)}]{Buballa:2003qv}%
  \BibitemOpen
  \bibfield  {author} {\bibinfo {author} {\bibfnamefont {M.}~\bibnamefont
  {Buballa}},\ }\href {\doibase 10.1016/j.physrep.2004.11.004} {\bibfield
  {journal} {\bibinfo  {journal} {Phys. Rept.}\ }\textbf {\bibinfo {volume}
  {407}},\ \bibinfo {pages} {205} (\bibinfo {year} {2005})},\ \Eprint
  {http://arxiv.org/abs/hep-ph/0402234} {arXiv:hep-ph/0402234} \BibitemShut
  {NoStop}%
\bibitem [{\citenamefont {Volkov}\ and\ \citenamefont
  {Radzhabov}(2006)}]{Volkov:2005kw}%
  \BibitemOpen
  \bibfield  {author} {\bibinfo {author} {\bibfnamefont {M.~K.}\ \bibnamefont
  {Volkov}}\ and\ \bibinfo {author} {\bibfnamefont {A.~E.}\ \bibnamefont
  {Radzhabov}},\ }\href {\doibase 10.1070/PU2006v049n06ABEH005905} {\bibfield
  {journal} {\bibinfo  {journal} {Phys. Usp.}\ }\textbf {\bibinfo {volume}
  {49}},\ \bibinfo {pages} {551} (\bibinfo {year} {2006})},\ \Eprint
  {http://arxiv.org/abs/hep-ph/0508263} {arXiv:hep-ph/0508263} \BibitemShut
  {NoStop}%
\bibitem [{\citenamefont {Bijnens}(1996)}]{Bijnens:1995ww}%
  \BibitemOpen
  \bibfield  {author} {\bibinfo {author} {\bibfnamefont {J.}~\bibnamefont
  {Bijnens}},\ }\href {\doibase 10.1016/0370-1573(95)00051-8} {\bibfield
  {journal} {\bibinfo  {journal} {Phys. Rept.}\ }\textbf {\bibinfo {volume}
  {265}},\ \bibinfo {pages} {369} (\bibinfo {year} {1996})},\ \Eprint
  {http://arxiv.org/abs/hep-ph/9502335} {arXiv:hep-ph/9502335} \BibitemShut
  {NoStop}%
\bibitem [{\citenamefont {Ratti}\ \emph {et~al.}(2006)\citenamefont {Ratti},
  \citenamefont {Thaler},\ and\ \citenamefont {Weise}}]{Ratti:2005jh}%
  \BibitemOpen
  \bibfield  {author} {\bibinfo {author} {\bibfnamefont {C.}~\bibnamefont
  {Ratti}}, \bibinfo {author} {\bibfnamefont {M.~A.}\ \bibnamefont {Thaler}}, \
  and\ \bibinfo {author} {\bibfnamefont {W.}~\bibnamefont {Weise}},\ }\href
  {\doibase 10.1103/PhysRevD.73.014019} {\bibfield  {journal} {\bibinfo
  {journal} {Phys. Rev. D}\ }\textbf {\bibinfo {volume} {73}},\ \bibinfo
  {pages} {014019} (\bibinfo {year} {2006})},\ \Eprint
  {http://arxiv.org/abs/hep-ph/0506234} {arXiv:hep-ph/0506234} \BibitemShut
  {NoStop}%
\bibitem [{\citenamefont {Ratti}\ \emph {et~al.}(2007)\citenamefont {Ratti},
  \citenamefont {Roessner}, \citenamefont {Thaler},\ and\ \citenamefont
  {Weise}}]{Ratti:2006wg}%
  \BibitemOpen
  \bibfield  {author} {\bibinfo {author} {\bibfnamefont {C.}~\bibnamefont
  {Ratti}}, \bibinfo {author} {\bibfnamefont {S.}~\bibnamefont {Roessner}},
  \bibinfo {author} {\bibfnamefont {M.}~\bibnamefont {Thaler}}, \ and\ \bibinfo
  {author} {\bibfnamefont {W.}~\bibnamefont {Weise}},\ }\href {\doibase
  10.1140/epjc/s10052-006-0065-x} {\bibfield  {journal} {\bibinfo  {journal}
  {Eur. Phys. J. C}\ }\textbf {\bibinfo {volume} {49}},\ \bibinfo {pages} {213}
  (\bibinfo {year} {2007})},\ \Eprint {http://arxiv.org/abs/hep-ph/0609218}
  {arXiv:hep-ph/0609218} \BibitemShut {NoStop}%
\bibitem [{\citenamefont {Chaudhuri}\ \emph {et~al.}(2020)\citenamefont
  {Chaudhuri}, \citenamefont {Ghosh}, \citenamefont {Sarkar},\ and\
  \citenamefont {Roy}}]{Chaudhuri:2020lga}%
  \BibitemOpen
  \bibfield  {author} {\bibinfo {author} {\bibfnamefont {N.}~\bibnamefont
  {Chaudhuri}}, \bibinfo {author} {\bibfnamefont {S.}~\bibnamefont {Ghosh}},
  \bibinfo {author} {\bibfnamefont {S.}~\bibnamefont {Sarkar}}, \ and\ \bibinfo
  {author} {\bibfnamefont {P.}~\bibnamefont {Roy}},\ }\href {\doibase
  10.1140/epja/s10050-020-00222-9} {\bibfield  {journal} {\bibinfo  {journal}
  {Eur. Phys. J. A}\ }\textbf {\bibinfo {volume} {56}},\ \bibinfo {pages} {213}
  (\bibinfo {year} {2020})},\ \Eprint {http://arxiv.org/abs/2003.05692}
  {arXiv:2003.05692 [nucl-th]} \BibitemShut {NoStop}%
\bibitem [{\citenamefont {Gatto}\ and\ \citenamefont
  {Ruggieri}(2010)}]{Gatto:2010qs}%
  \BibitemOpen
  \bibfield  {author} {\bibinfo {author} {\bibfnamefont {R.}~\bibnamefont
  {Gatto}}\ and\ \bibinfo {author} {\bibfnamefont {M.}~\bibnamefont
  {Ruggieri}},\ }\href {\doibase 10.1103/PhysRevD.82.054027} {\bibfield
  {journal} {\bibinfo  {journal} {Phys. Rev. D}\ }\textbf {\bibinfo {volume}
  {82}},\ \bibinfo {pages} {054027} (\bibinfo {year} {2010})},\ \Eprint
  {http://arxiv.org/abs/1007.0790} {arXiv:1007.0790 [hep-ph]} \BibitemShut
  {NoStop}%
\bibitem [{\citenamefont {Fukushima}(2008)}]{Fukushima:2008wg}%
  \BibitemOpen
  \bibfield  {author} {\bibinfo {author} {\bibfnamefont {K.}~\bibnamefont
  {Fukushima}},\ }\href {\doibase 10.1103/PhysRevD.77.114028} {\bibfield
  {journal} {\bibinfo  {journal} {Phys. Rev. D}\ }\textbf {\bibinfo {volume}
  {77}},\ \bibinfo {pages} {114028} (\bibinfo {year} {2008})},\ \bibinfo {note}
  {[Erratum: Phys.Rev.D 78, 039902 (2008)]},\ \Eprint
  {http://arxiv.org/abs/0803.3318} {arXiv:0803.3318 [hep-ph]} \BibitemShut
  {NoStop}%
\bibitem [{\citenamefont {Mattos}\ \emph
  {et~al.}(2021{\natexlab{a}})\citenamefont {Mattos}, \citenamefont
  {Frederico},\ and\ \citenamefont {Louren\c{c}o}}]{Mattos:2021alf}%
  \BibitemOpen
  \bibfield  {author} {\bibinfo {author} {\bibfnamefont {O.~A.}\ \bibnamefont
  {Mattos}}, \bibinfo {author} {\bibfnamefont {T.}~\bibnamefont {Frederico}}, \
  and\ \bibinfo {author} {\bibfnamefont {O.}~\bibnamefont {Louren\c{c}o}},\
  }\href {\doibase 10.1140/epjc/s10052-021-08827-0} {\bibfield  {journal}
  {\bibinfo  {journal} {Eur. Phys. J. C}\ }\textbf {\bibinfo {volume} {81}},\
  \bibinfo {pages} {24} (\bibinfo {year} {2021}{\natexlab{a}})},\ \Eprint
  {http://arxiv.org/abs/2101.07340} {arXiv:2101.07340 [hep-ph]} \BibitemShut
  {NoStop}%
\bibitem [{\citenamefont {Mattos}\ \emph
  {et~al.}(2021{\natexlab{b}})\citenamefont {Mattos}, \citenamefont
  {Frederico}, \citenamefont {Lenzi}, \citenamefont {Dutra},\ and\
  \citenamefont {Louren\c{c}o}}]{Mattos:2021tmz}%
  \BibitemOpen
  \bibfield  {author} {\bibinfo {author} {\bibfnamefont {O.~A.}\ \bibnamefont
  {Mattos}}, \bibinfo {author} {\bibfnamefont {T.}~\bibnamefont {Frederico}},
  \bibinfo {author} {\bibfnamefont {C.~H.}\ \bibnamefont {Lenzi}}, \bibinfo
  {author} {\bibfnamefont {M.}~\bibnamefont {Dutra}}, \ and\ \bibinfo {author}
  {\bibfnamefont {O.}~\bibnamefont {Louren\c{c}o}},\ }\href {\doibase
  10.1103/PhysRevD.104.116001} {\bibfield  {journal} {\bibinfo  {journal}
  {Phys. Rev. D}\ }\textbf {\bibinfo {volume} {104}},\ \bibinfo {pages}
  {116001} (\bibinfo {year} {2021}{\natexlab{b}})},\ \Eprint
  {http://arxiv.org/abs/2110.05602} {arXiv:2110.05602 [hep-ph]} \BibitemShut
  {NoStop}%
\bibitem [{\citenamefont {Wang}\ and\ \citenamefont
  {Wen}(2022)}]{Wang:2022xxp}%
  \BibitemOpen
  \bibfield  {author} {\bibinfo {author} {\bibfnamefont {Y.}~\bibnamefont
  {Wang}}\ and\ \bibinfo {author} {\bibfnamefont {X.-J.}\ \bibnamefont {Wen}},\
  }\href {\doibase 10.1103/PhysRevD.105.074034} {\bibfield  {journal} {\bibinfo
   {journal} {Phys. Rev. D}\ }\textbf {\bibinfo {volume} {105}},\ \bibinfo
  {pages} {074034} (\bibinfo {year} {2022})},\ \Eprint
  {http://arxiv.org/abs/2204.06135} {arXiv:2204.06135 [hep-ph]} \BibitemShut
  {NoStop}%
\bibitem [{\citenamefont {Bali}\ \emph
  {et~al.}(2012{\natexlab{b}})\citenamefont {Bali}, \citenamefont {Collins},
  \citenamefont {Deka}, \citenamefont {Glassle}, \citenamefont {Gockeler},
  \citenamefont {Najjar}, \citenamefont {Nobile}, \citenamefont {Pleiter},
  \citenamefont {Schafer},\ and\ \citenamefont {Sternbeck}}]{Bali:2012av}%
  \BibitemOpen
  \bibfield  {author} {\bibinfo {author} {\bibfnamefont {G.~S.}\ \bibnamefont
  {Bali}}, \bibinfo {author} {\bibfnamefont {S.}~\bibnamefont {Collins}},
  \bibinfo {author} {\bibfnamefont {M.}~\bibnamefont {Deka}}, \bibinfo {author}
  {\bibfnamefont {B.}~\bibnamefont {Glassle}}, \bibinfo {author} {\bibfnamefont
  {M.}~\bibnamefont {Gockeler}}, \bibinfo {author} {\bibfnamefont
  {J.}~\bibnamefont {Najjar}}, \bibinfo {author} {\bibfnamefont
  {A.}~\bibnamefont {Nobile}}, \bibinfo {author} {\bibfnamefont
  {D.}~\bibnamefont {Pleiter}}, \bibinfo {author} {\bibfnamefont
  {A.}~\bibnamefont {Schafer}}, \ and\ \bibinfo {author} {\bibfnamefont
  {A.}~\bibnamefont {Sternbeck}},\ }\href {\doibase 10.1103/PhysRevD.86.054504}
  {\bibfield  {journal} {\bibinfo  {journal} {Phys. Rev. D}\ }\textbf {\bibinfo
  {volume} {86}},\ \bibinfo {pages} {054504} (\bibinfo {year}
  {2012}{\natexlab{b}})},\ \Eprint {http://arxiv.org/abs/1207.1110}
  {arXiv:1207.1110 [hep-lat]} \BibitemShut {NoStop}%
\bibitem [{\citenamefont {Bandyopadhyay}\ and\ \citenamefont
  {Farias}(2021)}]{Bandyopadhyay:2020zte}%
  \BibitemOpen
  \bibfield  {author} {\bibinfo {author} {\bibfnamefont {A.}~\bibnamefont
  {Bandyopadhyay}}\ and\ \bibinfo {author} {\bibfnamefont {R.~L.~S.}\
  \bibnamefont {Farias}},\ }\href {\doibase 10.1140/epjs/s11734-021-00023-1}
  {\bibfield  {journal} {\bibinfo  {journal} {Eur. Phys. J. ST}\ }\textbf
  {\bibinfo {volume} {230}},\ \bibinfo {pages} {719} (\bibinfo {year}
  {2021})},\ \Eprint {http://arxiv.org/abs/2003.11054} {arXiv:2003.11054
  [hep-ph]} \BibitemShut {NoStop}%
\bibitem [{\citenamefont {Ferreira}\ \emph
  {et~al.}(2014{\natexlab{a}})\citenamefont {Ferreira}, \citenamefont {Costa},
  \citenamefont {Menezes}, \citenamefont {Provid\^encia},\ and\ \citenamefont
  {Scoccola}}]{Ferreira:2013tba}%
  \BibitemOpen
  \bibfield  {author} {\bibinfo {author} {\bibfnamefont {M.}~\bibnamefont
  {Ferreira}}, \bibinfo {author} {\bibfnamefont {P.}~\bibnamefont {Costa}},
  \bibinfo {author} {\bibfnamefont {D.~P.}\ \bibnamefont {Menezes}}, \bibinfo
  {author} {\bibfnamefont {C.}~\bibnamefont {Provid\^encia}}, \ and\ \bibinfo
  {author} {\bibfnamefont {N.}~\bibnamefont {Scoccola}},\ }\href {\doibase
  10.1103/PhysRevD.89.016002} {\bibfield  {journal} {\bibinfo  {journal} {Phys.
  Rev. D}\ }\textbf {\bibinfo {volume} {89}},\ \bibinfo {pages} {016002}
  (\bibinfo {year} {2014}{\natexlab{a}})},\ \bibinfo {note} {[Addendum:
  Phys.Rev.D 89, 019902 (2014)]},\ \Eprint {http://arxiv.org/abs/1305.4751}
  {arXiv:1305.4751 [hep-ph]} \BibitemShut {NoStop}%
\bibitem [{\citenamefont {Ferreira}\ \emph
  {et~al.}(2014{\natexlab{b}})\citenamefont {Ferreira}, \citenamefont {Costa},
  \citenamefont {Louren\c{c}o}, \citenamefont {Frederico},\ and\ \citenamefont
  {Provid\^encia}}]{Ferreira:2014kpa}%
  \BibitemOpen
  \bibfield  {author} {\bibinfo {author} {\bibfnamefont {M.}~\bibnamefont
  {Ferreira}}, \bibinfo {author} {\bibfnamefont {P.}~\bibnamefont {Costa}},
  \bibinfo {author} {\bibfnamefont {O.}~\bibnamefont {Louren\c{c}o}}, \bibinfo
  {author} {\bibfnamefont {T.}~\bibnamefont {Frederico}}, \ and\ \bibinfo
  {author} {\bibfnamefont {C.}~\bibnamefont {Provid\^encia}},\ }\href {\doibase
  10.1103/PhysRevD.89.116011} {\bibfield  {journal} {\bibinfo  {journal} {Phys.
  Rev. D}\ }\textbf {\bibinfo {volume} {89}},\ \bibinfo {pages} {116011}
  (\bibinfo {year} {2014}{\natexlab{b}})},\ \Eprint
  {http://arxiv.org/abs/1404.5577} {arXiv:1404.5577 [hep-ph]} \BibitemShut
  {NoStop}%
\bibitem [{\citenamefont {Farias}\ \emph {et~al.}(2017)\citenamefont {Farias},
  \citenamefont {Timoteo}, \citenamefont {Avancini}, \citenamefont {Pinto},\
  and\ \citenamefont {Krein}}]{Farias:2016gmy}%
  \BibitemOpen
  \bibfield  {author} {\bibinfo {author} {\bibfnamefont {R.~L.~S.}\
  \bibnamefont {Farias}}, \bibinfo {author} {\bibfnamefont {V.~S.}\
  \bibnamefont {Timoteo}}, \bibinfo {author} {\bibfnamefont {S.~S.}\
  \bibnamefont {Avancini}}, \bibinfo {author} {\bibfnamefont {M.~B.}\
  \bibnamefont {Pinto}}, \ and\ \bibinfo {author} {\bibfnamefont
  {G.}~\bibnamefont {Krein}},\ }\href {\doibase 10.1140/epja/i2017-12320-8}
  {\bibfield  {journal} {\bibinfo  {journal} {Eur. Phys. J. A}\ }\textbf
  {\bibinfo {volume} {53}},\ \bibinfo {pages} {101} (\bibinfo {year} {2017})},\
  \Eprint {http://arxiv.org/abs/1603.03847} {arXiv:1603.03847 [hep-ph]}
  \BibitemShut {NoStop}%
\bibitem [{\citenamefont {Avancini}\ \emph {et~al.}(2019)\citenamefont
  {Avancini}, \citenamefont {Farias},\ and\ \citenamefont
  {Tavares}}]{Avancini:2018svs}%
  \BibitemOpen
  \bibfield  {author} {\bibinfo {author} {\bibfnamefont {S.~S.}\ \bibnamefont
  {Avancini}}, \bibinfo {author} {\bibfnamefont {R.~L.}\ \bibnamefont
  {Farias}}, \ and\ \bibinfo {author} {\bibfnamefont {W.~R.}\ \bibnamefont
  {Tavares}},\ }\href {\doibase 10.1103/PhysRevD.99.056009} {\bibfield
  {journal} {\bibinfo  {journal} {Phys. Rev. D}\ }\textbf {\bibinfo {volume}
  {99}},\ \bibinfo {pages} {056009} (\bibinfo {year} {2019})},\ \Eprint
  {http://arxiv.org/abs/1812.00945} {arXiv:1812.00945 [hep-ph]} \BibitemShut
  {NoStop}%
\bibitem [{\citenamefont {Sheng}\ \emph {et~al.}(2022)\citenamefont {Sheng},
  \citenamefont {Wang},\ and\ \citenamefont {Yu}}]{Sheng:2021evj}%
  \BibitemOpen
  \bibfield  {author} {\bibinfo {author} {\bibfnamefont {B.-k.}\ \bibnamefont
  {Sheng}}, \bibinfo {author} {\bibfnamefont {X.}~\bibnamefont {Wang}}, \ and\
  \bibinfo {author} {\bibfnamefont {L.}~\bibnamefont {Yu}},\ }\href {\doibase
  10.1103/PhysRevD.105.034003} {\bibfield  {journal} {\bibinfo  {journal}
  {Phys. Rev. D}\ }\textbf {\bibinfo {volume} {105}},\ \bibinfo {pages}
  {034003} (\bibinfo {year} {2022})},\ \Eprint
  {http://arxiv.org/abs/2110.12811} {arXiv:2110.12811 [hep-ph]} \BibitemShut
  {NoStop}%
\bibitem [{\citenamefont {Mao}(2016)}]{Mao:2016fha}%
  \BibitemOpen
  \bibfield  {author} {\bibinfo {author} {\bibfnamefont {S.}~\bibnamefont
  {Mao}},\ }\href {\doibase 10.1016/j.physletb.2016.05.018} {\bibfield
  {journal} {\bibinfo  {journal} {Phys. Lett. B}\ }\textbf {\bibinfo {volume}
  {758}},\ \bibinfo {pages} {195} (\bibinfo {year} {2016})},\ \Eprint
  {http://arxiv.org/abs/1602.06503} {arXiv:1602.06503 [hep-ph]} \BibitemShut
  {NoStop}%
\bibitem [{\citenamefont {Fayazbakhsh}\ and\ \citenamefont
  {Sadooghi}(2014)}]{Fayazbakhsh:2014mca}%
  \BibitemOpen
  \bibfield  {author} {\bibinfo {author} {\bibfnamefont {S.}~\bibnamefont
  {Fayazbakhsh}}\ and\ \bibinfo {author} {\bibfnamefont {N.}~\bibnamefont
  {Sadooghi}},\ }\href {\doibase 10.1103/PhysRevD.90.105030} {\bibfield
  {journal} {\bibinfo  {journal} {Phys. Rev. D}\ }\textbf {\bibinfo {volume}
  {90}},\ \bibinfo {pages} {105030} (\bibinfo {year} {2014})},\ \Eprint
  {http://arxiv.org/abs/1408.5457} {arXiv:1408.5457 [hep-ph]} \BibitemShut
  {NoStop}%
\bibitem [{\citenamefont {Chaudhuri}\ \emph {et~al.}(2019)\citenamefont
  {Chaudhuri}, \citenamefont {Ghosh}, \citenamefont {Sarkar},\ and\
  \citenamefont {Roy}}]{Chaudhuri:2019lbw}%
  \BibitemOpen
  \bibfield  {author} {\bibinfo {author} {\bibfnamefont {N.}~\bibnamefont
  {Chaudhuri}}, \bibinfo {author} {\bibfnamefont {S.}~\bibnamefont {Ghosh}},
  \bibinfo {author} {\bibfnamefont {S.}~\bibnamefont {Sarkar}}, \ and\ \bibinfo
  {author} {\bibfnamefont {P.}~\bibnamefont {Roy}},\ }\href {\doibase
  10.1103/PhysRevD.99.116025} {\bibfield  {journal} {\bibinfo  {journal} {Phys.
  Rev.}\ }\textbf {\bibinfo {volume} {D99}},\ \bibinfo {pages} {116025}
  (\bibinfo {year} {2019})},\ \Eprint {http://arxiv.org/abs/1907.03990}
  {arXiv:1907.03990 [nucl-th]} \BibitemShut {NoStop}%
\bibitem [{\citenamefont {Chaudhuri}\ \emph
  {et~al.}(2021{\natexlab{a}})\citenamefont {Chaudhuri}, \citenamefont {Ghosh},
  \citenamefont {Sarkar},\ and\ \citenamefont {Roy}}]{Chaudhuri:2021skc}%
  \BibitemOpen
  \bibfield  {author} {\bibinfo {author} {\bibfnamefont {N.}~\bibnamefont
  {Chaudhuri}}, \bibinfo {author} {\bibfnamefont {S.}~\bibnamefont {Ghosh}},
  \bibinfo {author} {\bibfnamefont {S.}~\bibnamefont {Sarkar}}, \ and\ \bibinfo
  {author} {\bibfnamefont {P.}~\bibnamefont {Roy}},\ }\href {\doibase
  10.1103/PhysRevD.103.096021} {\bibfield  {journal} {\bibinfo  {journal}
  {Phys. Rev. D}\ }\textbf {\bibinfo {volume} {103}},\ \bibinfo {pages}
  {096021} (\bibinfo {year} {2021}{\natexlab{a}})},\ \Eprint
  {http://arxiv.org/abs/2104.11425} {arXiv:2104.11425 [hep-ph]} \BibitemShut
  {NoStop}%
\bibitem [{\citenamefont {Chaudhuri}\ \emph
  {et~al.}(2021{\natexlab{b}})\citenamefont {Chaudhuri}, \citenamefont
  {Mukherjee}, \citenamefont {Ghosh}, \citenamefont {Sarkar},\ and\
  \citenamefont {Roy}}]{Chaudhuri:2021lui}%
  \BibitemOpen
  \bibfield  {author} {\bibinfo {author} {\bibfnamefont {N.}~\bibnamefont
  {Chaudhuri}}, \bibinfo {author} {\bibfnamefont {A.}~\bibnamefont
  {Mukherjee}}, \bibinfo {author} {\bibfnamefont {S.}~\bibnamefont {Ghosh}},
  \bibinfo {author} {\bibfnamefont {S.}~\bibnamefont {Sarkar}}, \ and\ \bibinfo
  {author} {\bibfnamefont {P.}~\bibnamefont {Roy}},\ }\href@noop {} {\
  (\bibinfo {year} {2021}{\natexlab{b}})},\ \Eprint
  {http://arxiv.org/abs/2111.12058} {arXiv:2111.12058 [hep-ph]} \BibitemShut
  {NoStop}%
\bibitem [{\citenamefont {Ghosh}\ \emph
  {et~al.}(2020{\natexlab{a}})\citenamefont {Ghosh}, \citenamefont {Chaudhuri},
  \citenamefont {Sarkar},\ and\ \citenamefont {Roy}}]{Ghosh:2020xwp}%
  \BibitemOpen
  \bibfield  {author} {\bibinfo {author} {\bibfnamefont {S.}~\bibnamefont
  {Ghosh}}, \bibinfo {author} {\bibfnamefont {N.}~\bibnamefont {Chaudhuri}},
  \bibinfo {author} {\bibfnamefont {S.}~\bibnamefont {Sarkar}}, \ and\ \bibinfo
  {author} {\bibfnamefont {P.}~\bibnamefont {Roy}},\ }\href {\doibase
  10.1103/PhysRevD.101.096002} {\bibfield  {journal} {\bibinfo  {journal}
  {Phys. Rev. D}\ }\textbf {\bibinfo {volume} {101}},\ \bibinfo {pages}
  {096002} (\bibinfo {year} {2020}{\natexlab{a}})},\ \Eprint
  {http://arxiv.org/abs/2004.09203} {arXiv:2004.09203 [nucl-th]} \BibitemShut
  {NoStop}%
\bibitem [{\citenamefont {Xu}\ \emph {et~al.}(2021)\citenamefont {Xu},
  \citenamefont {Chao},\ and\ \citenamefont {Huang}}]{Xu:2020yag}%
  \BibitemOpen
  \bibfield  {author} {\bibinfo {author} {\bibfnamefont {K.}~\bibnamefont
  {Xu}}, \bibinfo {author} {\bibfnamefont {J.}~\bibnamefont {Chao}}, \ and\
  \bibinfo {author} {\bibfnamefont {M.}~\bibnamefont {Huang}},\ }\href
  {\doibase 10.1103/PhysRevD.103.076015} {\bibfield  {journal} {\bibinfo
  {journal} {Phys. Rev. D}\ }\textbf {\bibinfo {volume} {103}},\ \bibinfo
  {pages} {076015} (\bibinfo {year} {2021})},\ \Eprint
  {http://arxiv.org/abs/2007.13122} {arXiv:2007.13122 [hep-ph]} \BibitemShut
  {NoStop}%
\bibitem [{\citenamefont {Mei}\ and\ \citenamefont {Mao}(2020)}]{Mei:2020jzn}%
  \BibitemOpen
  \bibfield  {author} {\bibinfo {author} {\bibfnamefont {J.}~\bibnamefont
  {Mei}}\ and\ \bibinfo {author} {\bibfnamefont {S.}~\bibnamefont {Mao}},\
  }\href {\doibase 10.1103/PhysRevD.102.114035} {\bibfield  {journal} {\bibinfo
   {journal} {Phys. Rev. D}\ }\textbf {\bibinfo {volume} {102}},\ \bibinfo
  {pages} {114035} (\bibinfo {year} {2020})},\ \Eprint
  {http://arxiv.org/abs/2008.12123} {arXiv:2008.12123 [hep-ph]} \BibitemShut
  {NoStop}%
\bibitem [{\citenamefont {Aguirre}(2021)}]{Aguirre:2021ljk}%
  \BibitemOpen
  \bibfield  {author} {\bibinfo {author} {\bibfnamefont {R.~M.}\ \bibnamefont
  {Aguirre}},\ }\href {\doibase 10.1140/epja/s10050-021-00480-1} {\bibfield
  {journal} {\bibinfo  {journal} {Eur. Phys. J. A}\ }\textbf {\bibinfo {volume}
  {57}},\ \bibinfo {pages} {166} (\bibinfo {year} {2021})}\BibitemShut
  {NoStop}%
\bibitem [{\citenamefont {Ghosh}\ \emph {et~al.}(2021)\citenamefont {Ghosh},
  \citenamefont {Chaudhuri}, \citenamefont {Roy},\ and\ \citenamefont
  {Sarkar}}]{Ghosh:2021dlo}%
  \BibitemOpen
  \bibfield  {author} {\bibinfo {author} {\bibfnamefont {S.}~\bibnamefont
  {Ghosh}}, \bibinfo {author} {\bibfnamefont {N.}~\bibnamefont {Chaudhuri}},
  \bibinfo {author} {\bibfnamefont {P.}~\bibnamefont {Roy}}, \ and\ \bibinfo
  {author} {\bibfnamefont {S.}~\bibnamefont {Sarkar}},\ }\href {\doibase
  10.1103/PhysRevD.103.116008} {\bibfield  {journal} {\bibinfo  {journal}
  {Phys. Rev. D}\ }\textbf {\bibinfo {volume} {103}},\ \bibinfo {pages}
  {116008} (\bibinfo {year} {2021})},\ \Eprint
  {http://arxiv.org/abs/2104.14112} {arXiv:2104.14112 [hep-ph]} \BibitemShut
  {NoStop}%
\bibitem [{\citenamefont {Farias}\ \emph {et~al.}(2021)\citenamefont {Farias},
  \citenamefont {Tavares}, \citenamefont {Nunes},\ and\ \citenamefont
  {Avancini}}]{Farias:2021fci}%
  \BibitemOpen
  \bibfield  {author} {\bibinfo {author} {\bibfnamefont {R.~L.~S.}\
  \bibnamefont {Farias}}, \bibinfo {author} {\bibfnamefont {W.~R.}\
  \bibnamefont {Tavares}}, \bibinfo {author} {\bibfnamefont {R.~M.}\
  \bibnamefont {Nunes}}, \ and\ \bibinfo {author} {\bibfnamefont {S.~S.}\
  \bibnamefont {Avancini}},\ }\href@noop {} {\  (\bibinfo {year} {2021})},\
  \Eprint {http://arxiv.org/abs/2109.11112} {arXiv:2109.11112 [hep-ph]}
  \BibitemShut {NoStop}%
\bibitem [{\citenamefont {Chatterjee}\ \emph {et~al.}(2015)\citenamefont
  {Chatterjee}, \citenamefont {Elghozi}, \citenamefont {Novak},\ and\
  \citenamefont {Oertel}}]{Chatterjee:2014qsa}%
  \BibitemOpen
  \bibfield  {author} {\bibinfo {author} {\bibfnamefont {D.}~\bibnamefont
  {Chatterjee}}, \bibinfo {author} {\bibfnamefont {T.}~\bibnamefont {Elghozi}},
  \bibinfo {author} {\bibfnamefont {J.}~\bibnamefont {Novak}}, \ and\ \bibinfo
  {author} {\bibfnamefont {M.}~\bibnamefont {Oertel}},\ }\href {\doibase
  10.1093/mnras/stu2706} {\bibfield  {journal} {\bibinfo  {journal} {Mon. Not.
  Roy. Astron. Soc.}\ }\textbf {\bibinfo {volume} {447}},\ \bibinfo {pages}
  {3785} (\bibinfo {year} {2015})},\ \Eprint {http://arxiv.org/abs/1410.6332}
  {arXiv:1410.6332 [astro-ph.HE]} \BibitemShut {NoStop}%
\bibitem [{\citenamefont {Bali}\ \emph {et~al.}(2013)\citenamefont {Bali},
  \citenamefont {Bruckmann}, \citenamefont {Endrodi}, \citenamefont {Gruber},\
  and\ \citenamefont {Schaefer}}]{Bali:2013esa}%
  \BibitemOpen
  \bibfield  {author} {\bibinfo {author} {\bibfnamefont {G.~S.}\ \bibnamefont
  {Bali}}, \bibinfo {author} {\bibfnamefont {F.}~\bibnamefont {Bruckmann}},
  \bibinfo {author} {\bibfnamefont {G.}~\bibnamefont {Endrodi}}, \bibinfo
  {author} {\bibfnamefont {F.}~\bibnamefont {Gruber}}, \ and\ \bibinfo {author}
  {\bibfnamefont {A.}~\bibnamefont {Schaefer}},\ }\href {\doibase
  10.1007/JHEP04(2013)130} {\bibfield  {journal} {\bibinfo  {journal} {JHEP}\
  }\textbf {\bibinfo {volume} {04}},\ \bibinfo {pages} {130} (\bibinfo {year}
  {2013})},\ \Eprint {http://arxiv.org/abs/1303.1328} {arXiv:1303.1328
  [hep-lat]} \BibitemShut {NoStop}%
\bibitem [{\citenamefont {Canuto}\ and\ \citenamefont
  {Chiu}(1968)}]{Canuto:1968apg}%
  \BibitemOpen
  \bibfield  {author} {\bibinfo {author} {\bibfnamefont {V.}~\bibnamefont
  {Canuto}}\ and\ \bibinfo {author} {\bibfnamefont {H.~Y.}\ \bibnamefont
  {Chiu}},\ }\href {\doibase 10.1103/PhysRev.173.1210} {\bibfield  {journal}
  {\bibinfo  {journal} {Phys. Rev.}\ }\textbf {\bibinfo {volume} {173}},\
  \bibinfo {pages} {1210} (\bibinfo {year} {1968})}\BibitemShut {NoStop}%
\bibitem [{\citenamefont {Martinez}\ \emph {et~al.}(2003)\citenamefont
  {Martinez}, \citenamefont {Rojas},\ and\ \citenamefont
  {Mosquera~Cuesta}}]{Martinez:2003dz}%
  \BibitemOpen
  \bibfield  {author} {\bibinfo {author} {\bibfnamefont {A.~P.}\ \bibnamefont
  {Martinez}}, \bibinfo {author} {\bibfnamefont {H.~P.}\ \bibnamefont {Rojas}},
  \ and\ \bibinfo {author} {\bibfnamefont {H.~J.}\ \bibnamefont
  {Mosquera~Cuesta}},\ }\href {\doibase 10.1140/epjc/s2003-01192-6} {\bibfield
  {journal} {\bibinfo  {journal} {Eur. Phys. J. C}\ }\textbf {\bibinfo {volume}
  {29}},\ \bibinfo {pages} {111} (\bibinfo {year} {2003})},\ \Eprint
  {http://arxiv.org/abs/astro-ph/0303213} {arXiv:astro-ph/0303213} \BibitemShut
  {NoStop}%
\bibitem [{\citenamefont {Noronha}\ and\ \citenamefont
  {Shovkovy}(2007)}]{Noronha:2007wg}%
  \BibitemOpen
  \bibfield  {author} {\bibinfo {author} {\bibfnamefont {J.~L.}\ \bibnamefont
  {Noronha}}\ and\ \bibinfo {author} {\bibfnamefont {I.~A.}\ \bibnamefont
  {Shovkovy}},\ }\href {\doibase 10.1103/PhysRevD.76.105030,
  10.1103/PhysRevD.86.049901} {\bibfield  {journal} {\bibinfo  {journal} {Phys.
  Rev.}\ }\textbf {\bibinfo {volume} {D76}},\ \bibinfo {pages} {105030}
  (\bibinfo {year} {2007})},\ \bibinfo {note} {[Erratum: Phys.
  Rev.D86,049901(2012)]},\ \Eprint {http://arxiv.org/abs/0708.0307}
  {arXiv:0708.0307 [hep-ph]} \BibitemShut {NoStop}%
\bibitem [{\citenamefont {Huang}\ \emph {et~al.}(2010)\citenamefont {Huang},
  \citenamefont {Huang}, \citenamefont {Rischke},\ and\ \citenamefont
  {Sedrakian}}]{Huang:2009ue}%
  \BibitemOpen
  \bibfield  {author} {\bibinfo {author} {\bibfnamefont {X.-G.}\ \bibnamefont
  {Huang}}, \bibinfo {author} {\bibfnamefont {M.}~\bibnamefont {Huang}},
  \bibinfo {author} {\bibfnamefont {D.~H.}\ \bibnamefont {Rischke}}, \ and\
  \bibinfo {author} {\bibfnamefont {A.}~\bibnamefont {Sedrakian}},\ }\href
  {\doibase 10.1103/PhysRevD.81.045015} {\bibfield  {journal} {\bibinfo
  {journal} {Phys. Rev. D}\ }\textbf {\bibinfo {volume} {81}},\ \bibinfo
  {pages} {045015} (\bibinfo {year} {2010})},\ \Eprint
  {http://arxiv.org/abs/0910.3633} {arXiv:0910.3633 [astro-ph.HE]} \BibitemShut
  {NoStop}%
\bibitem [{\citenamefont {Ferrer}\ \emph {et~al.}(2010)\citenamefont {Ferrer},
  \citenamefont {de~la Incera}, \citenamefont {Keith}, \citenamefont
  {Portillo},\ and\ \citenamefont {Springsteen}}]{Ferrer:2010wz}%
  \BibitemOpen
  \bibfield  {author} {\bibinfo {author} {\bibfnamefont {E.~J.}\ \bibnamefont
  {Ferrer}}, \bibinfo {author} {\bibfnamefont {V.}~\bibnamefont {de~la
  Incera}}, \bibinfo {author} {\bibfnamefont {J.~P.}\ \bibnamefont {Keith}},
  \bibinfo {author} {\bibfnamefont {I.}~\bibnamefont {Portillo}}, \ and\
  \bibinfo {author} {\bibfnamefont {P.~L.}\ \bibnamefont {Springsteen}},\
  }\href {\doibase 10.1103/PhysRevC.82.065802} {\bibfield  {journal} {\bibinfo
  {journal} {Phys. Rev. C}\ }\textbf {\bibinfo {volume} {82}},\ \bibinfo
  {pages} {065802} (\bibinfo {year} {2010})},\ \Eprint
  {http://arxiv.org/abs/1009.3521} {arXiv:1009.3521 [hep-ph]} \BibitemShut
  {NoStop}%
\bibitem [{\citenamefont {Strickland}\ \emph {et~al.}(2012)\citenamefont
  {Strickland}, \citenamefont {Dexheimer},\ and\ \citenamefont
  {Menezes}}]{Strickland:2012vu}%
  \BibitemOpen
  \bibfield  {author} {\bibinfo {author} {\bibfnamefont {M.}~\bibnamefont
  {Strickland}}, \bibinfo {author} {\bibfnamefont {V.}~\bibnamefont
  {Dexheimer}}, \ and\ \bibinfo {author} {\bibfnamefont {D.~P.}\ \bibnamefont
  {Menezes}},\ }\href {\doibase 10.1103/PhysRevD.86.125032} {\bibfield
  {journal} {\bibinfo  {journal} {Phys. Rev. D}\ }\textbf {\bibinfo {volume}
  {86}},\ \bibinfo {pages} {125032} (\bibinfo {year} {2012})},\ \Eprint
  {http://arxiv.org/abs/1209.3276} {arXiv:1209.3276 [nucl-th]} \BibitemShut
  {NoStop}%
\bibitem [{\citenamefont {Dexheimer}\ \emph {et~al.}(2014)\citenamefont
  {Dexheimer}, \citenamefont {Menezes},\ and\ \citenamefont
  {Strickland}}]{Dexheimer:2012mk}%
  \BibitemOpen
  \bibfield  {author} {\bibinfo {author} {\bibfnamefont {V.}~\bibnamefont
  {Dexheimer}}, \bibinfo {author} {\bibfnamefont {D.~P.}\ \bibnamefont
  {Menezes}}, \ and\ \bibinfo {author} {\bibfnamefont {M.}~\bibnamefont
  {Strickland}},\ }\href {\doibase 10.1088/0954-3899/41/1/015203} {\bibfield
  {journal} {\bibinfo  {journal} {J. Phys. G}\ }\textbf {\bibinfo {volume}
  {41}},\ \bibinfo {pages} {015203} (\bibinfo {year} {2014})},\ \Eprint
  {http://arxiv.org/abs/1210.4526} {arXiv:1210.4526 [nucl-th]} \BibitemShut
  {NoStop}%
\bibitem [{\citenamefont {Sinha}\ \emph {et~al.}(2013)\citenamefont {Sinha},
  \citenamefont {Huang},\ and\ \citenamefont {Sedrakian}}]{Sinha:2013dfa}%
  \BibitemOpen
  \bibfield  {author} {\bibinfo {author} {\bibfnamefont {M.}~\bibnamefont
  {Sinha}}, \bibinfo {author} {\bibfnamefont {X.-G.}\ \bibnamefont {Huang}}, \
  and\ \bibinfo {author} {\bibfnamefont {A.}~\bibnamefont {Sedrakian}},\ }\href
  {\doibase 10.1103/PhysRevD.88.025008} {\bibfield  {journal} {\bibinfo
  {journal} {Phys. Rev. D}\ }\textbf {\bibinfo {volume} {88}},\ \bibinfo
  {pages} {025008} (\bibinfo {year} {2013})},\ \Eprint
  {http://arxiv.org/abs/1306.3300} {arXiv:1306.3300 [astro-ph.HE]} \BibitemShut
  {NoStop}%
\bibitem [{\citenamefont {Peres~Menezes}\ and\ \citenamefont
  {La\'ercio~Lopes}(2016)}]{PeresMenezes:2015ukv}%
  \BibitemOpen
  \bibfield  {author} {\bibinfo {author} {\bibfnamefont {D.}~\bibnamefont
  {Peres~Menezes}}\ and\ \bibinfo {author} {\bibfnamefont {L.}~\bibnamefont
  {La\'ercio~Lopes}},\ }\href {\doibase 10.1140/epja/i2016-16017-2} {\bibfield
  {journal} {\bibinfo  {journal} {Eur. Phys. J. A}\ }\textbf {\bibinfo {volume}
  {52}},\ \bibinfo {pages} {17} (\bibinfo {year} {2016})},\ \Eprint
  {http://arxiv.org/abs/1505.06714} {arXiv:1505.06714 [nucl-th]} \BibitemShut
  {NoStop}%
\bibitem [{\citenamefont {Menezes}\ \emph {et~al.}(2015)\citenamefont
  {Menezes}, \citenamefont {Pinto},\ and\ \citenamefont
  {Provid\^encia}}]{Menezes:2015fla}%
  \BibitemOpen
  \bibfield  {author} {\bibinfo {author} {\bibfnamefont {D.~P.}\ \bibnamefont
  {Menezes}}, \bibinfo {author} {\bibfnamefont {M.~B.}\ \bibnamefont {Pinto}},
  \ and\ \bibinfo {author} {\bibfnamefont {C.}~\bibnamefont {Provid\^encia}},\
  }\href {\doibase 10.1103/PhysRevC.91.065205} {\bibfield  {journal} {\bibinfo
  {journal} {Phys. Rev. C}\ }\textbf {\bibinfo {volume} {91}},\ \bibinfo
  {pages} {065205} (\bibinfo {year} {2015})},\ \Eprint
  {http://arxiv.org/abs/1503.08666} {arXiv:1503.08666 [hep-ph]} \BibitemShut
  {NoStop}%
\bibitem [{\citenamefont {Ferrer}\ \emph {et~al.}(2015)\citenamefont {Ferrer},
  \citenamefont {de~la Incera}, \citenamefont {Manreza~Paret}, \citenamefont
  {P\'erez~Mart\'\i{}nez},\ and\ \citenamefont {Sanchez}}]{Ferrer:2015wca}%
  \BibitemOpen
  \bibfield  {author} {\bibinfo {author} {\bibfnamefont {E.~J.}\ \bibnamefont
  {Ferrer}}, \bibinfo {author} {\bibfnamefont {V.}~\bibnamefont {de~la
  Incera}}, \bibinfo {author} {\bibfnamefont {D.}~\bibnamefont
  {Manreza~Paret}}, \bibinfo {author} {\bibfnamefont {A.}~\bibnamefont
  {P\'erez~Mart\'\i{}nez}}, \ and\ \bibinfo {author} {\bibfnamefont
  {A.}~\bibnamefont {Sanchez}},\ }\href {\doibase 10.1103/PhysRevD.91.085041}
  {\bibfield  {journal} {\bibinfo  {journal} {Phys. Rev. D}\ }\textbf {\bibinfo
  {volume} {91}},\ \bibinfo {pages} {085041} (\bibinfo {year} {2015})},\
  \Eprint {http://arxiv.org/abs/1501.06616} {arXiv:1501.06616 [hep-ph]}
  \BibitemShut {NoStop}%
\bibitem [{\citenamefont {Avancini}\ \emph {et~al.}(2018)\citenamefont
  {Avancini}, \citenamefont {Dexheimer}, \citenamefont {Farias},\ and\
  \citenamefont {Tim\'oteo}}]{Avancini:2017gck}%
  \BibitemOpen
  \bibfield  {author} {\bibinfo {author} {\bibfnamefont {S.~S.}\ \bibnamefont
  {Avancini}}, \bibinfo {author} {\bibfnamefont {V.}~\bibnamefont {Dexheimer}},
  \bibinfo {author} {\bibfnamefont {R.~L.~S.}\ \bibnamefont {Farias}}, \ and\
  \bibinfo {author} {\bibfnamefont {V.~S.}\ \bibnamefont {Tim\'oteo}},\ }\href
  {\doibase 10.1103/PhysRevC.97.035207} {\bibfield  {journal} {\bibinfo
  {journal} {Phys. Rev. C}\ }\textbf {\bibinfo {volume} {97}},\ \bibinfo
  {pages} {035207} (\bibinfo {year} {2018})},\ \Eprint
  {http://arxiv.org/abs/1709.02774} {arXiv:1709.02774 [hep-ph]} \BibitemShut
  {NoStop}%
\bibitem [{\citenamefont {Ferrer}\ and\ \citenamefont
  {Hackebill}(2019)}]{Ferrer:2019xlr}%
  \BibitemOpen
  \bibfield  {author} {\bibinfo {author} {\bibfnamefont {E.~J.}\ \bibnamefont
  {Ferrer}}\ and\ \bibinfo {author} {\bibfnamefont {A.}~\bibnamefont
  {Hackebill}},\ }\href {\doibase 10.1103/PhysRevC.99.065803} {\bibfield
  {journal} {\bibinfo  {journal} {Phys. Rev. C}\ }\textbf {\bibinfo {volume}
  {99}},\ \bibinfo {pages} {065803} (\bibinfo {year} {2019})},\ \Eprint
  {http://arxiv.org/abs/1903.08224} {arXiv:1903.08224 [nucl-th]} \BibitemShut
  {NoStop}%
\bibitem [{\citenamefont {Blandford}\ and\ \citenamefont
  {Hernquist}(1982)}]{Blandford_1982}%
  \BibitemOpen
  \bibfield  {author} {\bibinfo {author} {\bibfnamefont {R.~D.}\ \bibnamefont
  {Blandford}}\ and\ \bibinfo {author} {\bibfnamefont {L.}~\bibnamefont
  {Hernquist}},\ }\href {\doibase 10.1088/0022-3719/15/30/017} {\bibfield
  {journal} {\bibinfo  {journal} {Journal of Physics C: Solid State Physics}\
  }\textbf {\bibinfo {volume} {15}},\ \bibinfo {pages} {6233} (\bibinfo {year}
  {1982})}\BibitemShut {NoStop}%
\bibitem [{\citenamefont {Potekhin}\ and\ \citenamefont
  {Yakovlev}(2012)}]{Potekhin:2011eb}%
  \BibitemOpen
  \bibfield  {author} {\bibinfo {author} {\bibfnamefont {A.~Y.}\ \bibnamefont
  {Potekhin}}\ and\ \bibinfo {author} {\bibfnamefont {D.~G.}\ \bibnamefont
  {Yakovlev}},\ }\href {\doibase 10.1103/PhysRevC.85.039801} {\bibfield
  {journal} {\bibinfo  {journal} {Phys. Rev. C}\ }\textbf {\bibinfo {volume}
  {85}},\ \bibinfo {pages} {039801} (\bibinfo {year} {2012})},\ \Eprint
  {http://arxiv.org/abs/1109.3783} {arXiv:1109.3783 [astro-ph.SR]} \BibitemShut
  {NoStop}%
\bibitem [{\citenamefont {Ferrer}\ and\ \citenamefont {de~la
  Incera}(2012)}]{Ferrer:2011ze}%
  \BibitemOpen
  \bibfield  {author} {\bibinfo {author} {\bibfnamefont {E.~J.}\ \bibnamefont
  {Ferrer}}\ and\ \bibinfo {author} {\bibfnamefont {V.}~\bibnamefont {de~la
  Incera}},\ }\href {\doibase 10.1103/PhysRevC.85.039802} {\bibfield  {journal}
  {\bibinfo  {journal} {Phys. Rev. C}\ }\textbf {\bibinfo {volume} {85}},\
  \bibinfo {pages} {039802} (\bibinfo {year} {2012})},\ \Eprint
  {http://arxiv.org/abs/1110.0420} {arXiv:1110.0420 [astro-ph.SR]} \BibitemShut
  {NoStop}%
\bibitem [{\citenamefont {Huovinen}\ and\ \citenamefont
  {Petreczky}(2010)}]{Huovinen:2009yb}%
  \BibitemOpen
  \bibfield  {author} {\bibinfo {author} {\bibfnamefont {P.}~\bibnamefont
  {Huovinen}}\ and\ \bibinfo {author} {\bibfnamefont {P.}~\bibnamefont
  {Petreczky}},\ }\href {\doibase 10.1016/j.nuclphysa.2010.02.015} {\bibfield
  {journal} {\bibinfo  {journal} {Nucl. Phys. A}\ }\textbf {\bibinfo {volume}
  {837}},\ \bibinfo {pages} {26} (\bibinfo {year} {2010})},\ \Eprint
  {http://arxiv.org/abs/0912.2541} {arXiv:0912.2541 [hep-ph]} \BibitemShut
  {NoStop}%
\bibitem [{\citenamefont {Mamo}(2013)}]{Mamo:2012kqw}%
  \BibitemOpen
  \bibfield  {author} {\bibinfo {author} {\bibfnamefont {K.~A.}\ \bibnamefont
  {Mamo}},\ }\href {\doibase 10.1007/JHEP08(2013)083} {\bibfield  {journal}
  {\bibinfo  {journal} {JHEP}\ }\textbf {\bibinfo {volume} {08}},\ \bibinfo
  {pages} {083} (\bibinfo {year} {2013})},\ \Eprint
  {http://arxiv.org/abs/1210.7428} {arXiv:1210.7428 [hep-th]} \BibitemShut
  {NoStop}%
\bibitem [{\citenamefont {Guenther}\ \emph {et~al.}(2017)\citenamefont
  {Guenther}, \citenamefont {Bellwied}, \citenamefont {Borsanyi}, \citenamefont
  {Fodor}, \citenamefont {Katz}, \citenamefont {Pasztor}, \citenamefont
  {Ratti},\ and\ \citenamefont {Szab\'o}}]{Guenther:2017hnx}%
  \BibitemOpen
  \bibfield  {author} {\bibinfo {author} {\bibfnamefont {J.~N.}\ \bibnamefont
  {Guenther}}, \bibinfo {author} {\bibfnamefont {R.}~\bibnamefont {Bellwied}},
  \bibinfo {author} {\bibfnamefont {S.}~\bibnamefont {Borsanyi}}, \bibinfo
  {author} {\bibfnamefont {Z.}~\bibnamefont {Fodor}}, \bibinfo {author}
  {\bibfnamefont {S.~D.}\ \bibnamefont {Katz}}, \bibinfo {author}
  {\bibfnamefont {A.}~\bibnamefont {Pasztor}}, \bibinfo {author} {\bibfnamefont
  {C.}~\bibnamefont {Ratti}}, \ and\ \bibinfo {author} {\bibfnamefont {K.~K.}\
  \bibnamefont {Szab\'o}},\ }\href {\doibase 10.1016/j.nuclphysa.2017.05.044}
  {\bibfield  {journal} {\bibinfo  {journal} {Nucl. Phys. A}\ }\textbf
  {\bibinfo {volume} {967}},\ \bibinfo {pages} {720} (\bibinfo {year}
  {2017})},\ \Eprint {http://arxiv.org/abs/1607.02493} {arXiv:1607.02493
  [hep-lat]} \BibitemShut {NoStop}%
\bibitem [{\citenamefont {Bazavov}\ \emph {et~al.}(2018)\citenamefont
  {Bazavov}, \citenamefont {Petreczky},\ and\ \citenamefont
  {Weber}}]{Bazavov:2017dsy}%
  \BibitemOpen
  \bibfield  {author} {\bibinfo {author} {\bibfnamefont {A.}~\bibnamefont
  {Bazavov}}, \bibinfo {author} {\bibfnamefont {P.}~\bibnamefont {Petreczky}},
  \ and\ \bibinfo {author} {\bibfnamefont {J.~H.}\ \bibnamefont {Weber}},\
  }\href {\doibase 10.1103/PhysRevD.97.014510} {\bibfield  {journal} {\bibinfo
  {journal} {Phys. Rev. D}\ }\textbf {\bibinfo {volume} {97}},\ \bibinfo
  {pages} {014510} (\bibinfo {year} {2018})},\ \Eprint
  {http://arxiv.org/abs/1710.05024} {arXiv:1710.05024 [hep-lat]} \BibitemShut
  {NoStop}%
\bibitem [{\citenamefont {Bazavov}\ \emph {et~al.}(2017)\citenamefont {Bazavov}
  \emph {et~al.}}]{Bazavov:2017dus}%
  \BibitemOpen
  \bibfield  {author} {\bibinfo {author} {\bibfnamefont {A.}~\bibnamefont
  {Bazavov}} \emph {et~al.},\ }\href {\doibase 10.1103/PhysRevD.95.054504}
  {\bibfield  {journal} {\bibinfo  {journal} {Phys. Rev. D}\ }\textbf {\bibinfo
  {volume} {95}},\ \bibinfo {pages} {054504} (\bibinfo {year} {2017})},\
  \Eprint {http://arxiv.org/abs/1701.04325} {arXiv:1701.04325 [hep-lat]}
  \BibitemShut {NoStop}%
\bibitem [{\citenamefont {Roy}\ \emph {et~al.}(2015)\citenamefont {Roy},
  \citenamefont {Pu}, \citenamefont {Rezzolla},\ and\ \citenamefont
  {Rischke}}]{Roy:2015kma}%
  \BibitemOpen
  \bibfield  {author} {\bibinfo {author} {\bibfnamefont {V.}~\bibnamefont
  {Roy}}, \bibinfo {author} {\bibfnamefont {S.}~\bibnamefont {Pu}}, \bibinfo
  {author} {\bibfnamefont {L.}~\bibnamefont {Rezzolla}}, \ and\ \bibinfo
  {author} {\bibfnamefont {D.}~\bibnamefont {Rischke}},\ }\href {\doibase
  10.1016/j.physletb.2015.08.046} {\bibfield  {journal} {\bibinfo  {journal}
  {Phys. Lett. B}\ }\textbf {\bibinfo {volume} {750}},\ \bibinfo {pages} {45}
  (\bibinfo {year} {2015})},\ \Eprint {http://arxiv.org/abs/1506.06620}
  {arXiv:1506.06620 [nucl-th]} \BibitemShut {NoStop}%
\bibitem [{\citenamefont {Hernandez}\ and\ \citenamefont
  {Kovtun}(2017)}]{Hernandez:2017mch}%
  \BibitemOpen
  \bibfield  {author} {\bibinfo {author} {\bibfnamefont {J.}~\bibnamefont
  {Hernandez}}\ and\ \bibinfo {author} {\bibfnamefont {P.}~\bibnamefont
  {Kovtun}},\ }\href {\doibase 10.1007/JHEP05(2017)001} {\bibfield  {journal}
  {\bibinfo  {journal} {JHEP}\ }\textbf {\bibinfo {volume} {05}},\ \bibinfo
  {pages} {001} (\bibinfo {year} {2017})},\ \Eprint
  {http://arxiv.org/abs/1703.08757} {arXiv:1703.08757 [hep-th]} \BibitemShut
  {NoStop}%
\bibitem [{\citenamefont {Denicol}\ \emph {et~al.}(2018)\citenamefont
  {Denicol}, \citenamefont {Huang}, \citenamefont {Moln\'ar}, \citenamefont
  {Monteiro}, \citenamefont {Niemi}, \citenamefont {Noronha}, \citenamefont
  {Rischke},\ and\ \citenamefont {Wang}}]{Denicol:2018rbw}%
  \BibitemOpen
  \bibfield  {author} {\bibinfo {author} {\bibfnamefont {G.~S.}\ \bibnamefont
  {Denicol}}, \bibinfo {author} {\bibfnamefont {X.-G.}\ \bibnamefont {Huang}},
  \bibinfo {author} {\bibfnamefont {E.}~\bibnamefont {Moln\'ar}}, \bibinfo
  {author} {\bibfnamefont {G.~M.}\ \bibnamefont {Monteiro}}, \bibinfo {author}
  {\bibfnamefont {H.}~\bibnamefont {Niemi}}, \bibinfo {author} {\bibfnamefont
  {J.}~\bibnamefont {Noronha}}, \bibinfo {author} {\bibfnamefont {D.~H.}\
  \bibnamefont {Rischke}}, \ and\ \bibinfo {author} {\bibfnamefont
  {Q.}~\bibnamefont {Wang}},\ }\href {\doibase 10.1103/PhysRevD.98.076009}
  {\bibfield  {journal} {\bibinfo  {journal} {Phys. Rev. D}\ }\textbf {\bibinfo
  {volume} {98}},\ \bibinfo {pages} {076009} (\bibinfo {year} {2018})},\
  \Eprint {http://arxiv.org/abs/1804.05210} {arXiv:1804.05210 [nucl-th]}
  \BibitemShut {NoStop}%
\bibitem [{\citenamefont {Denicol}\ \emph {et~al.}(2019)\citenamefont
  {Denicol}, \citenamefont {Moln\'ar}, \citenamefont {Niemi},\ and\
  \citenamefont {Rischke}}]{Denicol:2019iyh}%
  \BibitemOpen
  \bibfield  {author} {\bibinfo {author} {\bibfnamefont {G.~S.}\ \bibnamefont
  {Denicol}}, \bibinfo {author} {\bibfnamefont {E.}~\bibnamefont {Moln\'ar}},
  \bibinfo {author} {\bibfnamefont {H.}~\bibnamefont {Niemi}}, \ and\ \bibinfo
  {author} {\bibfnamefont {D.~H.}\ \bibnamefont {Rischke}},\ }\href {\doibase
  10.1103/PhysRevD.99.056017} {\bibfield  {journal} {\bibinfo  {journal} {Phys.
  Rev. D}\ }\textbf {\bibinfo {volume} {99}},\ \bibinfo {pages} {056017}
  (\bibinfo {year} {2019})},\ \Eprint {http://arxiv.org/abs/1902.01699}
  {arXiv:1902.01699 [nucl-th]} \BibitemShut {NoStop}%
\bibitem [{\citenamefont {Hattori}\ and\ \citenamefont
  {Satow}(2016)}]{Hattori:2016cnt}%
  \BibitemOpen
  \bibfield  {author} {\bibinfo {author} {\bibfnamefont {K.}~\bibnamefont
  {Hattori}}\ and\ \bibinfo {author} {\bibfnamefont {D.}~\bibnamefont
  {Satow}},\ }\href {\doibase 10.1103/PhysRevD.94.114032} {\bibfield  {journal}
  {\bibinfo  {journal} {Phys. Rev. D}\ }\textbf {\bibinfo {volume} {94}},\
  \bibinfo {pages} {114032} (\bibinfo {year} {2016})},\ \Eprint
  {http://arxiv.org/abs/1610.06818} {arXiv:1610.06818 [hep-ph]} \BibitemShut
  {NoStop}%
\bibitem [{\citenamefont {Feng}(2017)}]{Feng:2017tsh}%
  \BibitemOpen
  \bibfield  {author} {\bibinfo {author} {\bibfnamefont {B.}~\bibnamefont
  {Feng}},\ }\href {\doibase 10.1103/PhysRevD.96.036009} {\bibfield  {journal}
  {\bibinfo  {journal} {Phys. Rev. D}\ }\textbf {\bibinfo {volume} {96}},\
  \bibinfo {pages} {036009} (\bibinfo {year} {2017})}\BibitemShut {NoStop}%
\bibitem [{\citenamefont {Ghosh}\ \emph
  {et~al.}(2020{\natexlab{b}})\citenamefont {Ghosh}, \citenamefont
  {Bandyopadhyay}, \citenamefont {Farias}, \citenamefont {Dey},\ and\
  \citenamefont {Krein}}]{Ghosh:2019ubc}%
  \BibitemOpen
  \bibfield  {author} {\bibinfo {author} {\bibfnamefont {S.}~\bibnamefont
  {Ghosh}}, \bibinfo {author} {\bibfnamefont {A.}~\bibnamefont
  {Bandyopadhyay}}, \bibinfo {author} {\bibfnamefont {R.~L.~S.}\ \bibnamefont
  {Farias}}, \bibinfo {author} {\bibfnamefont {J.}~\bibnamefont {Dey}}, \ and\
  \bibinfo {author} {\bibfnamefont {G.~a.}\ \bibnamefont {Krein}},\ }\href
  {\doibase 10.1103/PhysRevD.102.114015} {\bibfield  {journal} {\bibinfo
  {journal} {Phys. Rev. D}\ }\textbf {\bibinfo {volume} {102}},\ \bibinfo
  {pages} {114015} (\bibinfo {year} {2020}{\natexlab{b}})},\ \Eprint
  {http://arxiv.org/abs/1911.10005} {arXiv:1911.10005 [hep-ph]} \BibitemShut
  {NoStop}%
\bibitem [{\citenamefont {Kalikotay}\ \emph {et~al.}(2020)\citenamefont
  {Kalikotay}, \citenamefont {Ghosh}, \citenamefont {Chaudhuri}, \citenamefont
  {Roy},\ and\ \citenamefont {Sarkar}}]{Kalikotay:2020snc}%
  \BibitemOpen
  \bibfield  {author} {\bibinfo {author} {\bibfnamefont {P.}~\bibnamefont
  {Kalikotay}}, \bibinfo {author} {\bibfnamefont {S.}~\bibnamefont {Ghosh}},
  \bibinfo {author} {\bibfnamefont {N.}~\bibnamefont {Chaudhuri}}, \bibinfo
  {author} {\bibfnamefont {P.}~\bibnamefont {Roy}}, \ and\ \bibinfo {author}
  {\bibfnamefont {S.}~\bibnamefont {Sarkar}},\ }\href {\doibase
  10.1103/PhysRevD.102.076007} {\bibfield  {journal} {\bibinfo  {journal}
  {Phys. Rev. D}\ }\textbf {\bibinfo {volume} {102}},\ \bibinfo {pages}
  {076007} (\bibinfo {year} {2020})},\ \Eprint
  {http://arxiv.org/abs/2009.10493} {arXiv:2009.10493 [hep-ph]} \BibitemShut
  {NoStop}%
\bibitem [{\citenamefont {Tuchin}(2012)}]{Tuchin:2011jw}%
  \BibitemOpen
  \bibfield  {author} {\bibinfo {author} {\bibfnamefont {K.}~\bibnamefont
  {Tuchin}},\ }\href {\doibase 10.1088/0954-3899/39/2/025010} {\bibfield
  {journal} {\bibinfo  {journal} {J. Phys. G}\ }\textbf {\bibinfo {volume}
  {39}},\ \bibinfo {pages} {025010} (\bibinfo {year} {2012})},\ \Eprint
  {http://arxiv.org/abs/1108.4394} {arXiv:1108.4394 [nucl-th]} \BibitemShut
  {NoStop}%
\bibitem [{\citenamefont {Hattori}\ \emph {et~al.}(2017)\citenamefont
  {Hattori}, \citenamefont {Huang}, \citenamefont {Rischke},\ and\
  \citenamefont {Satow}}]{Hattori:2017qih}%
  \BibitemOpen
  \bibfield  {author} {\bibinfo {author} {\bibfnamefont {K.}~\bibnamefont
  {Hattori}}, \bibinfo {author} {\bibfnamefont {X.-G.}\ \bibnamefont {Huang}},
  \bibinfo {author} {\bibfnamefont {D.~H.}\ \bibnamefont {Rischke}}, \ and\
  \bibinfo {author} {\bibfnamefont {D.}~\bibnamefont {Satow}},\ }\href
  {\doibase 10.1103/PhysRevD.96.094009} {\bibfield  {journal} {\bibinfo
  {journal} {Phys. Rev. D}\ }\textbf {\bibinfo {volume} {96}},\ \bibinfo
  {pages} {094009} (\bibinfo {year} {2017})},\ \Eprint
  {http://arxiv.org/abs/1708.00515} {arXiv:1708.00515 [hep-ph]} \BibitemShut
  {NoStop}%
\bibitem [{\citenamefont {Kurian}\ and\ \citenamefont
  {Chandra}(2018)}]{Kurian:2018dbn}%
  \BibitemOpen
  \bibfield  {author} {\bibinfo {author} {\bibfnamefont {M.}~\bibnamefont
  {Kurian}}\ and\ \bibinfo {author} {\bibfnamefont {V.}~\bibnamefont
  {Chandra}},\ }\href {\doibase 10.1103/PhysRevD.97.116008} {\bibfield
  {journal} {\bibinfo  {journal} {Phys. Rev. D}\ }\textbf {\bibinfo {volume}
  {97}},\ \bibinfo {pages} {116008} (\bibinfo {year} {2018})},\ \Eprint
  {http://arxiv.org/abs/1802.07904} {arXiv:1802.07904 [nucl-th]} \BibitemShut
  {NoStop}%
\bibitem [{\citenamefont {Kurian}\ \emph {et~al.}(2019)\citenamefont {Kurian},
  \citenamefont {Mitra}, \citenamefont {Ghosh},\ and\ \citenamefont
  {Chandra}}]{Kurian:2018qwb}%
  \BibitemOpen
  \bibfield  {author} {\bibinfo {author} {\bibfnamefont {M.}~\bibnamefont
  {Kurian}}, \bibinfo {author} {\bibfnamefont {S.}~\bibnamefont {Mitra}},
  \bibinfo {author} {\bibfnamefont {S.}~\bibnamefont {Ghosh}}, \ and\ \bibinfo
  {author} {\bibfnamefont {V.}~\bibnamefont {Chandra}},\ }\href {\doibase
  10.1140/epjc/s10052-019-6649-z} {\bibfield  {journal} {\bibinfo  {journal}
  {Eur. Phys. J. C}\ }\textbf {\bibinfo {volume} {79}},\ \bibinfo {pages} {134}
  (\bibinfo {year} {2019})},\ \Eprint {http://arxiv.org/abs/1805.07313}
  {arXiv:1805.07313 [nucl-th]} \BibitemShut {NoStop}%
\bibitem [{\citenamefont {Kurian}\ \emph {et~al.}(2020)\citenamefont {Kurian},
  \citenamefont {Chandra},\ and\ \citenamefont {Das}}]{Kurian:2020kct}%
  \BibitemOpen
  \bibfield  {author} {\bibinfo {author} {\bibfnamefont {M.}~\bibnamefont
  {Kurian}}, \bibinfo {author} {\bibfnamefont {V.}~\bibnamefont {Chandra}}, \
  and\ \bibinfo {author} {\bibfnamefont {S.~K.}\ \bibnamefont {Das}},\ }\href
  {\doibase 10.1103/PhysRevD.101.094024} {\bibfield  {journal} {\bibinfo
  {journal} {Phys. Rev. D}\ }\textbf {\bibinfo {volume} {101}},\ \bibinfo
  {pages} {094024} (\bibinfo {year} {2020})},\ \Eprint
  {http://arxiv.org/abs/2002.03325} {arXiv:2002.03325 [nucl-th]} \BibitemShut
  {NoStop}%
\bibitem [{\citenamefont {Karmakar}\ \emph {et~al.}(2019)\citenamefont
  {Karmakar}, \citenamefont {Ghosh}, \citenamefont {Bandyopadhyay},
  \citenamefont {Haque},\ and\ \citenamefont {Mustafa}}]{Karmakar:2019tdp}%
  \BibitemOpen
  \bibfield  {author} {\bibinfo {author} {\bibfnamefont {B.}~\bibnamefont
  {Karmakar}}, \bibinfo {author} {\bibfnamefont {R.}~\bibnamefont {Ghosh}},
  \bibinfo {author} {\bibfnamefont {A.}~\bibnamefont {Bandyopadhyay}}, \bibinfo
  {author} {\bibfnamefont {N.}~\bibnamefont {Haque}}, \ and\ \bibinfo {author}
  {\bibfnamefont {M.~G.}\ \bibnamefont {Mustafa}},\ }\href {\doibase
  10.1103/PhysRevD.99.094002} {\bibfield  {journal} {\bibinfo  {journal} {Phys.
  Rev. D}\ }\textbf {\bibinfo {volume} {99}},\ \bibinfo {pages} {094002}
  (\bibinfo {year} {2019})},\ \Eprint {http://arxiv.org/abs/1902.02607}
  {arXiv:1902.02607 [hep-ph]} \BibitemShut {NoStop}%
\bibitem [{\citenamefont {Bali}\ \emph
  {et~al.}(2012{\natexlab{c}})\citenamefont {Bali}, \citenamefont {Bruckmann},
  \citenamefont {Constantinou}, \citenamefont {Costa}, \citenamefont {Endrodi},
  \citenamefont {Katz}, \citenamefont {Panagopoulos},\ and\ \citenamefont
  {Schafer}}]{Bali:2012jv}%
  \BibitemOpen
  \bibfield  {author} {\bibinfo {author} {\bibfnamefont {G.~S.}\ \bibnamefont
  {Bali}}, \bibinfo {author} {\bibfnamefont {F.}~\bibnamefont {Bruckmann}},
  \bibinfo {author} {\bibfnamefont {M.}~\bibnamefont {Constantinou}}, \bibinfo
  {author} {\bibfnamefont {M.}~\bibnamefont {Costa}}, \bibinfo {author}
  {\bibfnamefont {G.}~\bibnamefont {Endrodi}}, \bibinfo {author} {\bibfnamefont
  {S.~D.}\ \bibnamefont {Katz}}, \bibinfo {author} {\bibfnamefont
  {H.}~\bibnamefont {Panagopoulos}}, \ and\ \bibinfo {author} {\bibfnamefont
  {A.}~\bibnamefont {Schafer}},\ }\href {\doibase 10.1103/PhysRevD.86.094512}
  {\bibfield  {journal} {\bibinfo  {journal} {Phys. Rev. D}\ }\textbf {\bibinfo
  {volume} {86}},\ \bibinfo {pages} {094512} (\bibinfo {year}
  {2012}{\natexlab{c}})},\ \Eprint {http://arxiv.org/abs/1209.6015}
  {arXiv:1209.6015 [hep-lat]} \BibitemShut {NoStop}%
\bibitem [{\citenamefont {Bali}\ \emph {et~al.}(2020)\citenamefont {Bali},
  \citenamefont {Endr\H{o}di},\ and\ \citenamefont {Piemonte}}]{Bali:2020bcn}%
  \BibitemOpen
  \bibfield  {author} {\bibinfo {author} {\bibfnamefont {G.~S.}\ \bibnamefont
  {Bali}}, \bibinfo {author} {\bibfnamefont {G.}~\bibnamefont {Endr\H{o}di}}, \
  and\ \bibinfo {author} {\bibfnamefont {S.}~\bibnamefont {Piemonte}},\ }\href
  {\doibase 10.1007/JHEP07(2020)183} {\bibfield  {journal} {\bibinfo  {journal}
  {JHEP}\ }\textbf {\bibinfo {volume} {07}},\ \bibinfo {pages} {183} (\bibinfo
  {year} {2020})},\ \Eprint {http://arxiv.org/abs/2004.08778} {arXiv:2004.08778
  [hep-lat]} \BibitemShut {NoStop}%
\bibitem [{\citenamefont {Lin}\ \emph {et~al.}(2022)\citenamefont {Lin},
  \citenamefont {Xu},\ and\ \citenamefont {Huang}}]{Lin:2022ied}%
  \BibitemOpen
  \bibfield  {author} {\bibinfo {author} {\bibfnamefont {F.}~\bibnamefont
  {Lin}}, \bibinfo {author} {\bibfnamefont {K.}~\bibnamefont {Xu}}, \ and\
  \bibinfo {author} {\bibfnamefont {M.}~\bibnamefont {Huang}},\ }\href@noop {}
  {\  (\bibinfo {year} {2022})},\ \Eprint {http://arxiv.org/abs/2202.03226}
  {arXiv:2202.03226 [hep-ph]} \BibitemShut {NoStop}%
\bibitem [{\citenamefont {Roessner}\ \emph {et~al.}(2007)\citenamefont
  {Roessner}, \citenamefont {Ratti},\ and\ \citenamefont
  {Weise}}]{Roessner:2006xn}%
  \BibitemOpen
  \bibfield  {author} {\bibinfo {author} {\bibfnamefont {S.}~\bibnamefont
  {Roessner}}, \bibinfo {author} {\bibfnamefont {C.}~\bibnamefont {Ratti}}, \
  and\ \bibinfo {author} {\bibfnamefont {W.}~\bibnamefont {Weise}},\ }\href
  {\doibase 10.1103/PhysRevD.75.034007} {\bibfield  {journal} {\bibinfo
  {journal} {Phys. Rev. D}\ }\textbf {\bibinfo {volume} {75}},\ \bibinfo
  {pages} {034007} (\bibinfo {year} {2007})},\ \Eprint
  {http://arxiv.org/abs/hep-ph/0609281} {arXiv:hep-ph/0609281} \BibitemShut
  {NoStop}%
\bibitem [{\citenamefont {Fukushima}\ \emph {et~al.}(2010)\citenamefont
  {Fukushima}, \citenamefont {Ruggieri},\ and\ \citenamefont
  {Gatto}}]{Fukushima:2010fe}%
  \BibitemOpen
  \bibfield  {author} {\bibinfo {author} {\bibfnamefont {K.}~\bibnamefont
  {Fukushima}}, \bibinfo {author} {\bibfnamefont {M.}~\bibnamefont {Ruggieri}},
  \ and\ \bibinfo {author} {\bibfnamefont {R.}~\bibnamefont {Gatto}},\ }\href
  {\doibase 10.1103/PhysRevD.81.114031} {\bibfield  {journal} {\bibinfo
  {journal} {Phys. Rev. D}\ }\textbf {\bibinfo {volume} {81}},\ \bibinfo
  {pages} {114031} (\bibinfo {year} {2010})},\ \Eprint
  {http://arxiv.org/abs/1003.0047} {arXiv:1003.0047 [hep-ph]} \BibitemShut
  {NoStop}%
\bibitem [{\citenamefont {McLerran}\ and\ \citenamefont
  {Pisarski}(2007)}]{McLerran:2007qj}%
  \BibitemOpen
  \bibfield  {author} {\bibinfo {author} {\bibfnamefont {L.}~\bibnamefont
  {McLerran}}\ and\ \bibinfo {author} {\bibfnamefont {R.~D.}\ \bibnamefont
  {Pisarski}},\ }\href {\doibase 10.1016/j.nuclphysa.2007.08.013} {\bibfield
  {journal} {\bibinfo  {journal} {Nucl. Phys. A}\ }\textbf {\bibinfo {volume}
  {796}},\ \bibinfo {pages} {83} (\bibinfo {year} {2007})},\ \Eprint
  {http://arxiv.org/abs/0706.2191} {arXiv:0706.2191 [hep-ph]} \BibitemShut
  {NoStop}%
\bibitem [{\citenamefont {McLerran}\ \emph {et~al.}(2009)\citenamefont
  {McLerran}, \citenamefont {Redlich},\ and\ \citenamefont
  {Sasaki}}]{McLerran:2008ua}%
  \BibitemOpen
  \bibfield  {author} {\bibinfo {author} {\bibfnamefont {L.}~\bibnamefont
  {McLerran}}, \bibinfo {author} {\bibfnamefont {K.}~\bibnamefont {Redlich}}, \
  and\ \bibinfo {author} {\bibfnamefont {C.}~\bibnamefont {Sasaki}},\ }\href
  {\doibase 10.1016/j.nuclphysa.2009.04.001} {\bibfield  {journal} {\bibinfo
  {journal} {Nucl. Phys. A}\ }\textbf {\bibinfo {volume} {824}},\ \bibinfo
  {pages} {86} (\bibinfo {year} {2009})},\ \Eprint
  {http://arxiv.org/abs/0812.3585} {arXiv:0812.3585 [hep-ph]} \BibitemShut
  {NoStop}%
\bibitem [{\citenamefont {Abuki}\ \emph {et~al.}(2008)\citenamefont {Abuki},
  \citenamefont {Anglani}, \citenamefont {Gatto}, \citenamefont {Nardulli},\
  and\ \citenamefont {Ruggieri}}]{Abuki:2008nm}%
  \BibitemOpen
  \bibfield  {author} {\bibinfo {author} {\bibfnamefont {H.}~\bibnamefont
  {Abuki}}, \bibinfo {author} {\bibfnamefont {R.}~\bibnamefont {Anglani}},
  \bibinfo {author} {\bibfnamefont {R.}~\bibnamefont {Gatto}}, \bibinfo
  {author} {\bibfnamefont {G.}~\bibnamefont {Nardulli}}, \ and\ \bibinfo
  {author} {\bibfnamefont {M.}~\bibnamefont {Ruggieri}},\ }\href {\doibase
  10.1103/PhysRevD.78.034034} {\bibfield  {journal} {\bibinfo  {journal} {Phys.
  Rev. D}\ }\textbf {\bibinfo {volume} {78}},\ \bibinfo {pages} {034034}
  (\bibinfo {year} {2008})},\ \Eprint {http://arxiv.org/abs/0805.1509}
  {arXiv:0805.1509 [hep-ph]} \BibitemShut {NoStop}%
\bibitem [{\citenamefont {Hidaka}\ \emph {et~al.}(2008)\citenamefont {Hidaka},
  \citenamefont {McLerran},\ and\ \citenamefont {Pisarski}}]{Hidaka:2008yy}%
  \BibitemOpen
  \bibfield  {author} {\bibinfo {author} {\bibfnamefont {Y.}~\bibnamefont
  {Hidaka}}, \bibinfo {author} {\bibfnamefont {L.~D.}\ \bibnamefont
  {McLerran}}, \ and\ \bibinfo {author} {\bibfnamefont {R.~D.}\ \bibnamefont
  {Pisarski}},\ }\href {\doibase 10.1016/j.nuclphysa.2008.05.009} {\bibfield
  {journal} {\bibinfo  {journal} {Nucl. Phys. A}\ }\textbf {\bibinfo {volume}
  {808}},\ \bibinfo {pages} {117} (\bibinfo {year} {2008})},\ \Eprint
  {http://arxiv.org/abs/0803.0279} {arXiv:0803.0279 [hep-ph]} \BibitemShut
  {NoStop}%
\bibitem [{\citenamefont {Buisseret}\ and\ \citenamefont
  {Lacroix}(2012)}]{Buisseret:2011ms}%
  \BibitemOpen
  \bibfield  {author} {\bibinfo {author} {\bibfnamefont {F.}~\bibnamefont
  {Buisseret}}\ and\ \bibinfo {author} {\bibfnamefont {G.}~\bibnamefont
  {Lacroix}},\ }\href {\doibase 10.1103/PhysRevD.85.016009} {\bibfield
  {journal} {\bibinfo  {journal} {Phys. Rev. D}\ }\textbf {\bibinfo {volume}
  {85}},\ \bibinfo {pages} {016009} (\bibinfo {year} {2012})},\ \Eprint
  {http://arxiv.org/abs/1107.4672} {arXiv:1107.4672 [hep-ph]} \BibitemShut
  {NoStop}%
\bibitem [{\citenamefont {Hansen}\ \emph {et~al.}(2020)\citenamefont {Hansen},
  \citenamefont {Stiele},\ and\ \citenamefont {Costa}}]{Hansen:2019lnf}%
  \BibitemOpen
  \bibfield  {author} {\bibinfo {author} {\bibfnamefont {H.}~\bibnamefont
  {Hansen}}, \bibinfo {author} {\bibfnamefont {R.}~\bibnamefont {Stiele}}, \
  and\ \bibinfo {author} {\bibfnamefont {P.}~\bibnamefont {Costa}},\ }\href
  {\doibase 10.1103/PhysRevD.101.094001} {\bibfield  {journal} {\bibinfo
  {journal} {Phys. Rev. D}\ }\textbf {\bibinfo {volume} {101}},\ \bibinfo
  {pages} {094001} (\bibinfo {year} {2020})},\ \Eprint
  {http://arxiv.org/abs/1904.08965} {arXiv:1904.08965 [hep-ph]} \BibitemShut
  {NoStop}%
\bibitem [{\citenamefont {Yagi}\ \emph {et~al.}(2005)\citenamefont {Yagi},
  \citenamefont {Hatsuda},\ and\ \citenamefont {Miake}}]{Yagi:2005yb}%
  \BibitemOpen
  \bibfield  {author} {\bibinfo {author} {\bibfnamefont {K.}~\bibnamefont
  {Yagi}}, \bibinfo {author} {\bibfnamefont {T.}~\bibnamefont {Hatsuda}}, \
  and\ \bibinfo {author} {\bibfnamefont {Y.}~\bibnamefont {Miake}},\
  }\href@noop {} {\emph {\bibinfo {title} {{Quark-gluon plasma: From big bang
  to little bang}}}},\ Vol.~\bibinfo {volume} {23}\ (\bibinfo {year}
  {2005})\BibitemShut {NoStop}%
\bibitem [{\citenamefont {Letessier}\ and\ \citenamefont
  {Rafelski}(2002)}]{Letessier:2002gp}%
  \BibitemOpen
  \bibfield  {author} {\bibinfo {author} {\bibfnamefont {J.}~\bibnamefont
  {Letessier}}\ and\ \bibinfo {author} {\bibfnamefont {J.}~\bibnamefont
  {Rafelski}},\ }\href@noop {} {\emph {\bibinfo {title} {{Hadrons and quark -
  gluon plasma}}}}\ (\bibinfo  {publisher} {Cambridge University Press},\
  \bibinfo {year} {2002})\BibitemShut {NoStop}%
\end{thebibliography}%

\end{document}